%% file: main.tex
\documentclass[%
 reprint,
 superscriptaddress,
 amsmath,amssymb,
 aps,
 floatfix,
]{revtex4-1}

\usepackage{graphicx}
\usepackage{dcolumn}
\usepackage{bm}
\usepackage{physics}
\usepackage{mathtools}
\usepackage{amsfonts}
\usepackage{amssymb}
\usepackage{amsmath}
\usepackage[caption=false,labelformat=empty]{subfig}
\usepackage{diagbox}
\usepackage{dsfont}
\usepackage{hyperref}
\usepackage{placeins}
\usepackage{xspace}
\hypersetup{linktocpage,colorlinks,citecolor={blue},pdfdisplaydoctitle=true,pdfpagemode=UseOutlines,bookmarksnumbered=true}
\usepackage{tikz}
\usetikzlibrary{positioning,calc,fit}
\usetikzlibrary{mindmap,trees,shadows,arrows}
\usetikzlibrary{fpu}

\usetikzlibrary{shapes}
\usetikzlibrary{shapes.geometric}
\usetikzlibrary{automata}
\usetikzlibrary{decorations.markings}
\usetikzlibrary{patterns}
\usetikzlibrary{fit}
\usetikzlibrary{backgrounds}
\usetikzlibrary{matrix}
\usetikzlibrary{scopes}
\usetikzlibrary{shapes.symbols}
\usetikzlibrary{arrows.meta}
\usetikzlibrary{intersections}
\usetikzlibrary{chains}
\usetikzlibrary{angles}
\usetikzlibrary{quotes}
\usetikzlibrary{snakes}

\newcommand{\comf}{\ensuremath{\text{ComF}}\xspace}
\newcommand{\comsf}{\ensuremath{\text{ComSF}}\xspace}
\newcommand{\comav}{\ensuremath{\text{ComAV}}\xspace}
\newcommand{\jcal}{\ensuremath{\mathcal{J}}}
\newcommand{\dgr}{\ensuremath{^{\dagger}}}
\newcommand{\id}{\ensuremath{\mathds{1}}}
\newcommand\tran{\ensuremath{^T}}

\newcommand*\inv{\ensuremath{^{-1}}}
\newcommand{\nlr}{\ensuremath{\vec{n}_L,\vec{n}_R\,}}
\newcommand{\nf}{\ensuremath{\vec{n}_F}}
\newcommand{\bralr}{\ensuremath{\bra{\nlr}}}
\newcommand{\ketlr}{\ensuremath{\ket{\nlr}}}

\newcommand{\ketf}{\ensuremath{\ket{\nf}}}

\DeclarePairedDelimiter\floor{\lfloor}{\rfloor}

\begin{document}

\title{Reduced density matrix and entanglement in interacting quantum field theory with Hamiltonian truncation}

\author{Patrick Emonts}
\author{Ivan Kukuljan}
\affiliation{Max-Planck Institute of Quantum Optics, Hans-Kopfermann-Str. 1, 85748 Garching, Germany}
\affiliation{Munich Center for Quantum Science and Technology (MCQST), Schellingstr. 4, D-80799 München}

\date{\today}

\begin{abstract}
Entanglement is the fundamental difference between classical and quantum systems and has become one of the guiding principles in the exploration of high- and low-energy physics. The calculation of entanglement entropies in interacting quantum field theories, however, remains challenging.
Here, we present the first method for the explicit computation of reduced density matrices of interacting quantum field theories using truncated Hamiltonian methods.
The method is based on constructing an isomorphism between the Hilbert space of the full system and the tensor product of Hilbert spaces of sub-intervals.
This naturally enables the computation of the von Neumann and arbitrary Rényi entanglement entropies as well as mutual information, logarithmic negativity and other measures of entanglement.
Our method is applicable to equilibrium states and non-equilibrium evolution in real time.
It is model independent and can be applied to any Hamiltonian truncation method that uses a free basis expansion.
We benchmark the method on the free Klein-Gordon theory finding excellent agreement with the  analytic results.
We further demonstrate its potential on the interacting sine-Gordon model, studying the scaling of von Neumann entropy in ground states and real time dynamics following quenches of the model.
\end{abstract}

\maketitle

\section{\label{sec:intro}Introduction}
Entanglement is the key feature that distinguishes quantum from classical
systems and has as such been a central topic of study in the fields
of quantum information, condensed matter and high energy physics.
The interest has been recently stimulated by enormous progress in quantum technologies ranging from quantum communication to quantum computation and quantum simulation~\cite{chen_review_2021}.
There, entanglement plays the role of the main resource that these
technologies build upon. 
Furthermore, entanglement and its scaling has proven to be the underlying reason for the performance of tensor network based methods, arguably, one of the most successful computational tools for many body quantum physics~\cite{verstraete_renormalization_2004,orus_practical_2014,bridgeman_hand-waving_2017}.
Entanglement connects deeply also to fundamental aspects in high energy physics, including black hole physics~\cite{polchinski_black_2017} and holography~\cite{ryu_holographic_2006}.

Consequently, entanglement and its scaling in quantum systems has received a lot of attention in recent years. 
While formally defined through local operations and classical communication~\cite{horodecki_quantum_2009}, there is not a unique measure to quantify entanglement.
Some of the most commonly studied quantities are von~Neumann and Rényi entropies. 

If we split a quantum system whose state is described by the density matrix $\rho$ into two non-overlapping subsystems $A$ and $B$, then the important object of study is the \textit{reduced density matrix}
\begin{align}
    \rho_{A}\equiv\Tr_{B}\rho
\end{align}
which is obtained by tracing out the degrees of freedom in $B$. 
We can define the \textit{von~Neumann (vN) entropy} of subsystem $A$
\begin{align}
    S_A = -\Tr \left(\rho_A \log \rho_A\right)
\end{align}
which satisfies the properties of an entanglement monotone and for a pure state $\rho$ acts as a measure of the entanglement between $A$ and $B$. 
The logarithm of the density matrix is often difficult to compute.
Thus, it is convenient to study also \textit{Rényi entropies}
\begin{align}
    S^{(\alpha)}_A=\frac{1}{1-\alpha} \log\Tr \left(\rho_A^\alpha\right)
\end{align}
from which the $\alpha\rightarrow1$ limit recovers the vN entropy. 
More generally, the complete entanglement properties of a state are encoded in the \textit{entanglement Hamiltonian} $H_{A}$ defined
by $\rho_{A}=e^{-H_{A}}$,
and thus playing the role of a generating Hamiltonian for the reduced density matrix~\cite{laflorencie_quantum_2016}.

The computation of partial traces is very natural in lattice systems where the Hilbert space inherently has a local tensor product structure.
The same calculation is much more challenging in the case of field theories. 
Here, the number of degrees of freedom is continuous and the reduced density matrix can only be defined through the path integral formalism while entropies only become meaningfully defined after the introduction of an ultraviolet (UV) cutoff. 
A widely used approach to compute the entanglement entropies is the replica trick~\cite{calabrese_entanglement_2009}.
This procedure can be carried out for free theories~\cite{casini_entanglement_2009}, conformal field theories  (CFT)~\cite{calabrese_entanglement_2004} and integrable theories~\cite{cardy_form_2007}. 
It has been argued that some results apply also to the general non-integrable case~\cite{doyon_bipartite_2009}, yet it remains an open question. 
Another field theoretical approach consists of the covariance matrix formalism where entanglement entropies are computed from the correlation functions of Gaussian states~\cite{casini_entanglement_2009,serafini_quantum_2017}. 
Finally, the third widely used approach is to approximate a field theory with a lattice system and use discrete methods like the density matrix renormalisation group (DMRG) to compute the entanglement measures.
This has been extremely successful for computing equilibrium properties~\cite{schollwock_density-matrix_2011,orus_practical_2014,bridgeman_hand-waving_2017,carmen_banuls_review_2020} but is suffering from severe limitations when computing non-equilibrium dynamics due to the exponentially increasing bond dimension~\cite{schollwock_density-matrix_2011}. 
In this work, we develop a method for computing reduced density matrices and entanglement measures in interacting field theories both in and out of equilibrium.

A powerful class of numerical methods for quantum field theory (QFT) is based on Hamiltonian truncation~(HT)~\cite{james_non-perturbative_2018}. 
It has been successfully applied to study spectra of a range of different models, including both integrable and non-integrable models~\cite{lassig_scaling_1991,feverati_scaling_1998,bajnok_boundary_2001,bajnok_finite_2002,rychkov_hamiltonian_2015,rychkov_hamiltonian_2016,elias-miro_nlo_2017,konik_approaching_2021,horvath_chirally_2022} as well as gauge field theories~\cite{konik_studying_2015,azaria_particle_2016,kukuljan_continuum_2021}. 
It has also been used to study correlation functions~\cite{kukuljan_correlation_2018}, real time non-equilibrium dynamics~\cite{rakovszky_hamiltonian_2016,kukuljan_correlation_2018,hodsagi_quench_2018,horvath_nonequilibrium_2019,horvath_inhomogeneous_2021}, symmetry breaking~\cite{rychkov_hamiltonian_2015,rychkov_hamiltonian_2016}, Kibble-Zurek mechanism~\cite{hodsagi_kibblezurek_2020},  quantum chaos~\cite{brandino_energy_2010,srdinsek_signatures_2021}, confinement~\cite{lencses_confinement_2021}, spectral form factors~\cite{cubero_duality_2022} and genuinely field theoretical (continuum) phenomena~\cite{kukuljan_out--horizon_2020}. 
One advantage of HT is the direct formulation in the continuum. 
The method does not require to approximate the field theory with a lattice system and to take the continuum limit in the end.
HT is by construction applicable to any dimension but, due to the computational cost, has been so far successfully applied in 1+1 D and 2+1 D~\cite{hogervorst_truncated_2015,elias-miro_exploring_2020}. 

While HT is very successful at computing spectral properties and  non-equilibrium time evolution, it has not been the most convenient choice for computing entanglement related quantities. 
In preceding works~\cite{palmai_entanglement_2016,murciano_post-quantum_2021} HT was used to calculate matrix elements between higher excited states.
By using analytic replica techniques, correlation functions of twist fields and other entanglement related objects were calculated for those states. 
While such approaches proved useful at computing low Rényi entropies, the calculation of vN entropies and entanglement negativity remained out of reach.

In this work, we develop a general way to construct reduced density matrices with HT.  
The output of our method is the density matrix of a state explicitly represented in a computational basis which is a tensor product of the basis of the left and right subsystems. 
This enables the direct computation of almost any entanglement related quantity: vN entropy, entanglement negativity, mutual information and the direct study of the entanglement Hamiltonian and reduced density matrix of an interacting field theory itself. 
Our method is general and can be widely used on top of any HT code that uses expansions in free bases (see Section~\ref{sec:HT}) which is a common choice in modern applications~\cite{feverati_scaling_1998,bajnok_boundary_2001,rychkov_hamiltonian_2015,elias-miro_nlo_2017,horvath_chirally_2022}. 
This enables us to take full advantage of the power of HT for real time evolution of a wide range of interacting QFT and study the whole spectrum of entanglement related quantities without needing to approximate the theory with a lattice system.
It gives access to the entanglement properties of ground, excited and time dependent non-equilibrium states as well as thermodynamic entropies of thermal states. 
Additionally, the method has the potential to work also in $D>1+1$.

The manuscript is organised as follows: in sec.~\ref{sec:HT}, we briefly introduce the basic concepts of HT. 
In sec.~\ref{sec:method} we present our method for computing reduced density matrices. 
We begin in sec.~\ref{sec:method_splitting} with the theoretical construction and continue in sec.~\ref{sec:method_algorithm}  outlining an efficient algorithm for the numerical implementation. 
In sec.~\ref{sec:models} we introduce the QFT models that we test the method on. 
In sec.~\ref{sec:results} we present the results and a comparison against analytical predictions. 
Sec.~\ref{sec:results_equil} focuses on equilibrium states while sec. 
\ref{sec:results_dynamics} covers non-equilibrium dynamics. 
We conclude in sec~\ref{sec:conclusion} with an overview, discussion and the scope for the future work. 
The appendices cover the more technical details of the method.

\section{\label{sec:HT} Hamiltonian Truncation}

HT is a numerical method for strongly interacting QFT first introduced in the 90's by Yurov and Zamolodchikov~\cite{yurov_truncated_1990,yurov_truncated-fermionic-space_1991}. 
It is based on the Hamiltonian formalism and the idea is to represent the Hamiltonian of a field theory defined on a compact domain as
\begin{align}
    H=H_{\text{solv}}+\Phi_{\text{pert}}
\end{align}
where $H_{\text{solv}}$ is the solvable part of the Hamiltonian and $\Phi_{\text{pert}}$ is a perturbing potential. 
Traditionally, the CFT of the UV fixed point of the theory was used as $H_{\text{solv}}$ but more modern approaches consist of using other solvable theories like free massless and massive theories. 
The perturbing potential $\Phi_{\text{pert}}$ does not need to be small which gives HT the power to capture non-perturbative effects. 
The method proceeds with representing the potential $\Phi_{\text{pert}}$ as a matrix in the Hilbert space of $H_{\text{solv}}$, the space $\mathcal{H}_{\text{solv}}$. 
The crucial step of HT is to introduce a high energy cutoff, keeping only the low energy states of $H_{\text{solv}}$ which renders the matrices finite and enables numerical computation. 
The method converges if $\Phi_{\text{pert}}$ does not mix significantly the low energy sector of $\mathcal{H}_{\text{solv}}$ with the higher energy sectors. 
In case of an expansion around the CFT point, this is guaranteed by the renormalisation group theory for relevant perturbations $\Phi_{\text{pert}}$. 
If computed for several high enough cutoffs the results of a HT simulation can often be extrapolated to obtain the infinite cutoff value. 
Alternatively, a numerical renormalisation group algorithm can be used~\cite{konik_numerical_2007}. 

Computing entanglement related quantities has been challenging for HT. 
The HT Hilbert space is usually spanned either by primary and descendant CFT states or free model eigenstates in momentum basis.
It does not allow for an easy bipartition in position space.
Earlier approaches ~\cite{palmai_entanglement_2016,murciano_post-quantum_2021} were based on mapping the problem to the CFT calculation of entanglement related CFT objects for descendant fields. 
While conceptually elegant, such calculations for higher descendant states are often tedious and are associated with several restrictions. 
They have so far been limited to the first few Rényi entropies.  
In this work we want to overcome this problem and construct a more general and robust approach which can be exploited for the calculation of almost any entanglement related quantity.

\section{\label{sec:method} Method}
Our main goal is the construction of reduced density matrices with Hamiltonian truncation. 
We consider a field theory defined on a finite interval $F=[0,L]$ (full) with open boundary conditions and are interested in computing the entanglement between a subsystem  $L=[0,\ell]$ (left) and its complement $R=[\ell,L]$ (right).
Note that we use $L$ for the full system size and as a label for the left subsystem. 
The interpretation is clear from the context.
Using HT, we can compute the density matrix $\rho$ of a ground, excited, thermal or non-equilibrium state of the theory expressed in the Hilbert space $\mathcal{H}_F$ of the full interval $F$. 
However, tracing out a spatial part of the system is difficult in the momentum basis of $\mathcal{H}_F$.
If we express $\rho$ as $\rho_{LR}$ in a Hilbert space $\mathcal{H}_L\otimes\mathcal{H}_R$ built out of Hilbert spaces $\mathcal{H}_{L}$ and $\mathcal{H}_R$ on the intervals $L$ and $R$, we can trace one part of the system directly.
We thus need to construct  $\mathcal{H}_F$, $\mathcal{H}_L$ and $\mathcal{H}_R$ and find the unitary transformation
\begin{align}
    U_T:\mathcal{H}_F\rightarrow \mathcal{H}_L\otimes\mathcal{H}_R
    \label{eq:def_U_T}
\end{align}
to compute
\begin{align}
    \rho_{LR} = U_T \rho U_T^\dagger.
\end{align}
The idea of the method is visualized in Figure~\ref{fig:system}.

\subsection{\label{sec:method_splitting} Splitting the System}
\subsubsection{Fields and Hilbert spaces}\label{sec:method_split_Hilbert}
We start with the construction of the Hilbert space on the full interval $\mathcal{H}_F$.
We expand the fields of the free theory (massless or massive) in terms of momentum modes.
The Fock space generated by the mode creation operators serves as the computational basis. 
For concreteness, we choose an expansion around a massless free bosonic field theory with Dirichlet boundary conditions~($\phi(0)=\phi(L)=0$) at the edges. 
The procedure is easily generalisable to expansions around massive free theories and other boundary conditions and could by construction be applied also in $D>1+1$. 

The field expansion can be written as
\begin{align}
    \phi(x,t)=\frac{1}{\sqrt{L}}\sum_{k=1}^\infty \frac{1}{\sqrt{p_k}} \left( A_k e^{-i p_k t}+A_{k}\dgr e^{i p_k t} \right)\sin(p_k x),
    \label{eq:def_phi_full}
\end{align}
with $p_k=k\frac{\pi}{L}$ and $A_k$ the bosonic modes on the full interval  fulfilling the commutation relations $\comm{A_k}{A_l}=\comm{A_k^\dagger}{A_l^\dagger}=0$ and $\comm{A_k}{A_l^\dagger}=\delta_{k,l}$.
We refer to the modes $A_k$ as full modes in the rest of the text.

The full modes $A_k$ span the Hilbert space $\mathcal{H}_F$ 
\begin{align}
    \ket{\vec{n}_F}\equiv \ket{n_1,n_2,\ldots}\equiv \frac{1}{N_F}\prod_{k>0}\left( A_k^\dagger\right)^{n_k}\ket{0}
    \label{eq:state_full}
\end{align}
with $n_k$ the bosonic occupation numbers, the normalisation $N_F=\prod_{k>0}\sqrt{n_k!}$ and $\ket{0}$ the vacuum of the massive free boson theory.

\begin{figure}
    \centering
    \includegraphics[width=\columnwidth]{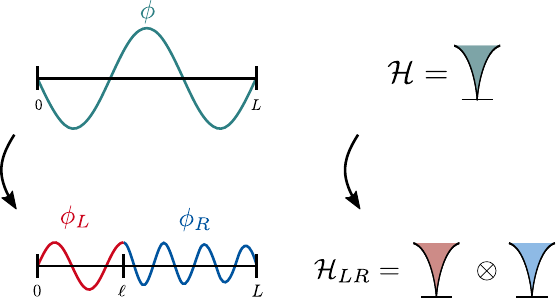}
    \caption{Schematic drawing of the algorithm.
     \emph{Left} -- We split the system on the top by representing the field $\phi$ living on the full interval $[0,L]$ with an equivalent setting: a pair of fields $\phi_L$ and $\phi_R$ living on subintervals $[0,\ell]$ and $[\ell,L]$ with an additional boundary condition at $\ell$.
     The figure depicts the case of Neumann boundary conditions at the cut.
     The boundary conditions at the edges are chosen to be Dirichlet.
     \emph{Right} Quantization of the fields gives rise to two isomorphic Hilbert spaces $\mathcal{H}$ and $\mathcal{H}_{LR}=\mathcal{H}_L\otimes\mathcal{H}_R$ and a unitary map between them. 
     This maps density matrices to a form suitable for taking partial traces. 
     The cones in the figure represent the exponentially growing number of states with the energy above the ground state.
     $\mathcal{H}$ and $\mathcal{H}_{LR}$ are generated on top of different vacua (horizontal line below the cone).
    }
    \label{fig:system}
\end{figure}

A cut at position $\ell$ divides the system into two subsystems $L$ and $R$ (left and right), as shown in Figure~\ref{fig:system}.
In a similar fashion as $\mathcal{H}$, we construct $\mathcal{H}_L\otimes\mathcal{H}_R$ by defining two fields $\phi_L$ and $\phi_R$ (split fields) living on the two subintervals and quantizing them.

The formulation of the fields on the intervals depends on the additional boundary conditions that we introduce at the cut. 
We choose to study Neumann~($\partial_x\phi_L(\ell)=\partial_x\phi_R(\ell)=0$) or Dirichlet~($\phi_L(\ell)=\phi_R(\ell)=0$) boundary conditions at the cut.
In the main text, we focus on Neumann boundary conditions at the cut.
The treatment of Dirichlet boundary conditions at the cut is detailed in appendix~\ref{sec:app_dirichlet}.
The outer edges of the system are always fixed to be Dirichlet boundary conditions~($\phi(0)=\phi(L)=0$).

For Neumann boundary conditions at the cut, the fields on the intervals are defined as 
\begin{align}
    \begin{split}
        \phi_{L}(x,t) & =\frac{1}{\sqrt{\ell}}\sum_{m=1}^{\infty}\frac{1}{\sqrt{q_{m}^{(\ell)}}}\times\\
        &\hspace{1cm}\left(a_{m}^{L}e^{-i q_{m}^{(\ell)}t}+a_{m}^{L,\dagger}e^{i q_{m}^{(\ell)}t}\right)\sin\left(q_{m}^{(\ell)}x\right) 
        \label{eq:def_phi_l_neu}
    \end{split}\\
    \begin{split}
    \phi_{R}(x,t) & =\frac{1}{\sqrt{L-\ell}}\sum_{m=1}^{\infty}\frac{1}{\sqrt{q_{m}^{(L-\ell)}}}\times\\
    &\left(a_{m}^{R}e^{-i q_{m}^{(L-\ell)}t}+a_{m}^{R,\dagger}e^{i q_{m}^{(L-\ell)}t}\right)\sin\left(q_{m}^{(L-\ell)}(L-x)\right) 
    \label{eq:def_phi_r_neu}
    \end{split} 
\end{align}
where $q_{m}^{(\ell)}=(m-\frac{1}{2})\frac{\pi}{\ell}$, $q_{m}^{(L-\ell)}=(m-\frac{1}{2})\frac{\pi}{L-\ell}$ and $a_m^{\sigma}$ are the bosonic annihilation operators on the two partitions for $\sigma\in\{L,R\}$.
Both, the fields $\phi_{L}$, $\phi_{R}$ and the modes $a_{L/R}$ defined in~\eqref{eq:def_phi_l_neu} and~\eqref{eq:def_phi_r_neu} fulfill the respective bosonic commutation relations.
Fields and modes on different subintervals commute.
In analogy to the full fields, the modes on the sub-intervals span their respective Hilbert spaces $\mathcal{H}_{L}$ and $\mathcal{H}_R$.
The computational bases for the two sub-intervals are
\begin{align}
    \ket{\vec{n}_\sigma}\equiv \ket{n_{1,\sigma},n_{2,\sigma},\ldots}\equiv \frac{1}{N_\sigma}\prod_{m>0}\left( a_m^{\sigma,\dagger}\right)^{n_m}\ket{0}_\sigma,
    \label{eq:state_split}
\end{align}
with the normalisation $N_\sigma=\prod_{m>0}\sqrt{n_{m,\sigma}!}$.
The choice of mixed boundary conditions (Dirichlet on the edges and Neumann at the cut) has the advantage that no zero-modes appear in the system.
The vacua of the sub-intervals are not equal to each other and in particular they are not equal to the full system vacuum $\ket{0}_L\neq\ket{0}_R\neq \ket{0}$.
The product space $\mathcal{H}_L\otimes\mathcal{H}_R$ is then generated by $\ket{\vec{n}_L,\vec{n}_R}\equiv\ket{\vec{n}_L}\otimes \ket{\vec{n}_R}$ on top of the vacuum $\ket{0,0}\equiv\ket{0}_L\otimes\ket{0}_R \neq \ket{0}$.

Before we continue, a couple of words on conventions and notation. 
Throughout the paper, we will use $k$ as an index for the full modes $A_k$ and $l,m$ as indices for partial modes $a_m^\sigma$.
Greek indices always indicate a left or a right partition, $\sigma\in\{L,R\}$.

\subsubsection{Bogoliubov transformation \label{sec:method_split_bogoliubov}}

At first glance it might not be obvious that the the descriptions of the system in terms of the full field and split fields are equivalent.
From an intuitive point of view: for any given field configuration of $\phi$, one can find a configuration of $\phi_L$ and $\phi_R$ that is arbitrarily close to $\phi$ and still respects the boundary condition at the cut. 
Indeed, we can choose $\phi_L$ and $\phi_R$ to be equal to $\phi$ everywhere except for a small neighborhood of the cut. 
There, they have to deviate in order to satisfy the boundary condition. 
But we can make this neighborhood arbitrarily small while still preserving the continuity of $\phi_L$ and $\phi_R$ and the boundary conditions.
Later, we give a more detailed argument for the correspondence on the algebraic level.

We now formally construct the unitary mapping between $\mathcal{H}_F$ and $\mathcal{H}_L\otimes\mathcal{H}_R$ proposed in~\eqref{eq:def_U_T}
\begin{align}
    \left(U_T\right)_{\vec{n}_L\vec{n}_R;\vec{n}_F}=\braket{\vec{n}_L,\vec{n}_R}{\vec{n}_F}.
    \label{eq:def_U_T_mat_elements}
\end{align}

In order to compute the matrix elements~\eqref{eq:def_U_T_mat_elements}, we take two distinct steps.
Firstly, we express the full modes $A_k$ in terms of the partial modes $\{a_m^{\sigma}\}_{m,\sigma}$ and $\{a_n^{\sigma\dagger}\}_{n,\sigma}$.
Secondly, we rewrite the full vacuum in terms of the partial vacua and partial modes.
The latter is particularly important because neglecting that the Hilbert spaces are not defined on top of the same vacuum will lead to wrong results. 

We rewrite the full modes as
\begin{align}
 A_k=&\sum_{m}\gamma_{km}^{+,L} a_{m}^{L} +\sum_{m} \gamma_{km}^{-,L}a_{m}^{L,\dagger}\nonumber\\
     &+\sum_{m}\gamma_{km}^{+,R} a_{m}^{R} +\sum_{m} \gamma_{km}^{-,R}a_{m}^{R,\dagger},
 \label{eq:full_modes_as_split_modes}
\end{align}
where the coefficients $\gamma$ are to be determined.
The coefficients $\gamma$ are the result of identifying the fields on the full interval with the split fields
\begin{align}
    \phi(x,t)= \begin{cases}
    \phi_L(x,t) &\text{if } x<\ell,\\
    \phi_R(x,t) &\text{if } \ell<x<L
    \end{cases}.
    \label{eq:phi_system_split}
\end{align}
This identification, the continuity condition, is the core of the unitary map between the full and the split Hilbert spaces. 

We can express the full modes $A_k$ in terms of the fields $\phi(x,t)$ and the canonical momentum operator $\pi(x,t)=\pdv{t}\phi(x,t)$
\begin{align}
  A_{k} = \sqrt{\frac{p_k}{L}} \int_{0}^{L} \dd{x} \left[\phi(x,t)+\frac{i}{p_k} \pi(x,t)\right] \sin\left(p_k x\right).
    \label{eq:full_modes_integral}
\end{align}
Combining~\eqref{eq:full_modes_integral} with the continuity condition~\eqref{eq:phi_system_split} links the full system modes $A_k$ to the partial modes $a_m^\sigma$.
Evaluating the integral, we obtain the coefficients $\gamma$ for Neumann boundary conditions at the cut as 
\begin{align}
    \gamma_{km}^{+,L}&=\begin{cases}
        \frac{(-1)^{m}\sqrt{p_{k}}\cos\left(p_{k}\ell\right)}{\sqrt{L\ell}\sqrt{q_{m}^{(\ell)}}(p_{k}-q_{m}^{(\ell)})} & p_{k}\neq q_{m}^{(\ell)}\\
        \sqrt{\frac{\ell}{L}} & p_{k}=q_{m}^{(\ell)}
    \end{cases}\label{eq:gamma_l_plus_neu}\\
    \gamma_{km}^{-,L}&=\frac{(-1)^{m}\sqrt{p_{k}}\cos\left(p_{k}\ell\right)}{\sqrt{L\ell}\sqrt{q_{m}^{(\ell)}}(p_{k}+q_{m}^{(\ell)})}\label{eq:gamma_l_minus_neu}\\
    \gamma_{km}^{+,R}&=\begin{cases}
        \frac{(-1)^{m+1}\sqrt{p_{k}}\cos\left(p_{k}\ell\right)}{\sqrt{L(L-\ell)}\sqrt{q_{m}^{(\ell)}}(p_{k}-q_{m}^{(L-\ell)})} & p_{k}\neq q_{m}^{(L-\ell)}\\
        \frac{(-1)^{k+1}\sqrt{L-\ell}}{\sqrt{L}}\left(1+\frac{\sin\left(p_{k}\ell\right)}{2p_{k}(L-\ell)}\right) & p_{k}=q_{m}^{(L-\ell)}
    \end{cases}\label{eq:gamma_r_plus_neu}\\
    \gamma_{km}^{-,R}&=\frac{(-1)^{m+1}\sqrt{p_{k}}\cos\left(p_{k}\ell\right)}{\sqrt{L(L-\ell)}\sqrt{q_{m}^{(\ell)}}(p_{k}+q_{m}^{(L-\ell)})}.\label{eq:gamma_r_minus_neu}
\end{align}
The special cases in equations~\eqref{eq:gamma_l_plus_neu} and~\eqref{eq:gamma_r_plus_neu} are divergences of the integrand.
The detailed calculation as well as the expressions for the Dirichlet boundary conditions at the cut can be found in appendix~\ref{sec:app_dirichlet}.

\subsubsection{Multimode squeezed coherent vacuum\label{sec:coherent_vacuum}}
When expressing the full modes as a superposition of partial modes, we also have to re-express the vacuum of the system.
In order to find a formulation of the full system vacuum in terms of the partial modes, we identify Equation~\eqref{eq:full_modes_as_split_modes} as a multi-mode Bogoliubov transformation~\cite{qin_general_2001}
\begin{align}
  \mqty[ A\\ A^{\dagger} ]=M\mqty[ a\\ a^{\dagger}]
  \label{eq:bgbtrafo_modes}
\end{align}
with $A=(A_1, \dots, A_{N_F})$, $a=(a_1^{L},\dots,a_{N_L}^{L},\dots,a_{N_R}^R)$ and
\begin{align}
  M=\mqty[ u & v\\ v & u]=\mqty[ \gamma^{L,+} & \gamma^{R,+} & \gamma^{L,-} & \gamma^{R,-}\\ \gamma^{L,-} & \gamma^{R,-} & \gamma^{L,+} & \gamma^{R,+}].
 \label{eq:bgbtrafo}
\end{align}
Note that $M$ is not an operator here, but a matrix of numbers. 
Since all the coefficients $\gamma$ in equations~\eqref{eq:gamma_l_plus_neu}-\eqref{eq:gamma_r_minus_neu} are real, we focus on the case of real $u$ and $v$.
We use the same symbols as in~\eqref{eq:gamma_l_minus_neu}-\eqref{eq:gamma_r_plus_neu} without the subscript indices to refer to matrices of coefficients.
For ease of notation, we still express equations in terms of $u=\mqty[\gamma^{L,+} & \gamma^{R,+}]$ and $v=\mqty[\gamma^{L,-} & \gamma^{R,-}]$.

The transformation~\eqref{eq:bgbtrafo} expresses bosonic modes $A$ in terms of different bosonic modes $a$.
Thus, the transformation must preserve the commutation relations.
These are encoded in the symplectic structure of $M$
\begin{align}
    M^{-1}=KM^{\dagger}K \qqtext{with} K=\mqty[ \id\\ & -\id ].
    \label{eq:bgb_restriction}
\end{align}
It can be verified that $\gamma$ coefficients in eq.~\eqref{eq:gamma_l_plus_neu}-\eqref{eq:gamma_r_minus_neu} obey the symplectic structure of the Bogoliubov transformation.

The Bogoliubov transform~\eqref{eq:bgbtrafo} is equivalent to a unitary transformation~\cite{qin_general_2001}
\begin{align}
   U\mqty[ a\\ a\dgr]U\dgr =M\mqty[ a\\ a\dgr]
\end{align}
with
\begin{align}
  U=\exp\left(-\frac{1}{2} \mqty[a^{\dagger T} & a\tran ] K \ln M \mqty[ a\\ a\dgr] \right).
\end{align}
In contrast to $M$, $U$ is an operator and not a matrix of numbers.
Thus, the vacuum of the full modes $\ket{0}$ can be expressed in terms of the vacuum of the partial modes $\ket{0,0}$ as
\begin{align}
   \ket{0}=U\ket{0,0}.
\end{align}
because then $A_k\ket{0} = U^{\dagger} a U U^{\dagger} \ket{0,0} = 0$. 

$U$ can be written in a more convenient form for actual computations, the so-called disentangling form
 \begin{align}
    \begin{split}
          U=&\exp\left(-\Tr(\sigma)\right)\exp\left(-a^{\dagger T}\chi a^{\dagger}\right)\times\\
          &\exp\left(-2a^{\dagger T}\sigma a\right)\exp\left(a^{T}\tau a\right) 
    \end{split}
    \label{eq:bgb_disentangling_form}
\end{align}
with 
\begin{align}
    \chi=\frac{1}{2}u\inv v,\quad \sigma=\frac{1}{2}\ln u,\quad \tau=\frac{1}{2} v^* u\inv.
    \label{eq:def_rho_sigma_tau}
\end{align}

Since we have an expression of the full modes in terms of the partial modes and an expression of the full vacuum in terms of the partial vacua, we can compute the matrix elements of $U_T$.
The elements of $U_T$ are overlaps between states in the full basis $\ketf$ and the split basis $\ketlr$.
\begin{align}
    \begin{split}
    \braket{\nlr}{\nf}= &\frac{1}{N} \bra{0,0} \left[\prod_{m>0}\left(a_{m}^{R}\right)^{n_{m,R}}\left(a_{m}^{L}\right)^{n_{m,L}}\right]\times\\
    &\left[\prod_{k>0}\left[\sum_{\sigma}\sum_{l>0}\left(\gamma_{kl}^{\sigma,-}a_{l}^{\sigma}+\gamma_{kl}^{\sigma,+}a_{l}^{\sigma \dagger}\right)\right]^{n_{k}}\right]\times\\
    &\left[\exp\left(-\sum_{ij}\sum_{\sigma,\chi}a_{i}^{\sigma \dagger}\chi_{ij}^{\sigma,\xi}a_{j}^{\xi\dagger}\right)\right]\ket{0,0},
    \end{split}
    \label{eq:def_matrix_elements_basic}
\end{align}
with 
\begin{align}
    N=\frac{1}{\exp\left(-\Tr\left(\sigma\right)\right)}\left[\prod_{m>0}\sqrt{\left(n_{m,R}!\right)\left(n_{m,L}!\right)}\right]\left[\prod_{k>0}\sqrt{\left(n_{k}!\right)}\right].
    \label{eq:norm_matrix_elements_basic}
\end{align}
The first bracket in~\eqref{eq:def_matrix_elements_basic} builds the occupation number state $\bralr$ from the partial vacuum $\bra{0,0}$.
The order of left and right creation operators does not matter here, since they commute as they act on different partitions.
The second bracket represents the operators $A_k\dgr$ which build $\ketf$ on top of the vacuum of the full modes $\ket{0}$.
We choose to express the full modes in terms of the partial modes [cf. Equation~\eqref{eq:full_modes_as_split_modes}].
The opposite way of expressing the partial modes in terms of full modes would work as well.
The last bracket in~\eqref{eq:def_matrix_elements_basic} expresses the vacuum of full modes in terms of the split modes.
Looking at formulation of the norm in~\eqref{eq:norm_matrix_elements_basic}, we recognize the first term from the transformation of the vacuum; it is constant for all matrix elements.
The second and third term are the norms of the partial and the full occupation number states.
The contributions of the exponentials with $\sigma$ and $\tau$ in~\eqref{eq:bgb_disentangling_form} vanish due to the order of operators in the exponentials and their action on the vacuum to the right.

\subsubsection{Equivalence of Hilbert spaces\label{sec:equiv_hilbert}}
We have now fully exposed the unitary transformation between $\mathcal{H}$ and $\mathcal{H}_{L}\otimes\mathcal{H}_{R}$. 
We are thus ready to justify why the two descriptions of a physical state, in terms of $\rho$ and in terms of $\rho_{LR}$, are equivalent despite the additional boundary condition at the cut. 

The way to think of the transformation is in terms of Fourier analysis. 
The sine modes of the split intervals in eqs.~\eqref{eq:def_phi_l_neu} and \eqref{eq:def_phi_r_neu} serve as a functional basis in which a field configuration of the full interval field is expanded. 
The fact that it is indeed a basis, is a simple consequence of Carleson's theorem for convergence of Fourier series~\cite{trembinska_variations_1985,titchmarsh_theory_2002}. 

However, we are dealing with quantum fields not just simple scalar valued functions. 
Upon quantization, the Fourier coefficients in the expansion become operator valued as indicated in eqs.~\eqref{eq:def_phi_l_neu} and \eqref{eq:def_phi_r_neu}.
The symplectic structure of the Bogoliubov transform between the two algebras~\eqref{eq:bgb_restriction} guarantees that the two quantizations of the system, the one in terms of the full field in eq.~\eqref{eq:def_phi_full} and the one in terms of the split fields, are equivalent.

Thus in the complete, infinite dimensional Hilbert spaces, the unitary map between $\mathcal{H}$ and $\mathcal{H}_{L}\otimes\mathcal{H}_{R}$ is exact. 
However, once we introduce a truncation, it becomes an approximation in the same fashion as HT is always an approximation of the quantum state using the low energy sector of the Hilbert space. 
Using the partial field expansion \eqref{eq:def_phi_r_neu} then becomes in spirit very similar to using a truncated Fourier series to approximate a function. 
In section \ref{sec:results} we demonstrate that such an approximation indeed performs excellently at computing entanglement entropies. 

\subsubsection{Truncation}

\begin{figure}
    \centering
    \includegraphics[width=\columnwidth]{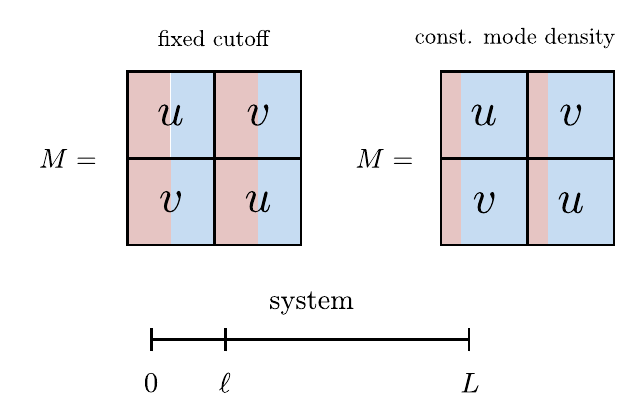}
    \caption{Illustration of the $M$ matrix for different cut-off schemes.
        The amount of coefficients for the left and the right partition in the matrices $u$ and $v$ differ depending on the cut-off scheme.
        The coefficients belonging to the left(right) side are displayed in red(blue).
    }
    \label{fig:cutoff_schemes}
\end{figure}

The mapping between the full system and the partitioned system is so far formulated without considering the truncation.
In order to be able to implement the method with a finite amount of computer memory, we have to consider a finite dimensional approximation of the Hilbert spaces.
In HT, it is often important to choose the most suitable truncation scheme for the problem.
In our case, we have to introduce three truncations: of $\mathcal{H}$,  $\mathcal{H}_{L}$ and $\mathcal{H}_{R}$.
For convenience, we choose the cutoffs such that the Bogoliubov transformation~\eqref{eq:bgbtrafo} becomes a square matrix. 
This leads to the restriction $s_F=s_L+s_R$, where $s_F$, $s_L$ and $s_R$ are the number of momentum modes kept in the full system, the left and the right partition, respectively.
We truncate all Hilbert spaces, $\mathcal{H}$,  $\mathcal{H}_{L}$ and $\mathcal{H}_{R}$ with an energy cutoff such that the cutoff energy is equal to the energy of the single excitation of the largest momentum mode kept, $s_F$, $s_L$ and $s_R$ respectively.
For a free massless boson Hamiltonian, the energy of the state $\ket{\vec{n}_F}$ defined in eq.~\eqref{eq:state_full} above the ground state  is $\epsilon(\vec{n}_F)-\epsilon(0)=\frac{\pi}{L} \sum_{k=1}^{\infty}kn_k$.
Similarly for the states in the split Hilbert spaces~\eqref{eq:state_split}, where $L$ is replaced by $\ell$ and $L-\ell$, respectively.

In infinite dimensional Hilbert spaces, before the truncation, the symplectic structure~\eqref{eq:bgb_restriction} is fulfilled exactly.
By introducing the cut-off this property can be compromised.
We use the preservation of the symplectic structure as a guiding principle to select a cut-off scheme.

We investigate two different cut-off schemes.
In the first one, we distribute the number of partial modes equally across the left and the right partition independent of the position of the cut (fixed cutoff): $s_L=s_R=s_F/2$.
This leads to an easy implementation but is questionable from a physical standpoint.
A short interval with many modes leads to a greater resolution in position space than a large interval with the same number of modes.
Furthermore, such a non-uniform UV cut-off leads to a position-dependent non-universal constant in the entanglement entropy that obscures the true functional dependence.
The second cut-off scheme takes into account the length of the partition.
The number of modes are distributed proportionally to the length of the interval $s_L=\frac{\ell}{L}s_F$ and $s_R=s_F-s_L$ (constant mode density).
This scheme keeps a constant density of momentum modes ($s_L/\ell=s_R/(L-\ell)$) and thus a homogeneous UV cut-off.
Both cut-off schemes are displayed in Figure~\ref{fig:cutoff_schemes}. 

All further considerations use the constant mode density cut-off scheme.
It indeed reproduces the bosonic commutation relations more faithfully as shown in appendix~\ref{sec:app_bgb_trafo} where also further details of the cutoff scheme are discussed.

\subsection{\label{sec:method_algorithm} Algorithm}
The evaluation of the overlap as given in~\eqref{eq:def_matrix_elements_basic} is difficult because the expression is not normal ordered and the sums in the expression are taken to the powers $n_k$.
It would require an overwhelming number of commutations of individual mode operators.

\subsubsection{Generating functional formulation\label{sec:gen_func}}
The expression simplifies if we express it in the spirit of a generating functional. 
The repeated application of a mode $a$ is equivalent to 
\begin{align}
    a^{n}=\left[\dv[n]{\jcal}e^{\jcal a}\right]_{\jcal=0},
    \label{eq:mode_power_as_exp}
\end{align}
where $\jcal$ is a scalar variable.

Inserting the identity~\eqref{eq:mode_power_as_exp} in the formulation of the matrix elements of $U_T$ in~\eqref{eq:def_matrix_elements_basic}, we obtain an expression resembling a generating functional  
\begin{widetext}
\begin{align}
    \begin{split}
        \braket{\nlr}{\nf}=\frac{1}{N} \left.\prod_{m>0}\prod_{\sigma}\dv[n_{m,\sigma}]{j_{m,\sigma}}\prod_{k>0}\dv[n_{k}]{J_{L}}\bra{0,0}
        e^{S}e^{F}e^{V}\ket{0,0}\right|_{J_{k}=0,j_{m,\sigma}=0},
    \end{split}
    \label{eq:def_matrix_elements_unordered}
\end{align}
\end{widetext}
with
\begin{align}
    S&=\sum_{m>0}\sum_{\sigma}j_{m,\sigma}a_{m}^{\sigma}\\
    F&=\sum_{k,m>0}\sum_{\xi}J_{k}\left(\gamma_{k,m}^{+,\xi}a_{m}^{\xi\dagger}+\gamma_{k,m}^{-,\xi}a_{m}^{\xi}\right)\\
    V&=-\sum_{\kappa,\lambda}\sum_{m,n>0}a_{-m}^{\kappa}\chi_{m,n}^{\kappa,\lambda}a_{-n}^{\lambda}.
\end{align}
Here, $S$ are all terms related to the split modes, $F$ is the term that is associated with the full system modes and $V$ are the terms that build the vacuum.
We introduced two kinds of additional, scalar variables. 
$J_k$ are the additional variables for the full modes and $j_{m,\sigma}$ are defined for the partial modes. 

Using the Baker-Campbell-Hausdorff (BCH) relations~\cite{hausdorff_symbolische_1906}, we can bring this expression into normal order and evaluate the expectation value.
The series of commutators in the BCH relations terminate at most at the second order since the highest power of mode operators in the exponent is two.
We can write the normal ordered expression as 
\begin{align}
    e^{S}e^{F}e^{V} = Z \,e^{\left[A,V\right]}\,:e^{S}e^{F}e^{V}:
    \label{eq:gen_exp_normal_ordering}
\end{align}
with
\begin{align}
    Z&=\exp\left[\comf+\comsf+\comav\right]\, .
    \label{eq:def_z}
\end{align}
The terms \comf, \comsf and \comav are results of the BCH operations.
They are defined as 
\begin{align}
    \comf =\frac{1}{2}\sum_{k,k'=1}^{\infty}J_{k}J_{k'}\sum_{\sigma}\sum_{n>0}\gamma_{k,n}^{+,\sigma}\gamma_{k',n}^{-,\sigma}
    \label{eq:def_comf}
\end{align}
\begin{align}
    \comsf = \sum_{m>0}\sum_{\sigma}j_{m,\sigma}\sum_{k>0}J_{k}\gamma_{k,m}^{+,\sigma}
    \label{eq:def_comsf}
\end{align}
\begin{align}
    \begin{split}
        \comav =-\frac{1}{2} &\sum_{m>0} \sum_{\sigma} \left \{ j_{m,\sigma} + \sum_{k>0} J_{k} \gamma_{k,m}^{-,\sigma} \right\}\\ 
        &\sum_{m'>0} \sum_{\mu} \left\{ j_{m',\mu} + \sum_{k'>0} J_{k'} \gamma_{k',m'}^{-,\mu} \right\}\\
        &\left(\chi_{m',m}^{\mu,\sigma} + \chi_{m,m'}^{\sigma,\mu}\right)
    \end{split}
    \label{eq:def_comav}
\end{align}
The commutator $\left[A,V\right]$ is linear in creation operators.
The derivation of the commutators is detailed in appendix~\ref{sec:app_commutators}.

The problem of computing the overlap reduces to computing multiple derivatives of a scalar expression if we express~\eqref{eq:def_matrix_elements_basic} as the derivative of a generating functional 
\begin{align}
    \braket{\nlr}{\nf} &=\frac{1}{N}\left.\prod_{m>0}\prod_{\sigma}\dv[n_{m,\sigma}]{j_{m,\sigma}}\prod_{k>0}\dv[n_{k}]{J_{L}}Z\right|_{J_{k}=0,j_{m,\sigma}=0}.
    \label{eq:def_matrix_elements_gen_func}
\end{align}

\subsubsection{Tackling the exponential complexity of differentiation\label{sec:exponential_complexity}}
The next goal is the efficient calculation of all derivatives in~\eqref{eq:def_matrix_elements_gen_func}.
The pure symbolic evaluation of derivatives becomes prohibitively expensive with increasing cut-off.
The number of terms grows as $n!/\left[2^{n/2}(n/2)!\right]$ with the number of derivatives $n$. 
For large $n$, this scales as $\exp\left[(n/2)\left(\log(n/2)-1/4\right)\right]$.
In the following, we show how the structure of the generating functional helps us to make the evaluation more efficient.

From an algorithmic point of view, there are two distinct exponentially scaling problems involved in the computation.
On the one hand, the size of the unitary transformation grows with the size of the Hilbert space.
We have to evaluate exponentially many terms in order to fill the matrix.
It is impossible to circumvent this exponential since the method is based on exact diagonalization. 
On the other hand, each matrix element needs an increasing number of derivatives with increasing occupation numbers.
The number of terms in the derivation scales also exponentially with the occupation number (and thus the cut-off).
In the following section, we describe an algorithm to make the evaluation of the derivatives feasible for relevant cut-offs.
We are not able to reduce the exponential growth of terms to a polynomial growth.
However, we derive a procedure which largely reduces the exponential growth. 
Thus, it is possible to reach HT cut-offs that provide reasonable approximations of interesting physics. 

A commonly used alternative to symbolic differentiation is automatic differentiation (AD)~\cite{wengert_simple_1964,bartholomew-biggs_automatic_2000}.
The algorithm of AD tracks the computation of the function and uses predefined derivatives of elementary functions to evaluate the derivative numerically.
In our case, it is hard to use AD directly since we have to compute possibly very high derivatives of the function and that we do not need the actual function value.
Furthermore, we can exploit the structure of the function to determine which terms must be 0 without computing them.
Therefore, we take a more specialized approach and do not rely on AD.

By inspecting the structure of the expressions in Equation~\eqref{eq:def_matrix_elements_gen_func}, we note that the derivative always acts on an expression of the form
\begin{align}
Z=e^{T},
\end{align}
with $T=\comf+\comsf+\comav$, a shorthand for all terms in the exponent of $Z$ in~\eqref{eq:def_z}.
For the ensuing discussion, we introduce a shorthand notation for the derivatives of $Z$
\begin{align}
    \begin{split}
        \dv{\jcal_i}Z&=\dv{\jcal_i}e^{T}\\
        &=T\left[\jcal_{i},\bullet\right]e^{T}.
    \end{split}
\end{align}
The expression $T\equiv T[\bullet,\bullet]$ has two arguments because the commutators in $T$ are always quadratic in $J_k$ and $j_{m,\sigma}$.
For the rest of the discussion of the algorithm, we will not distinguish $j_{m,\sigma}$ and $J_k$.
We can always write them in terms of a general $\jcal_i$ by using $i$ as a multi-index.

This new notation helps us to demonstrate that many terms are 0 and we can drop them.
Due to the commutativity of the derivatives and the step of setting $\jcal_k=0$ in the end [cf. eq.~\eqref{eq:def_matrix_elements_gen_func}], we find
\begin{align}
    \begin{split}
        T\left[\jcal_{i},\jcal_{i'}\right]&=T\left[\jcal_{i'},\jcal_{i}\right]\\
        \left.T\left[\bullet,\bullet\right]\right|_{\jcal_{i}=0\,\forall i} &=0\\
        \left.T\left[\jcal_{k},\bullet\right]\right|_{\jcal_{i}=0\,\forall i} &=0\,\forall k.
    \end{split}
    \label{eq:t_identities}
\end{align}
All expressions that are not derived twice must be zero if we set all $\jcal_{k}=0$ in the end because $\jcal_k$ appears quadratically in each commutator.
Thus, the number of $T$s for each derivative is given by $N_T=\frac{\sum_{i}n_{i}}{2}$, where $n_i$ are the occupation numbers of the full-system state and the partitioned state.

The restrictions described above lead to a more efficient algorithm in comparison to symbolic derivation of the full expression.
Considering the restrictions in equation~\eqref{eq:t_identities}, the result of the derivatives in~\eqref{eq:def_matrix_elements_gen_func} is heavily constrained.
Every term must be derived twice (otherwise it is 0).
Furthermore, we only sum over unique combinations since we can freely exchange the arguments of $T$ and the order of the $T[\jcal_{l_1},\jcal_{l_2}]$ in the product over $l$.

The input of the algorithm is a list of $\jcal_{i}$ with corresponding powers $n_i$ and we only compute combinations of fully derived $T$
\begin{align}
    \left.\prod_{i}\frac{d^{n_{i}}}{d\jcal_{i}^{n_{i}}}e^{T}\right|_{\jcal_{i}=0}=\sideset{}{^{'}}\sum_{k}c_{k}\prod_{l}T^{p_{kl}}\left[\jcal_{l_{1}},\jcal_{l_{2}}\right],
    \label{eq:derivative_unique_configs}
\end{align}
where $c_{k}$ are the multiplicities of the terms in $T$. 
The sum $\sideset{}{^{'}}\sum$ runs over all unique combinations of $T$. 
A combination is unique if it cannot be transformed into another combination of $T$s by swapping the arguments of $T$ or commuting $T$s.
This corresponds to iterating over all pairwise lexicographically ordered tuples of $\jcal$.
The exponents $p_{kl}$ are the powers of certain terms $T$ if the same arguments $(\jcal_{l_1}, \jcal_{l_2})$ appear multiple times in the same sequence.
Thus, the number of terms in the product over $l$ can vary depending on the number of individual combinations of $(\jcal_{l_1},\jcal_{l_2})$.
Since we are not considering the full system and the split modes separately at the moment, the indices $k,m$, and $l$ are used without further implications here.
For a more detailed discussion of the prefactors $c_k$, we refer to appendix~\ref{sec:app_proof_prefactors}.

For concreteness, we consider a simple example of three modes $\jcal_1$, $\jcal_2$ and $\jcal_3$ to illustrate the procedure.
We are interested in the second derivative with respect to each $\jcal_i$.
In terms of occupation numbers, we can write the configuration as $\vec{n}=(2,2,2)$, where $\vec{n}$ is the vector of occupations numbers.
The primed sum in~\eqref{eq:derivative_unique_configs} runs over all unique configurations of strings.
In our example, there are five distinct configurations
\begin{align}
    \begin{split}
        1: &\left(\jcal_1,\jcal_1\right),\left(\jcal_2,\jcal_2\right),\left(\jcal_3,\jcal_3\right)\\ 
        2: &\left(\jcal_1,\jcal_2\right),\left(\jcal_1,\jcal_2\right),\left(\jcal_3,\jcal_3\right)\\ 
        3: &\left(\jcal_1,\jcal_1\right),\left(\jcal_2,\jcal_3\right),\left(\jcal_2,\jcal_3\right)\\ 
        4: &\left(\jcal_1,\jcal_3\right),\left(\jcal_1,\jcal_3\right),\left(\jcal_2,\jcal_2\right)\\ 
        5: &\left(\jcal_1,\jcal_2\right),\left(\jcal_1,\jcal_3\right),\left(\jcal_2,\jcal_3\right).
    \end{split}
    \label{eq:example_t_strings}
\end{align}
The tuples represent the arguments of $T\left[ \bullet,\bullet \right]$.
All other combinations other than those listed in eq.~\eqref{eq:example_t_strings} can either be generated by swapping tuples or by exchanging the arguments inside of a tuple.
A swap of two tuples is allowed due to the commutativity of multiplication in~\eqref{eq:derivative_unique_configs}.
The exchange of arguments is equivalent to exchanging the derivatives of a single $T$ which corresponds to one of the identities in~\eqref{eq:t_identities}.
Since there are six derivatives in total, we must have three distinct $T$ terms in each string.
We can express the five combinations in~\eqref{eq:example_t_strings} more compactly with powers~$p_{kl}$
\begin{align}
    \begin{split}
        1: &\left(\jcal_1,\jcal_1\right),\left(\jcal_2,\jcal_2\right),\left(\jcal_3,\jcal_3\right)\\ 
        2: &\left(\jcal_1,\jcal_2\right)^2,\left(\jcal_3,\jcal_3\right)\\ 
        3: &\left(\jcal_1,\jcal_1\right),\left(\jcal_2,\jcal_3\right)^2\\ 
        4: &\left(\jcal_1,\jcal_3\right)^2,\left(\jcal_2,\jcal_2\right)\\ 
        5: &\left(\jcal_1,\jcal_2\right),\left(\jcal_1,\jcal_3\right),\left(\jcal_2,\jcal_3\right).
    \end{split}
    \label{eq:example_t_strings_powers}
\end{align}
Here, $k$ is the index of the overall combination of all pairs $\jcal$ and $l$ is the index of the tuple in the string. 
More concretely, $p_{2,1}=2$ because the second string contains $(\jcal_1,\jcal_2)^2$ as first pair.

Some of the configurations in~\eqref{eq:example_t_strings} may appear multiple times during the application of the product rule in~\eqref{eq:derivative_unique_configs}.
Thus, we have to take care of the multiplicities in front of the terms. 
In our simple example, we can just list them as $\vec{c}=(1,2,2,2,8)$.
Here, they are calculated by explicitly performing the derivatives on the left side of~\eqref{eq:derivative_unique_configs}.
The primed sum in~\eqref{eq:derivative_unique_configs} can be evaluated given all configurations in~\eqref{eq:example_t_strings} and the vector $\vec{c}$.
All terms of the form $T[\jcal_{l_1},\jcal_{l_2}]$ are numbers that can be evaluated by summing the derivatives of commutators in~\eqref{eq:def_z} explicitly.

As demonstrated in the example, the computation of~\eqref{eq:derivative_unique_configs} can be divided into two subproblems.
Firstly, we have to determine all unique combinations of pairs $\left(\jcal_{l_{1}},\jcal_{l_{2}}\right)$ for a given $\vec{n}$.
Secondly, we have to compute the coefficients $c_{k}$ given $p_{kl}$ and the tuples $(\jcal_{l_{1}},\jcal_{l_{2}})$.

The first task can be solved with a tree-based algorithm that is described in detail in appendix~\ref{sec:app_tree}.
The idea is to build only the combinations of tuples $(\jcal_{l_1},\jcal_{l_2})$ that adhere to the uniqueness condition defined for the primed sum, i.e. lexicographical ordering of all index tuples.
The condition can be checked locally at every node of the tree.
Thus, only nodes that can still build valid configurations are expanded in subsequent operations.
The trivial approach of listing all combinations of $\jcal_i$ for a given $\vec{n}$ and filtering for the unique ones gets prohibitively costly already for low cut-offs.

The coefficients $c_k$ have a closed form expression and are given by
\begin{align}
    c_{k}=\frac{\prod_{i}\left(n_{i}!\right)}{2^{N_{\text{diag},k}}\prod_{l}\left(p_{kl}!\right)},
    \label{eq:def_multiplicities}
\end{align}
where $n_i$ are the occupation numbers and $N_{\text{diag},k}$ is the number of identical arguments for T in the string with index $k$.
In our example of $\vec{n}=(2,2,2)$, $N_{\text{diag},1}=3$ and $N_{\text{diag},3}=1$.
The proof of the equation is given in appendix~\ref{sec:app_proof_prefactors}.

Finally, we can put all the pieces together.
An element of $U_T$ corresponds to the calculation of an overlap of the form $\braket{\nlr}{\nf}$.
Each of the states is given as an occupation number vector.
Equation~\eqref{eq:def_matrix_elements_gen_func} connects the occupation numbers to derivatives of a scalar function.
These derivatives can be computed explicitly by first enumerating all unique configurations of the primed sum in~\eqref{eq:derivative_unique_configs}.
Each of the tuples in a configuration represents the arguments of $T$.
The derivatives of $T$ for some tuple $(\jcal_{l_1},\jcal_{l_2})$ can be evaluated explicitly. 
The product of all $T$ values in a string is weighed by a factor~\eqref{eq:def_multiplicities} and summed to yield the final value of the matrix element.
The explicit expressions for the derivatives of the commutators are given in appendix~\ref{sec:app_commutators}.

\section{\label{sec:models}Models}
\subsection{\label{sec:models_kg}Klein-Gordon model}

For concreteness, we demonstrate the potential of our method on two well known QFT models. The first example is the \textit{Klein-Gordon} (KG) \textit{model}, the massive free boson theory, described by the Hamiltonian
\begin{align}
    H_{\text{mFB}}=\frac{1}{2}\int_0^{L}\dd x \left[ (\partial_t\phi(x))^2 + (\partial_x\phi(x))^2 +m^2 \phi^2(x)\right],
    \label{eq:kg_hamiltonian}
\end{align}
where $\phi(x)$ is a real scalar field and $m$ is the boson mass.
This free model serves as a perfect test bed for our method.
Its entanglement properties are known analytically both from  replica trick techniques~\cite{calabrese_entanglement_2004,casini_entanglement_2009} and from covariance matrix methods~\cite{casini_entanglement_2009,serafini_quantum_2017} including the equilibrium states and the non-equilibrium dynamics.
For the massless case the following well known result has been derived~\cite{calabrese_entanglement_2004}:
\begin{align}
    S(\ell)=\frac{c}{6}\log\left(\frac{L}{\pi a}\sin\left(\frac{\pi \ell}{L}\right)\right)+2g+U(a)\label{eq:log_law_entanglement}
\end{align}
where the central charge of the CFT $c=1$, $a$ is a UV cutoff, $g$ is the Affleck-Ludwig boundary entropy~\cite{affleck_universal_1991} and $U(a)$ is a non-universal constant dependent on the precise form of the cutoff. 

Although being a free theory, the massive case of the KG model is the first nontrivial test example of our method. 
While the massless case is diagonal in our computational basis, the massive case is fully non-diagonal. 
Due to a finite correlation length $\xi\sim\frac{1}{m}$ the entanglement for $m>\frac{1}{L}$  saturates to an area law plateau where $S(\ell)=\text{const}$.
At distance closer than $\xi$ to the boundaries, the curve $S_N$ interpolates smoothly to the zero value at the boundaries. 
For $m<\frac{1}{L}$, there is a smooth crossover from a log law to an area law scaling of the entanglement entropy.

For thermal states, the vN entanglement entropy becomes the thermodynamic entropy and there is a smooth crossover with increasing temperature to a volume law $S(\ell)\propto\ell$. 
In non-equilibrium dynamics in the massless case, the vN entropy is expected to grow linearly in time~\cite{calabrese_entanglement_2009}. 
In case of a finite system, the growth stops when excitations from the splitting point reach the boundaries of the system and one expects recurrent dynamics.
In the massive case, the linear growth is superposed with an oscillatory component with a frequency given by the boson mass~\cite{alba_entanglement_2018}. 

To generate analytical predictions for the KG model to compare our numerical method against, we employ the covariance matrix formalism~\cite{casini_entanglement_2009,serafini_quantum_2017}.
It is a convenient framework because of its simplicity and because it also enables to model cutoff effects. 
Further details are outlined in appendix~\ref{sec:app_covmat}.

\subsection{\label{sec:models_sg}Sine Gordon model}
A paradigmatic model of strongly interacting QFT is the \textit{sine-Gordon} (sG) \textit{model} 
\begin{align}
    \begin{split}
    H_{\text{sG}}&=\int \dd x \Big[ \frac{1}{2}\left\{ (\partial_t\phi(x))^2+ (\partial_x\phi(x))^2 \right\}\\
    &\hspace{3cm}- \frac{m^2}{\beta^2}\cos(\beta \phi(x)) \Big]
    \end{split}
    \label{eq:sg_hamiltonian}
\end{align}
with the mass parameter $m$ and the interaction parameter $\beta$. 
The sG model is one of the simplest models displaying confinement and is an integrable model solvable by S-matrix bootstrap techniques~\cite{mussardo_statistical_2020}. 
The model has solitonic topological excitations and a rich phase diagram. 
For $\beta<\sqrt{4\pi}$ the interaction is attractive and the solitons form bound states - breathers. 
For $\sqrt{4\pi}<\beta<\sqrt{8\pi}$ the interaction is repulsive, for the separating line $\beta=\sqrt{4\pi}$, the model can be mapped to a free Dirac fermion and at $\beta\sim\sqrt{8\pi}$ the model undergoes a Berezinskii–Kosterlitz–Thouless phase transition to a free model~\cite{mussardo_statistical_2020}.

A convenient way to parameterize the sG interaction parameter in the attractive regime $\beta<\sqrt{4\pi}$ is
\begin{align}
    \beta^2=\frac{8\pi}{1+\lambda}\label{eq:sg_lambda}
\end{align}
where the parameter $\lambda$ is convenient because $\left\lfloor\lambda \right\rfloor$ equals number of breathers present in the sG spectrum. The mass $m_n$ of the $n$-th breather is given by
\begin{align}
    m_n=2M\sin\left(\frac{n\pi}{2\lambda}\right),
    \label{eq:breather_mass}
\end{align}
where $M$ is the soliton mass. In particular, the mass of the lightest particle, the first breather, $m_1$ determines the gap of the system. 
In finite system size, these masses get modified and can be computed using the form factor and boundary bootstrap formalism~\cite{ghoshal_boundary_1994,ghoshal_bound_1994,mattsson_boundary_2000,bajnok_finite_2002}.
Each of the breathers has a tower of excited states as a result of acquiring a nonzero momentum. 
The set of allowed momentum values is discrete in finite volume.
The expressions for finite volume breather energies are given in appendix~\ref{sec:app_sg_breather}.

The entanglement properties of the sG model in the repulsive regime have been studied by spectral form factor and corner transfer matrix techniques ~\cite{castro-alvaredo_bi-partite_2008,ercolessi_exact_2010} and predict the height of the vN entropy area law plateau
\begin{align}
    \begin{split}
        S=&\frac{1}{6}\log\left(\frac{1}{Ma}\right)+\frac{1}{6}\log\left(\frac{\sin\left[\pi\left(1-\frac{\beta^2}{8\pi}\right)\right]}{1-\frac{\beta^2}{8\pi}}\right)\\
        &+O\left(\frac{1}{\log(a)}\right),
    \end{split}
\end{align}
where $M$ is the soliton mass which is a function of  $m$ and $\beta$. 
In the attractive regime, the entanglement properties are less understood. 
Based on general arguments for gapped systems, the vN entropy plateau is expected to follow the  form ~\cite{doyon_bipartite_2009}:
\begin{align}
    S=\frac{c}{3}\log(\xi_1)+U-\frac{1}{8}\sum_{\alpha=1}^{\ell}K_0(2\ell m_{\alpha})+O(e^{-3rm_1}),
    \label{eq:vN_entropy_massive_general}
\end{align}
where $K_0$ is the modified Bessel function, $c$ is the central charge of the UV critical point, $m_{\alpha}$ are the masses of the particles in the spectrum (breathers in the sG case), $\xi_1$ the correlation length corresponding to the lightest particle and $U$ a constant.

Concerning the non-equilibrium dynamics, it has been recently shown using form factor techniques for small quenches of the sG model that the entanglement entropy exhibits non damped oscillations in time with frequencies corresponding to even breather masses~\cite{castro-alvaredo_branch_2021}.

Here we implement a HT for the sG as developed in~\cite{feverati_scaling_1998, bajnok_finite_2002}. 
We list all the HT matrix elements used in appendix~\ref{sec:app_ht_matrix_elements}.

\section{\label{sec:results}Results}
The following section is structured in two main parts.
In the first part, we show results of the method for the Klein Gordon model in equilibrium.
These results are compared to covariance matrix calculations and serve as a benchmark.
Additionally, we show results for the interacting sine Gordon model to demonstrate that our method works beyond the free regime.
The second part of the result section contains non-equilibrium evolution of the von Neumann entropy in real time for Klein Gordon and sine Gordon models.

All results shown in this section are computed for a finite cut-off $s_F=18$. 
This corresponds to 1597 states. 
The boundary conditions at the cut are chosen to be Neumann while the (physical) boundary conditions at the outer edges are Dirichlet.
A constant mode density truncation scheme is used in all computations.
Further details on the cut-off scheme are described in appendix~\ref{sec:app_bgb_trafo}.
Normalisations (like the prefactor in~\eqref{eq:norm_matrix_elements_basic}) are enforced by normalising the reduced density matrix numerically.

\subsection{\label{sec:results_equil}Equilibrium}
The Klein Gordon model~\eqref{eq:kg_hamiltonian} represents a non-trivial check for HT because its Hamiltonian is non-diagonal when expanded in the massless (CFT) basis for any mass $m\neq0$.

\begin{figure}
    \centering
    \includegraphics[width=\columnwidth]{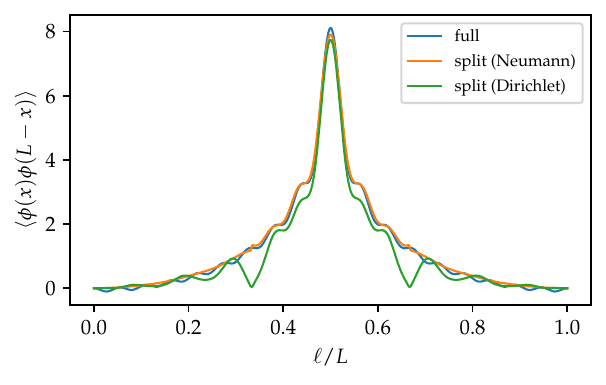}
    \caption{Correlations $\expval{\phi(x)\phi(L-x)}$ of the Klein-Gordon model with mass $m=1.0$. 
    The plot shows the correlations of the system computed with the original (full) mode decomposition of the fields as a reference.
    The system is cut at $\ell/L=1/3$.}
    \label{fig:correlations}
\end{figure}

As an initial check for the splitting procedure, we reproduce the correlations of the Klein Gordon theory in terms of the split modes.
The correlations $\expval{\phi(x)\phi(L-x)}$ can be either calculated in terms of the full fields $\phi$ acting on the full interval density matrix $\rho$ or in terms of the split fields of the left and right partition $\phi_L$ and $\phi_R$ acting on the partitioned density matrix $\rho_{LR}$:
\begin{align}
    \begin{split}
        \expval{\phi(x)\phi(L-x)}&=\Tr(\phi(x)\phi(L-x)\rho)\\
        &=\Tr(\phi_{L/R}(x)\phi_{L/R}(L-x)\rho_{LR}).
    \end{split}
\end{align}
Here, we use the notation $\phi_{L/R}$ to refer to the field on the sub-interval that $x$ belongs to.
Figure~\ref{fig:correlations} compares the correlations across the full range of the system for a cut at position $\ell/L=1/3$ for Neumann and Dirichlet boundary conditions at the cut.
Both of the split field curves agree well with the correlations of the full system.
In the case of Neumann boundary conditions at the cut, we only observe deviations at the cut.
A plateau forms around the split at $\ell$, since we impose $\partial_x\phi=0$.
The Dirichlet boundary conditions enforce $\phi=0$ at $\ell$ and we notice that the correlations drop to zero as expected.
The figure is symmetric around $\ell/L=0.5$ due to choice of arguments in the correlator.
The overall wavy features in the curve for the full and the partial modes are a feature of the finite cut-off in HT. 
With an increase in the cut-off, we expect these features to reduce in amplitude.
Since correlations with Neumann boundary conditions at the cut agree better with the full correlations, we choose Neumann boundary conditions at the cut for all further entropy computations.
We expect Dirichlet boundary conditions at the cut to be eventually equivalent to the choice of Neumann boundary conditions for higher cut-offs (cf. section~\ref{sec:equiv_hilbert}).

\begin{figure}
    \centering
   \includegraphics[width=\columnwidth]{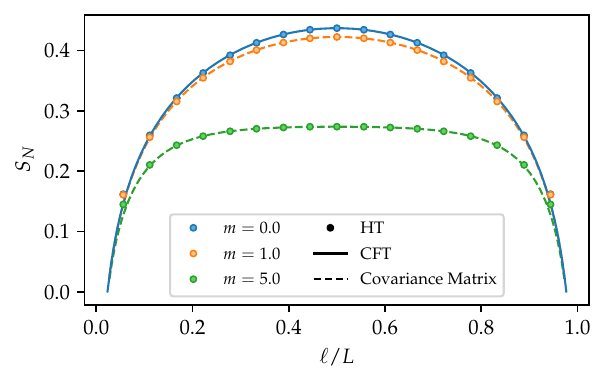}
    \caption{Spatially resolved von Neumann entropy for the Klein-Gordon model at different masses $m$ (displayed in different colors).
    Different methods are encoded in the linestyle.
    The massless case is compared to the CFT result while the massive cases are compared to covariance matrices computations.
    }
    \label{fig:entropy_kg}
\end{figure}

We continue checking the performance of our method by calculating the von Neumann entropy.
We compare the von Neumann entropy with an analytic calculation using the covariance matrix approach (cf. Figure~\ref{fig:entropy_kg}).
The formalism is explained in detail in section~\ref{sec:app_covmat}.
All covariance matrix computations in this section are performed using 200 momentum modes.
HT entropies are calculated at all points $\ell/L=n/s_F$, $n=1,\dots,s_F-1$ since the bosonic commutation relations in the truncated split basis are fulfilled best at these points (cf. appendix~\ref{sec:app_bgb_trafo}).
The calculation of the entropy at other points is possible, but will result in more significant errors due to the truncation effect leading to worse preservation of the canonical commutation relation by the splitting procedure.
The covariance matrix results (dashed lines) and the CFT results (solid lines) in the figure are shifted by a constant to coincide with the HT curves at $\ell/L=0.5$ for ease of comparison. 
This accounts for the non-universal cutoff dependent constant (see for example eq.~\eqref{eq:log_law_entanglement}) which is slightly different in the analytic and the HT case due to the different truncation schemes.  

In all cases, our method agrees excellently with the analytic predictions.
The massless boson shows the expected logarithmic growth in entropy. This agrees perfectly with  the CFT prediction~\cite{calabrese_entanglement_2004}, eq. \eqref{eq:log_law_entanglement}.

With increasing mass, the curve develops a flat plateau in the central region, transitioning to the area law regime as expected for a massive boson.
For distances less than a correlation length away from the boundary, the curve undergoes a non-linear behaviour before it reaches 0 at the boundaries due to finite size effects.

Similar data can be obtained for Dirichlet boundary conditions at the cut.
We present the results for Neumann boundary conditions here since they show better agreement with the expectation.
The Dirichlet data has slightly stronger deviations close to the boundaries.

\begin{figure}
    \centering
    \includegraphics[width=\columnwidth]{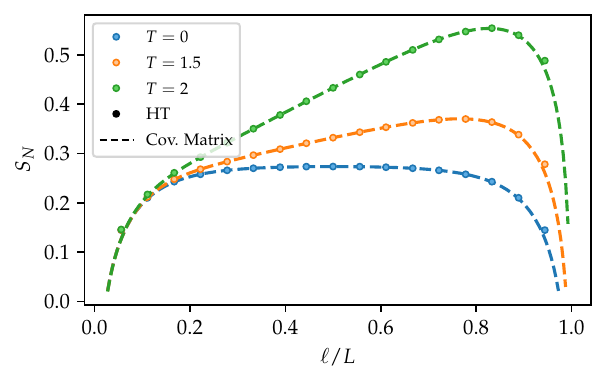}
    \caption{Von Neumann entropy of thermal states of the Klein-Gordon system with $m=5$.
    The dots are the results of HT and the dashed lines are covariance matrix computations.
    Entropies at finite temperature are computed from a Boltzmann distribution at temperature $T$.
    The curve at zero temperature uses the groundstate of the system.}
    \label{fig:therm_kg}
\end{figure}

HT provides access to the reduced density matrix at arbitrary temperatures so in addition to ground state properties, we can also access the von Neumann entropy of thermal states. For $T>0$, the vN entropy coincides with the classical thermodynamic entropy. 
Figure~\ref{fig:therm_kg} shows the entanglement entropy ($T=0$) and the thermodynamic entropy ($T>0$) of a massive ($m=5$) free boson at different spatial positions.
As before, the results obtained by our method agree well with the covariance matrix computation.
The dashed curves are again shifted to coincide at $\ell/L=0.5$ to account for cut-off dependent constants.
At $T=0$, the we see the expected plateau of the area law of the entanglement entropy.
At a finite temperature, the entropy becomes extensive and grows linear with the system size.

\begin{figure}
    \centering
    \includegraphics[width=\columnwidth]{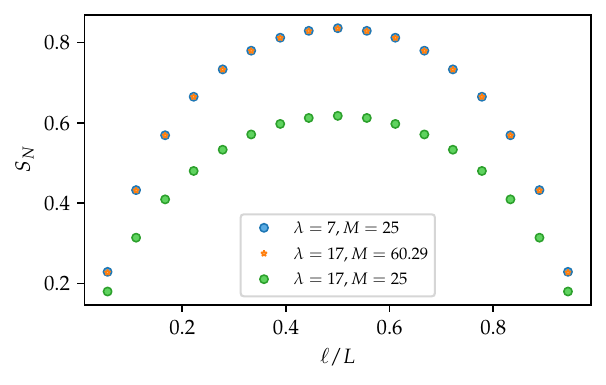}
    \caption{Spatially resolved von Neumann entropy of the sine Gordon model for Neumann boundary conditions at the cut.
    The curves are plotted for the interaction parameter $\lambda=7$ and soliton mass $M=25$ and for $\lambda=17$ for two different mass values, $M=25$ at $M=60.29$. 
    At $M=60.29$ the gap (mass of the first breather) agrees with that of the $\lambda=7$ case and is $m_1=11.13L$. 
    Thus, the correlation length is less than one tenth of the system size.
    The curves display a logarithmic scaling of the entanglement entropy.}
    \label{fig:entropy_sg}
\end{figure}

The basis transformation from a full to a split system is independent of the model.
Thus, we apply the same methodology to the interacting sine-Gordon Hamiltonian~\eqref{eq:sg_hamiltonian} as shown in Figure~\ref{fig:entropy_sg}. 
We compare the curves for two different values of the coupling parameter $\lambda$ and different values of the soliton mass $M$. 
The cases of $\lambda=7$, $M=25$ and $\lambda=17$, $M=60.29$ are chosen such that the gap of the model (the mass of the first breather) matches.
In comparison to the Klein-Gordon model, we do not see the onset of a plateau in the middle of the curve. 
For the matching breather mass case,  the gap is $m_1=11.13L$ meaning that the correlation length is less than one tenth of the system size. 
At such a short correlation length, an area law plateau would generally be expected. 
The log-like deviation from that could be indicative of longer range entanglement in the sG case which could be a consequence of the topological nature of solitons or a subtlety of the continuum missed by discrete calculations.
It would be interesting to further understand this surprising scaling with analytical tools. 

The perfect overlap of the curves $\lambda=7$, $M=25$ and $\lambda=17$, $M=60.29$ indicates that the vN entropy scaling in the attractive regime of the sG model is dominated solely by the first breather and not by the higher particles in the spectrum. 
This is consistent with the general expression~\eqref{eq:vN_entropy_massive_general}. 
At large volumes, the $K_0$ corrections are highly suppressed, resulting in the value of vN entropy depending only on the correlation length.

\subsection{\label{sec:results_dynamics}Real time Dynamics}

We continue by using our method to study the real time dynamics of the vN entropy following quenches.

\begin{figure*}
    \centering
    \includegraphics[width=\textwidth]{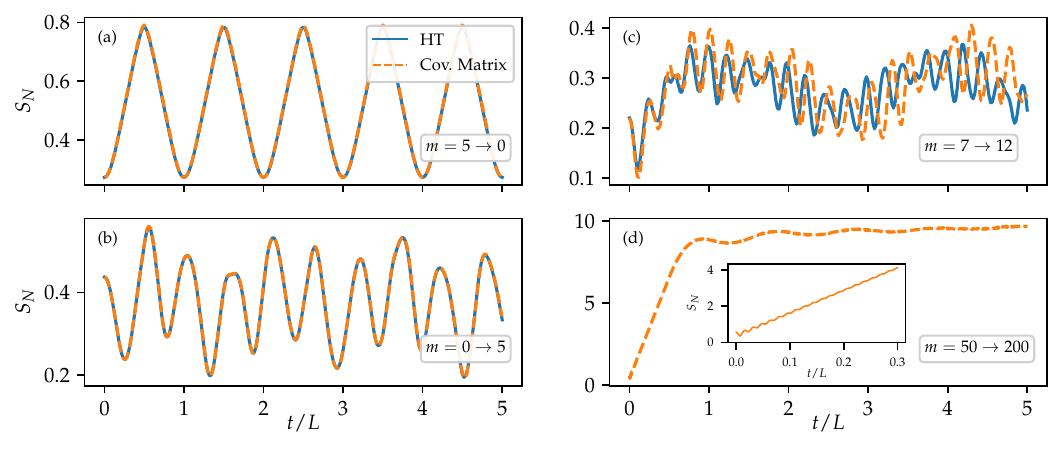}
    \caption{
    Time evolution of the von Neumann entropy after quenches in the Klein Gordon model.
    The system is split at $\ell/L=0.5$.
    Different panels show quenches from and to different masses (as indicated in the insets).
    Solid lines are results from HT and dashed lines represent the covariance matrix results.
    The inset in panel~(d) details the behaviour of the quench from  $m_0=50$ to $m=200$ at short times.
    }
    \label{fig:kg_to_kg_quench}
\end{figure*}

We begin with the analytically tractable KG model. 
The mass quenches with several increasing post-quench masses are shown in Figure~\ref{fig:kg_to_kg_quench}. 
We study the dynamics of the vN entropy between the left and the right half of the system ($\ell=0.5$) and compare the HT results with the analytical results from the covariance matrix formalism.
We displace the curves by a constant such that they start from the same point. 
This is to account for the non-universal cutoff dependent constant resulting form the difference in truncation schemes in the two methods. 

In the quenches to the massless post-quench Hamiltonian, we observe the expected CFT linear growth of the vN entropy. 
The linear growth is interrupted at $t=L/2$  by a reflection when the quasiparticles from the cut reach the system boundaries. 
At $t=L$ this results in a recurrence and the free dispersionless nature of the model leads to periodic dynamics.
This is shown in panel~(a).

At nonzero mass, the vN entropy develops an oscillatory component with a frequency proportional to the boson mass $m$. 
For a thermodynamically large system $L\gg1/m$, oscillatory dynamics are expected to be on top of a linear growth before reaching a plateau. 
This is indeed what we observe in panel~(d) with the largest mass case. 

At intermediate masses [panels (b) and (c)], the vN entropy is influenced by both factors - the massive particle and the finite system size.
For masses of the order of the system size, the oscillations driven by the mass are visible but the oscillations coming from the reflections from the boundaries are still prominent. Thus, the linear growth becomes obscured by them which is what we see in panel~(b). 
For intermediate masses for which the correlation length is an order of magnitude but not more smaller than the system size, the linear growth becomes visible as shown in panel~(c). 
However, the plateau keeps undergoing significant oscillations due to reflections from the boundary.

The KG plots in Figure~\ref{fig:kg_to_kg_quench} expose the limitations of the truncated Hamiltonian approximation.
At smaller masses [panels (a) and (b)], the HT results with our method match perfectly the analytic prediction up to times several times longer than the system size. 
This shows that the HT calculation of real time dynamics can be very reliable up to considerably long times. 
At higher masses [panel (c)], the real time dynamics start to deviate from the analytic curve for late times and the curve develops a phase shift. 
This is due to truncation effects -- at higher masses the low energy part of the Hilbert space becomes too small to accommodate all the relevant modes for the dynamics. 
The quality of the time evolution depends also on the amplitude of the quench, the difference between the pre- and the postquench mass. 
For small quenches, the HT evolution is reliable even at large masses and for bigger quenches it gets less reliable also at smaller masses. 
This is because a large quench generates excitations high up in the spectrum, exceeding the HT truncation. 
The very high masses shown in panel~(d) cannot be reliably simulated with our current implementation of the HT and we show only the analytic curve to support the discussions in the previous paragraphs.
Such high masses could be implemented also with our methods, though, if a massive basis was chosen for the HT expansion instead of the massless basis.
In this case, HT becomes exact also at nonzero masses. 

\begin{figure}
    \centering
    \includegraphics[width=\columnwidth]{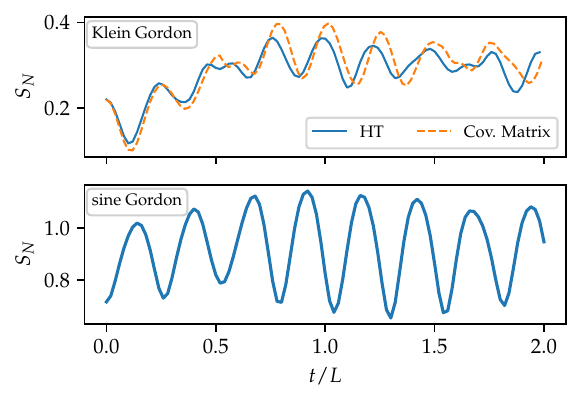}
    \caption{
    Time evolution of the von Neumann entropy after quenches. 
    The top panel shows the Klein-Gordon evolution of the entropy at $\ell/L=0.5$ for a quench from boson mass $m_0=7$ to $m=12$.
    The bottom panel depicts a quench of the sine Gordon model for $\lambda=7$ from a soliton mass of $M_0=15.73$ to $M=26.96$. The masses are chosen such that the first breather masses $m_1$ of the sG model agree with the KG boson masses.}
    \label{fig:kg_to_sg_quench}
\end{figure}

\begin{figure}
    \centering
    \includegraphics[width=\columnwidth]{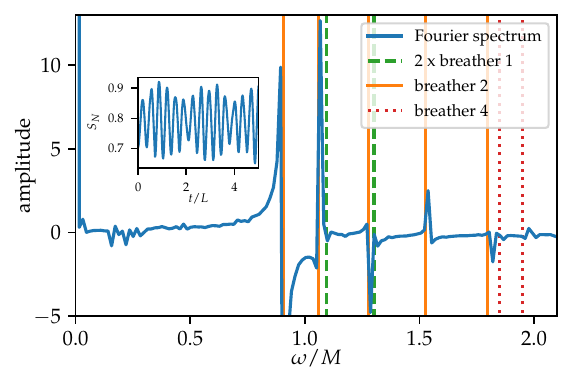}
    \caption{
    Frequency spectrum of the von Neumann entropy $S_N$ time evolution after a sine-Gordon mass quench from soliton mass $M_0=15$ to $M=20$ at $\lambda=7$. 
    The spectrum is obtained using a discrete Fourier transform. 
    The amplitude at frequencies $\omega$ is compared against energy levels of (moving) breathers. 
    The breather energies are computed using reflection factors~\cite{ghoshal_boundary_1994,ghoshal_bound_1994,mattsson_boundary_2000,bajnok_finite_2002}, the expressions are listed in Appendix~\ref{sec:app_sg_breather}.
    Due to the charge conjugation symmetry, only $C$-even breather state frequencies appear.
    The inset shows the original time evolution of the quench.}
    \label{fig:fourier_sg_quench}
\end{figure}

We turn to studying the interacting sG dynamics. 
The vN entropy dynamics of the sG quench is shown in 
Figure~\ref{fig:kg_to_sg_quench} and compared with the KG quench. The comparison is done at such choices of the parameters that the gaps of the two systems agree.  
We observe an oscillatory motion as predicted by recent work by Castro-Alvaredo~\cite{castro-alvaredo_branch_2021}. 
As known previously in the literature~\cite{alba_entanglement_2018} and also demonstrated here in Figure~\ref{fig:kg_to_kg_quench}, the oscillating dynamics is a generic consequence of the gap and in not a special feature of interaction.  
From our present results it is however not yet possible to determine whether the oscillations in sG quenches remain undamped at longer times as predicted by~\cite{castro-alvaredo_branch_2021}.  
In Figure~\ref{fig:fourier_sg_quench} we perform the Fourier analysis of the time series and compare the frequency spectra with breather energy levels. 
In order to have a more reliable time evolution at longer times, we study a small quench in mass - a quench generated by a moderate change of the soliton mass.  
The analytical breather energies for comparison are computed using the form factor and boundary bootstrap formalism in references~\cite{ghoshal_boundary_1994,ghoshal_bound_1994,mattsson_boundary_2000,bajnok_finite_2002}, for completeness the  expressions are listed also in Appendix~\ref{sec:app_sg_breather}. 
The sG ground states which are the prequench states are even under the charge conjugation $C:\phi\rightarrow-\phi$ which interchanges solitons with anti-solitons. 
Therefore, as predicted by~\cite{castro-alvaredo_branch_2021} and confirmed also by our results, only $C$ even states get populated during the quench. 
These include even breather states and even multiples of odd breathers.

Interesting questions remain whether the oscillations in the sG dynamics of vN entropy are damped and whether there is a linear growth superposed to oscillations as in the KG case or not. 
As shown in fig.~\ref{fig:kg_to_sg_quench}, there seems to be a slight growth at early times but it is hard to determine reliably whether it is not just an oscillation of another slower frequency superposed on top of the higher frequency oscillations. 
In order to study both questions, the quenches would have to be computed at a much larger post quench soliton mass, allowing to explore larger times before the reflection from the boundaries. 
Recently, an advanced HT implementation of the sG model has been developed~\cite{horvath_chirally_2022} allowing for calculations with Hilbert space sizes of several hundred thousand states. 
It would be interesting to combine our method with such an approach to study sG quenches in large volume. 
This would, however, require even more efficient approaches to deal with the exponential complexity of derivatives discussed in Section~\ref{sec:exponential_complexity}.

\section{\label{sec:conclusion} Conclusion}
We presented a method to compute a reduced density matrix of a quantum field theory within the Hamiltonian truncation framework. 
Our method constructs a unitary transformation between the Hilbert space of the full system and a tensor product of Hilbert spaces corresponding to the subsystems.
This maps the density matrix of a state into a form which is convenient for taking partial traces.

The method allows for the direct evaluation of a wide spectrum of entanglement related quantities, including von Neumann and Rényi entropies, mutual information, entanglement negativity and entanglement Hamiltonians. 
Our method makes it possible to study entanglement in ground, excited and thermal states as well as the real time evolution in non-equilibrium dynamics.
Furthermore, our method is model-independent and can be applied for any HT that is based on an expansion around a free (massive or massless) theory, which is a common choice in modern implementations.
By construction, this method could in principle be applied in dimensions $D>1+1$.  

We benchmarked the method using the massive free boson. 
Despite being a free theory, it represents a nontrivial test of the method because its Hamiltonian is a non-diagonal perturbation of the massless free theory.
The exact solutions for this model can be obtained using covariance matrix methods~\cite{serafini_quantum_2017} making it suitable as a benchmarking model. 
We found excellent agreement of the von Neumann entropy with theoretical predictions for ground and thermal states as well as for dynamics after quenches. 
We have demonstrated that the method is capable of a reliable time evolution up to times several times longer than the system size.

We proceeded by studying an interacting system, the sine-Gordon field theory in the attractive regime. 
For the scaling of the ground state von Neumann entropy,  we found sG ground states to be much more long-range entangled than KG ground states, exhibiting a logarithmic scaling.
In the large volume regime, the vN entropy depends only on the gap of the system but not the higher particle content.
Studying the quench dynamics of the sG model, we found an oscillating behavior, as predicted by~\cite{castro-alvaredo_branch_2021}. 
The resonances in the frequency spectrum of the time series matched the masses of the lowest breather states even under charge conjugation.  
The questions whether the oscillations are damped and superposed with a linear growth remains unanswered. 
In order to study that, a more sophisitcated implementation of the sG model would be required that would allow for higher cutoffs. 
The recently developed chirally factorised approach~\cite{horvath_chirally_2022} could be a suitable candidate. 

Our method opens the doors to many interesting explorations and can be extended in several directions. 
It would be interesting to explore further the oscillatory time dependence of the vN entropy dynamics following quenches. 
Here, a possible direction could be to explore the role of integrability and the effects of integrability breaking. 
To do that, non-integrable perturbations of the sG model could be considered, like the double sG model and the massive sG model.
Another more fundamental possibility would be the $\phi^4$ theory which is a canonical non-integrable QFT model and has already been successfully implemented in the HT framework~\cite{rychkov_hamiltonian_2015,rychkov_hamiltonian_2016}. 
Our method can be adapted for the $\phi^4$ model with a straightforward step of computing the Bogoliubov coefficients for the massive field HT expansion. 

Furthermore, it would be interesting to study the entanglement Hamiltonian and the Bisognano-Wichmann theorem~\cite{bisognano_duality_1975,bisognano_duality_1976}. 
Several interesting properties have been established for the CFT case~\cite{cardy_entanglement_2016,wen_entanglement_2018,giudici_entanglement_2018,roy_entanglement_2020} and it would be important to explore how they extend to the interacting gapped QFT~\cite{dalmonte_quantum_2018,kokail_entanglement_2021}.
The explicit representation of the reduced density matrix in a computational basis makes our method naturally suited to such a study.

Much attention has been recently devoted to symmetry resolved entanglement (see~\cite{goldstein_symmetry-resolved_2018,lukin_probing_2019,bonsignori_symmetry_2019,fraenkel_symmetry_2020,azses_symmetry-resolved_2020,parez_exact_2021,weisenberger_symmetry-resolved_2021,calabrese_symmetry-resolved_2021}). 
It would be an interesting extension of our method to make it sensitive to the symmetry charge of the subintervals and thus resolve the entanglement per sectors. 

An implementation of our method in $D=2+1$ would be a very interesting step because of the lack of methods in dimensions higher than  $D=1+1$.
By construction, our method can be easily generalised to any dimension.
The main obstacle would be the quickly growing size of the full and the split system Hilbert spaces. 
However, HT has already been successfully applied in higher dimensions~\cite{hogervorst_truncated_2015} and since our dimension of the split Hilbert space in practice does not exceed the dimension of the full Hilbert space, such an undertaking seems possible.

Finally, in case of free theories, massless and massive, our construction yields an exact construction of the reduced density matrix of the theory. 
It could be a fruitful direction to use that to get further analytical insights into the entanglement structure of QFT.

\begin{acknowledgments}
We would like to thank Spyros Sotiriadis, Ignacio Cirac, Gabor Takacs and Mari Carmen Ba{\~{n}}uls for many fruitful discussions.
Furthermore, we thank Teo Kukuljan for providing the proof of equation~\eqref{eq:def_multiplicities}. 
We thank Albert Gasull and David Horvath for comments on an early version of the manuscript.
The work of I.K. was supported by the Max-Planck-Harvard Research Center for Quantum Optics (MPHQ).
Patrick Emonts acknowledges support from the International Max-Planck Research School for Quantum Science and Technology (IMPRS-QST).
\end{acknowledgments}

\bibliography{references.bib}
\clearpage

\newpage
\appendix
\onecolumngrid

\FloatBarrier
\section{Cut-off effects and symplectic structure\label{sec:app_bgb_trafo}}
The main approximation in our method is the representation of the full system modes in terms of a finite number of partial modes.
We have to ensure that the approximation conserves basic properties of the system like the bosonic commutation relations.
Equation~\eqref{eq:bgb_restriction} can be reformulated to test the transformation as
\begin{align}
    M K M^{\dagger}&=K\label{eq:bgb_test_full_modes}\\
    M^\dagger KM&=K.\label{eq:bgb_test_split_modes}
\end{align}
Equation~\eqref{eq:bgb_test_full_modes} evaluates the commutation relations of the full modes $A_k$ expressed in terms of the partial modes.
The structure of $K$ on the diagonal of $\id$ on the first half of the diagonal and $-\id$ on the second half reflects the anti-symmetric nature of the commutator upon exchanging its arguments.
The commutation relations of the reverse transformation, partial modes expressed in full modes, are tested in~\eqref{eq:bgb_test_split_modes}.
The two equations provide us with an objective quality criterion of our truncated method.
If the commutation relations of the bosonic modes are not fulfilled, the transformation is invalid. 

\begin{figure}[h]
    \centering
    \includegraphics{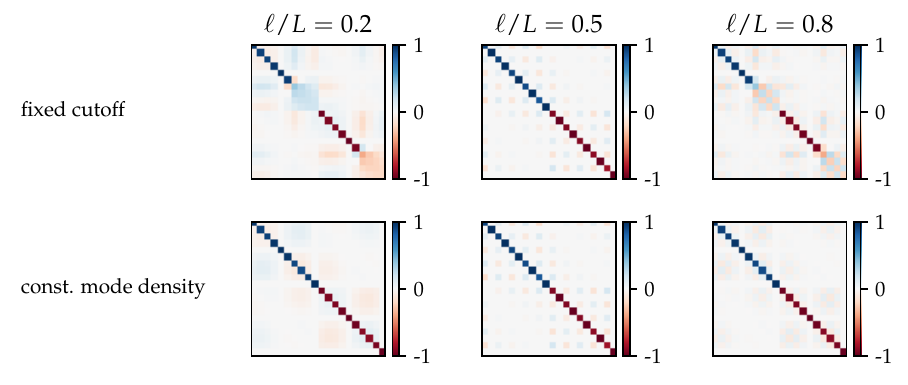}
    \caption{Check of the symplectic properties of the transformation from full modes to partial modes, eq.~\eqref{eq:bgb_test_full_modes}, for $s_F=10$.
    The fixed cut-off scheme does not faithfully reproduce the bosonic commutation relations except for the case of $x=0.5$.
    In this case, the two schemes coincide. }
    \label{fig:app_commutation_relations}
\end{figure}

As described in the main text, we consider two cut-off schemes.
The \textit{fixed cut-off} scheme distributes the partial modes symmetrically across both intervals ($s_L=s_R=\frac{s_F}{2}$).
The second scheme, \textit{constant mode density}, distributes the modes proportionally to the size of the intervals ($s_L=\frac{\ell}{L}s_F$, $s_R=s_F-s_L$).
The quality of the two schemes can be assessed by checking the commutation relations of the transformed modes.
Figure~\ref{fig:app_commutation_relations} shows the result of the calculation of~\eqref{eq:bgb_test_full_modes}.
The top row shows that the fixed cut-off scheme does not reproduce the bosonic commutation relations if the full modes are expressed in terms of partial modes for splits that are not at $x=0.5$.
A cut in the middle represents a special case.
Here, the constant mode density cut-off scheme and the fixed cut-off scheme coincide.
The bottom row illustrates that the constant mode density cut-off scheme reproduces the correct commutation relations for different cuts.
All the computations in the main text are performed with this cut-off scheme.

The constant mode density can be realised exactly only for a finite number of splittings $\ell$.
If we assume $s_F$ full modes, we can split the system at multiples of $1/s_F$, that is $\ell\in\{1/s_F,2/s_F,\ldots,L-1/s_F\}$ such that we have the exact same density of modes on the left  and the right side of the cut ($s_L/\ell=s_R/(L-\ell)$).
We call those values commensurate cuts.
For other points, the mode densities cannot be chosen to be the same on the two partitions. 
A possible choice could be to pick a rounding scheme for the distribution of the partial cutoffs, for example $s_L=\text{round}(\frac{\ell}{L}s_F)$. 
Unfortunately, this leads to imperfect realisation of the symplectic structure~\eqref{eq:bgb_test_split_modes}.

\begin{figure}[h]
    \centering
\includegraphics{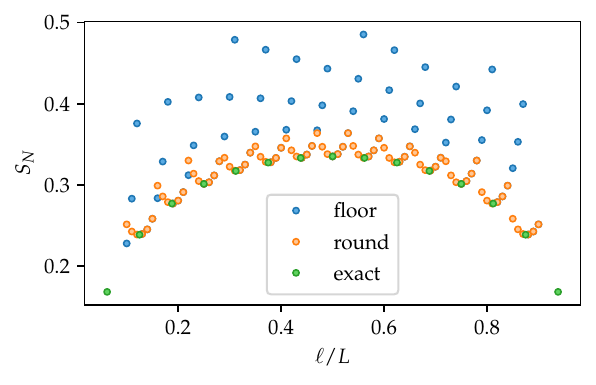}
    \caption{Effect of rounding schemes for the number of modes $s_L$ and $s_R$ in the left and right partition for Dirichlet boundary conditions at the cut.
    The system is a Klein-Gordon model with mass $m=1$.
    The transformation yields the correct bosonic commutation on commensurate cuts (green) with distance $1/s_F$.
    If we do not cut at commensurate splits, the $s_L$ can be either floored to the next integer (blue) or rounded (orange).}
    \label{fig:app_rounding}
\end{figure}

As expected the errors in the symplectic structure in non-commensurate points lead to incorrect values of the vN entropy.
We show in Figure~\ref{fig:app_rounding} the influence of different rounding schemes.
The points in blue floor the number of modes in the left partition $s_L=\floor*{\frac{\ell}{L}s_F}$.
This leads to increasingly bad results as we move to the right between commensurate cuts.
If we round the number of left modes instead ($s_L=\text{round}(\frac{\ell}{L}s_F)$, depicted in orange), the problems get less severe and obtain a symmetric structure around the middle of the intervals.
The method only gives correct results for commensurate cuts of the system which are drawn in green in Figure~\ref{fig:app_rounding}. 
All results presented in the main text are computed at commensurate splittings.
For these points, the equilibrium state results for the Klein-Gordon model converge to the results  predictions already at very modest cutoffs $s_F$.

\section{\texorpdfstring{$\gamma$}{gamma} coefficients for Dirichlet boundary conditions at \texorpdfstring{$\ell$}{l}\label{sec:app_dirichlet}}

In this Appendix, we first show into more detail how to recover \eqref{eq:full_modes_integral} in the main text. 
For completeness we also outline the $\gamma$ coefficients for the case of Dirichlet boundary conditions at the cut~($\phi_L(\ell)=\phi_R(\ell)=0$). 
Via the Bogoliubov transform, eq.~\eqref{eq:bgbtrafo_modes} in the main text, they relate full system modes $A_k$  in terms of the partial modes $a_{m}^{R}$ and $a_{m}^{L}$

The starting point is the mode expansion of the scalar field of the full system
\begin{align}
    \phi(x,t)=\frac{1}{\sqrt{L}}\sum_{k=1}^\infty \frac{1}{\sqrt{p_k}} \left( A_k e^{-i p_k t}+A_{k}\dgr e^{i p_k t} \right)\sin(p_k x),
    \label{eqApp:def_phi_full}
\end{align}
with $p_k=k\frac{\pi}{L}$ and  $\comm{A_k}{A_l}=\comm{A_k^\dagger}{A_l^\dagger}=0$ and $\comm{A_k}{A_l^\dagger}=\delta_{k,l}$.
The expression can be inverted with the help of the canonical conjugate momentum field
\begin{align}
    \pi(x,t)=\partial_{t}\phi(x,t) = -\frac{ i}{\sqrt{L}} \sum_{k=1}^{\infty} \sqrt{p_k} \left(A_{k} e^{-i p_k t}-A_{k}^{\dagger} e^{i p_k t}\right) \sin\left(p_k x\right).
    \label{eq:def_momentum_dir}
\end{align}
By taking a linear combination of the scalar field~\eqref{eqApp:def_phi_full} and its momentum~\eqref{eq:def_momentum_dir}, we obtain for each $k$ and $t=0$:
\begin{align}
    \begin{split}
        \left[\phi(x,t)+\frac{i}{p_k}\pi(x,t)\right]_{k,t=0}= \frac{2}{\sqrt{L p_k}}A_{k}\sin\left(p_k x\right).
    \end{split}
    \label{eq:substitution_single_k}
\end{align}
Projecting out by multiplying the expression on both sides with $\sin\left(p_k x\right)$ and integrating completes the inversion of the mode expansion of the field
\begin{align}
  A_{k} = \sqrt{\frac{p_k}{L}} \int_{0}^{L} \dd{x} \left[\phi(x,t)+\frac{i}{p_k} \pi(x,t)\right] \sin\left(p_k x\right).
    \label{eqApp:full_modes_integral}
\end{align}

Our aim is to express the field operator on the full interval in~\eqref{eqApp:def_phi_full} by the fields defined on the sub-intervals. 
For Dirichlet boundary conditions at the cut~($\phi_L(\ell)=\phi_R(\ell)=0$), their mode expansion is given by:
\begin{align}
\phi_{L}(x,t) & =\frac{1}{\sqrt{\ell}}\sum_{m=1}^{\infty}\frac{1}{\sqrt{p_m^{(\ell)}}}\left(a_{m}^{L}e^{-i p_m^{(\ell)}t}+a_{m}^{L,\dagger}e^{i p_m^{(\ell)}t}\right)\sin\left(p_m^{(\ell)}x\right)\label{eqApp:split_fields_dirichlet_l}\\
\phi_{R}(x,t) & =\frac{1}{\sqrt{L-\ell}}\sum_{m=1}^{\infty}\frac{1}{\sqrt{p_m^{(L-\ell)}}}\left(a_{m}^{R}e^{-i p_m^{(L-\ell)}t}+a_{m}^{R,\dagger}e^{i p_m^{(L-\ell)}t}\right)\sin\left(p_m^{(L-\ell)}\left(x-\ell\right)\right)\label{eqApp:split_fields_dirichlet_r},
\end{align}
where we have defined $p_m^{(d)}=m\frac{\pi}{d}$.

The relationship between the full fields~\eqref{eqApp:def_phi_full} and split fields~(\ref{eqApp:split_fields_dirichlet_l}-\ref{eqApp:split_fields_dirichlet_l}) is given by the continuity condition, eq.~\eqref{eq:phi_system_split} in the main text.
Plugging it together with the split field expansions~(\ref{eqApp:split_fields_dirichlet_l}-\ref{eqApp:split_fields_dirichlet_r})  into~\eqref{eqApp:full_modes_integral} and performing the integrals gives the desired Bogoliubov transformation between the full and the split modes, eq.~\eqref{eq:bgbtrafo_modes} in the main text. 

The resulting $\gamma$ coefficients for Dirichlet boundary conditions at the cut are
\begin{align}
    \gamma_{km}^{+,L}&=\begin{cases}
\frac{(-1)^{m}\sqrt{p_{m}^{(\ell)}}\sin\left(p_{k}\ell\right)}{\sqrt{L\ell}\sqrt{p_{k}}(p_{k}-p_{m}^{(\ell)})} & p_{k}\neq p_{m}^{(\ell)}\\
\sqrt{\frac{\ell}{L}} & p_{k}=p_{m}^{(\ell)}
\end{cases}\\\gamma_{km}^{-,L}&=\frac{(-1)^{m}\sqrt{p_{m}^{(\ell)}}\sin\left(p_{k}\ell\right)}{\sqrt{L\ell}\sqrt{p_{k}}(p_{k}+p_{m}^{(\ell)})}\\\gamma_{km}^{+,R}&=\begin{cases}
-\frac{\sqrt{p_{m}^{(L-\ell)}}\sin\left(p_{k}\ell\right)}{\sqrt{L(L-\ell)}\sqrt{p_{k}}(p_{k}-p_{m}^{(L-\ell)})} & p_{k}\neq p_{m}^{(L-\ell)}\\
\frac{\sin\left(p_{k}\ell\right)+p_{k}(L-\ell)\cos\left(p_{k}\ell\right)}{p_{k}\sqrt{L}\sqrt{L-\ell}} & p_{k}=p_{m}^{(L-\ell)}
\end{cases}\\\gamma_{km}^{-,R}&=-\frac{\sqrt{p_{m}^{(L-\ell)}}\sin\left(p_{k}\ell\right)}{\sqrt{L(L-\ell)}\sqrt{p_{k}}(p_{k}+p_{m}^{(L-\ell)})}
\end{align}

The case of Neumann boundary conditions at the cut~($\partial_x\phi_L(\ell)=\partial_x\phi_R(\ell)=0$) which was used for most of the results presented in this work is outlined in the main text. 
The corresponding $\gamma$ coefficients are given in eqs.~(\ref{eq:gamma_l_plus_neu}-\ref{eq:gamma_r_minus_neu}).

\section{Algorithm\label{sec:app_algorithm}}

\subsection{Derivation and Derivatives of Commutators\label{sec:app_commutators}}
In order to use the scalar formulation for the matrix elements of $U_T$ in~\eqref{eq:def_matrix_elements_gen_func}, we have to bring the terms into normal order.
Our starting point is~\eqref{eq:def_matrix_elements_unordered}
\begin{align*}
    \begin{split}
        \braket{\nlr}{\nf}=\frac{1}{N} \prod_{m>0}\prod_{\sigma}\dv[n_{m,\sigma}]{j_{m,\sigma}}\prod_{k>0}\dv[n_{k}]{J_{L}}\left.\bra{0,0}
        e^{S}e^{F}e^{V}\ket{0,0}\right|_{J_{k}=0,j_{m,\sigma}=0}.
    \end{split}
\end{align*}
with
\begin{align*}
    S&\equiv\sum_{m>0}\sum_{\sigma}j_{m,\sigma}a_{m}^{\sigma}\\
    F&\equiv\sum_{k,m>0}\sum_{\xi}J_{k}\left(\gamma_{k,m}^{+,\xi}a_{m}^{\xi\dagger}+\gamma_{k,m}^{-,\xi}a_{m}^{\xi}\right)\\
    V&\equiv-\sum_{\xi,\lambda}\sum_{m,n>0}a_{m}^{\xi,\dagger}\rho_{m,n}^{\xi,\lambda}a_{n}^{\lambda,\dagger}.
\end{align*}
To avoid unnecessary jumping back and forth between the main text and the appendix, we will repeat some of the equations here. 
As mentioned in the main text, $S$ creates the excitations of the split modes on the partial vacuum according to the occupation numbers in $\ket{\nlr}$.
$F$ represents the creation operators of the full modes according to $\ket{\nf}$ expressed in the split modes.
Finally, $V$ transforms the full vacuum into a squeezed state on top of the split vacuum.

We normal order the expression in three steps.
Firstly, we normal order the exponential $e^{F}$ which contains both creation and annihilation operators.
Then, we commute $e^{S}$, which consists of annihilation operators only, past the creation operators of $e^{F}$.
Finally, we commute all annihilation operators of $e^{S}$ and $e^{F}$ past the vacuum transformation $e^{V}$.

Normal ordering or $e^{F}$ is achieved by the application of  the Baker-Campbell-Hausdorff formula $e^{X}e^{Y}=e^{X+Y+\frac{1}{2}[X,Y]}$ for $[[X,Y],X]=[[X,Y],Y]=0$. We get
\begin{align}
    \begin{split}
        e^{F} & =e^{F^{+}+F^{-}}\\
         & =e^{\comf}e^{F^{+}}e^{F^{-}}
     \end{split}
     \label{eq:app_comf}
\end{align}
where we have defined
\begin{align*}
    F_{k}^{+} & \equiv\sum_{k,n>0}\sum_{\xi}J_{k}\gamma_{k,n}^{+,\xi}a_{n}^{\xi,\dagger}\\
    F_{k}^{-} & \equiv\sum_{k,n>0}\sum_{\xi}J_{k}\gamma_{k,n}^{-,\xi}a_{n}^{\xi}
\end{align*}
to be the parts containing creation/annihilation operators respectively.
The commutator in~\eqref{eq:app_comf} evaluates to
\begin{align}
    \comf&\equiv-\frac{1}{2}\left[F^{+},F^{-}\right]\nonumber\\
    &=\frac{1}{2}\sum_{k,k'=1}^{\infty}J_{k}J_{k'}\sum_{\sigma}\sum_{n>0}\gamma_{k,n}^{+,\sigma}\gamma_{k',n}^{-,\sigma}.
    \label{eqApp:comf}
\end{align}

In a second step, we commute annihilation operators of partial modes in $e^{S}$ past $e^{F^{+}}$, using BCH in the form ${\displaystyle e^{X}e^{Y}=e^{Y+\left[X,Y\right]}e^{X}}$  for $[[X,Y],X]=[[X,Y],Y]=0$. 
We get
\begin{align}
    e^{S}e^{F^{+}} 
     & =e^{\comsf}e^{F^{+}}e^{S}
\end{align}
with the commutator
\begin{align}
    \comsf&\equiv\left[S,F^{+}\right]\nonumber\\
    &=\sum_{k,m>0}\sum_{\sigma}j_{m,\sigma}J_{k}\gamma_{k,m}^{+,\sigma}.
    \label{eqApp:comsf}
\end{align}

Finally, we commute the annihilation operators in $e^{S+F^{-}}$ and $F^{-}$ past the vacuum transformation $e^{V}$. 
For convenience, we denote the annihilation operators in $S$ and $F^{-}$ as $A$
\begin{align}
    \begin{split}
    A & \equiv S+F^{-}\\
     & =\sum_{m>0}\sum_{\sigma}\left\{ j_{m,\sigma}+\sum_{k>0}J_{k}\gamma_{k,m}^{-,\sigma}\right\} a_{m}^{\sigma}.
     \end{split}
\end{align}
In case of this commutation, the terms in the exponent of the BCH formula only vanish after the second commutator $\comm{A}{\comm{A}{V}}$.
Thus, using ${\displaystyle e^{X}e^{Y}=e^{(Y+\left[X,Y\right]+\frac{1}{2!}[X,[X,Y]])}~e^{X}}$, we find
\begin{align}
    e^{A}e^{V}=e^{\comav}e^{V+\left[A,V\right]}e^{A}
\end{align}
with the commutators
\begin{align}
    \left[A,V\right] 
     & =-\sum_{m>0}\sum_{\sigma}\left\{ j_{m,\sigma}+\sum_{k>0}J_{k}\gamma_{k,m}^{-,\sigma}\right\} \sum_{\xi}\sum_{l>0}\left(\chi_{m,l}^{\sigma,\xi}+\chi_{l,m}^{\xi,\sigma}\right)a_{l}^{\xi,\dagger},
\end{align}
\begin{align}
    \comav&\equiv\frac{1}{2}\left[A,\left[A,V\right]\right]\nonumber\\
    &=-\frac{1}{2}\sum_{m>0}\sum_{\sigma}\left\{ j_{m,\sigma}+\sum_{k>0}J_{k}\gamma_{k,m}^{-,\sigma}\right\} \sum_{l>0}\sum_{\xi}\left\{ j_{l,\xi}+\sum_{k'>0}J_{k'}\gamma_{k',l}^{-,\xi}\right\} \left(\chi_{m,l}^{\sigma,\xi}+\chi_{l,m}^{\xi,\sigma}\right).
    \label{eqApp:comav}
\end{align}

Finally, the full expression in the normal ordered form is
\begin{align}
    e^{S}e^{F}e^{V}=\exp\left[\comf+\comsf+\comav\right]\,e^{\left[A,V\right]}\,:e^{S}e^{F}e^{V}:.
\end{align}

When computing the expectation value in the split vacuum, only the zeroth order in the power expansion of $e^{\left[A,V\right]}\,:e^{S}e^{F}e^{V}:$ survives and we get
\begin{align}
    \bra{0,0}
        e^{S}e^{F}e^{V}\ket{0,0}&=\exp\left[\comf+\comsf+\comav\right]\nonumber\\
        &\equiv e^T .
\end{align}
The computation of the matrix elements of $U_T$ in~\eqref{eq:def_matrix_elements_gen_func} does not depend on the form of $T$ directly, but on the second derivatives $T[J_i,J_k]$.
All terms that are not derived twice will vanish once we set $J_k=0$.

The second derivatives are:

\comf, given in~\eqref{eqApp:comf}:
\begin{align}
    \dv{J_{i}}\dv{J_{p}}\comf=\frac{1}{2}\sum_{\sigma}\sum_{n>0}\left(\gamma_{i,n}^{+,\sigma}\gamma_{p,n}^{-,\sigma}+\gamma_{p,n}^{+,\sigma}\gamma_{i,n}^{-,\sigma}\right)
\end{align}
and other derivatives vanish.

\comsf, given in~\eqref{eqApp:comsf}: 
\begin{align}
    \dv{j_{l,\xi}}\dv{J_{i}}\comsf&=\gamma_{i,l}^{+,\xi}
\end{align}
and other derivatives vanish.

\comav, given in~\eqref{eqApp:comav}: 
\begin{align}
    \dv{J_{p}}\dv{J_{i}}\comav&=-\sum_{m,m'>0}\sum_{\sigma,\mu}\gamma_{i,m'}^{-,\sigma}\gamma_{p,m}^{-,\mu}\left(\chi_{m,m'}^{\mu,\sigma}+\chi_{m',m}^{\sigma,\mu}\right)\\
    \dv{j_{p,\alpha}}\dv{j_{l,\xi}}\comav&=-\left(\chi_{l,p}^{\xi,\alpha}+\chi_{p,l}^{\alpha,\xi}\right)\\
    \dv{J_{i}}\dv{j_{l,\xi}}\comav&=-\sum_{m>0}\sum_{\sigma}\gamma_{i,m}^{-,\sigma}\left(\chi_{l,m}^{\xi,\sigma}+\chi_{m,l}^{\sigma,\xi}\right).
\end{align}

\subsection{Tree building algorithm\label{sec:app_tree}}
The primed sum in eq.~\eqref{eq:derivative_unique_configs} runs over all lexicographically unique configurations of the arguments $(\jcal_1, \jcal_2)$ of $T$.
Lexicographically unique implies that all tuples are sorted internally $\jcal_1<\jcal_2$ and the string of tuples is sorted as well.
Two tuples are sorted by sorting them first by their first entry and then by second entry.

The generation of all unique pairs can be approached in at least two ways. 
We could take all $\jcal$s independently and find all possible ways of distributing them as pairs.
The available $\jcal$s are determined by the occupation numbers in the states that are determining the matrix elements in the unitary transformation matrix.
If the occupation number vector is $\vec{n}=(2,2,2)$ (as in the example), the available $\jcal$s are $(\jcal_1, \jcal_1, \jcal_2, \jcal_2, \jcal_3, \jcal_3)$, i.e. we derive each twice with respect to each \jcal.
In the case of independent generation of all combinations, we would have to filter out all the repeated configurations due to ordering in the arguments of $T$ and in the string.

Alternatively, we can incrementally create all orderings in a tree-like structure.
By tracking the ordering as we progress, we can avoid the generation of forbidden configurations.
We will only consider this second alternative since the generation of all permutations scales with $n!$ where $n=\sum_i n_i$, i.e. the number derivatives and the task of computing all combinations is unnecessary.

We start with vector $\vec{n}=(n_{1},n_{2},\dots,r_{N})$.
Since we consider only sorted tuples and a globally sorted string of tuples, we build all valid pairs $(\jcal_i,\jcal_k)$ with the first $\jcal_i$ corresponding to the smallest non-zero $n_{i}$. 
We proceed in a recursive manner and modify the vector $\vec{n}$ by subtracting one from $n_i$ and $n_k$ and select again all valid pairs in the next step of the tree.
A pair is valid if the $\jcal_i<\jcal_k$ for a pair $\left(\jcal_{i},\jcal_{k}\right)$ and the pair is greater or equal to the previous selected pair.

\begin{figure}
    \centering
    \input{figures/derivation_tree_222.tikz}
    \caption{Tree to build all tuples of configurations of $\vec{n}=(2,2,2)$.
    The expression in parentheses are the $\jcal_i$ that still have to be distributed.
    The tuple that is added to the configuration at every step is noted in brackets.
    The final configuration of every path can be assembled by following the arrows and collecting the entries of all brackets.
    The missing leaf in the last layer on the right indicates a configuration that cannot be build due to the restrictions on the tuples.}
    \label{fig:app_derivation_tree_222}
\end{figure}
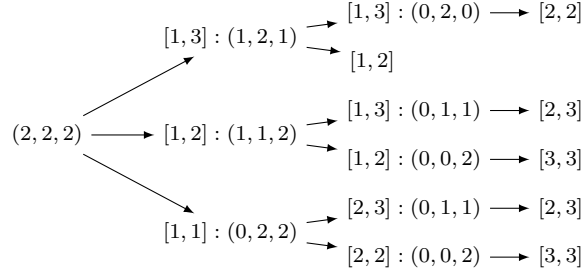

In total, we build a tree of the form in figure~\ref{fig:app_derivation_tree_222}.
We start on the left with the full string of the example that is also used in the main text $\vec{n}=(2,2,2)$.
Lists in round brackets describe the occupation numbers of the state.
Each level of the tree represents one level of recursion. 
In each level, the vectors in round parentheses represent the remaining vector $\vec{n}$ after picking the tuple in brackets.
The topmost entry on the second level describes the case of picking $[1,3]$ from the tuple as first argument for $T$.
Thus, the first and the third entry are decreased by one.
All valid combinations of $T$ can be enumerated by following the branches of the tree.
If we pick up all tuples in brackets, we obtain the full string of arguments for $T$.
The algorithm stops if no valid pair can be found for a given vector $\vec{n}$ or if $\vec{n}=\vec{0}$.
In the example, the second condition is met after $3$ iterations.
Some branches are not continued because the following tuple is smaller than the previous one (first termination condition).
In the example, $[1,2]<[1,3]$ in the third level and we abort the branch.
This step would not result in a sorted combination of tuples.
In total, the tree in fig~\ref{fig:app_derivation_tree_222} has five leafs.
Following all the paths leading to those leafs, we obtain the five configurations that are listed in eq.~\ref{eq:example_t_strings}.

As we can see in the tree, the generated strings depend only on the initial occupation number vector $\vec{n}$.
The allowed occupation number vectors, in turn, depend only on the cut-offs that are chosen for the subsystems.
Thus, we can generate all the strings once and use them for different splittings along the interval.

\subsection{Combinatorial multiplicities\label{sec:app_proof_prefactors}}

Let us show how to obtain the combinatorial multiplicity $c_{k}$ stated in eq.~\eqref{eq:def_multiplicities} in the main text. 
The prefactor counts how many terms in the derivative expansion of the generative functional~\eqref{eq:derivative_unique_configs} are equal to a given unique combination of $T$s - a lexicographically ordered string of derivative indices.

Recall that we are differentiating the generative functional of the form
\begin{align}
Z=e^{T(\jcal_1,\jcal_2,\ldots,\jcal_N)},
\end{align}
where $T$ is an expression quadratic in multipliers $\jcal_i$. As before, for generality, $\jcal$'s can be either $J_i$ or $j_{l,\sigma}$ depending on the context. 
By denoting 
\begin{align}
    T\left[i,j\right]\equiv \dv{ \jcal_i}\dv{ \jcal_j}T(\jcal_1,\jcal_2,\ldots,\jcal_N)
\end{align}
we have
\begin{align}
    T\left[i,j\right]&=T\left[j,i\right]
\end{align}
as a consequence of the commutatvity of derivatives. 
We also have
\begin{align}
    \begin{split}
        \left.T\right|_{J_{i}=0\,\forall i} &=0\\
        \left.\dv{ \jcal_k}T\right|_{J_{i}=0\,\forall i} &=0\,\forall k.
    \end{split}\label{eqApp:Tderivatives_rules}
\end{align}
For convenience, let us introduce the following notation. 
Let us define  $\nu$ to be the set of following singletons $e_i^{(j)}$:
\begin{align}
    \nu&\equiv(n_1,n_2,\ldots,n_N)\\
    &\equiv \left\{e_1^{(1)},e_1^{(2)},\ldots ,e_1^{(n_1)},e_2^{(1)},\ldots ,e_2^{(n_2)},\ldots\,\ldots,e_N^{(n_N)} \right\}
\end{align}
where we distinguish $e_i^{(j)}$ from $e_i^{(k)}$ for $j\neq k$. 
Then it is convenient to define
\begin{align}
    \text{d}^{e_i^{(k)}}f&\equiv \dv{ \jcal_i}f
\end{align}
so that
\begin{align}
    \text{d}^{\nu}f\equiv \frac{\text{d}^{n_1}}{\text{d} \jcal_1^{n_1}}\frac{\text{d}^{n_2}}{\text{d} \jcal_2^{n_2}}\cdots\frac{\text{d}^{n_N}}{\text{d} \jcal_N^{n_N}}f.
\end{align}

Expressed in this notation, we want to calculate the combinatorial coefficient $c_{k}$ such that
\begin{align}
    \left.\text{d}^{\nu}e^{T}\right|_{J_{i}=0\,\forall i}&=\sideset{}{^{'}}\sum_{k}c_{k}\prod_{l}T^{p_{kl}}\left[{l_{1}},{l_{2}}\right]\label{eqApp:derivative_functional}\\
    &=\sum_{\begin{array}{c}
I_{k}=(i_{1},i_{2};i_{3},i_{4};\ldots;i_{|\nu|-1},i_{|\nu|})\\
(i_{1},i_{2})\leq(i_{3},i_{4})\leq\ldots\leq(i_{|\nu|-1},i_{|\nu|})\\
i_{2n+1}\leq i_{2n+2},\,
\left\{ e_{i_{1}},e_{i_{2}},\ldots,e_{i_{|\nu|}}\right\} =\nu
\end{array}}c_{k}\,T[i_1,i_2]T[i_3,i_4]\cdots T[i_{|\nu|-1},i_{|\nu|}]\nonumber
\end{align}
where  sum $\sideset{}{^{'}}\sum_{k}$ runs over all unique combinations of $T$. 
A combination is unique if it cannot be transformed into another combination of $T$s by swapping the arguments of $T$ or commuting $T$s.
The exponents $p_{kl}$ are the number of repetitions of terms $T$ with same arguments $(J_{l_1}, J_{l_2})$. 
All unique combinations of $T$ are precisely generated by all pairwise  lexicographically index sets, which we denoted in the second line. 
A pair $(i_{1},i_{2})$ is lexicographically ordered if $i_{1}\leq i_{2}$. 
Two pairs are lexicographically ordered $(i_{1},i_{2})\leq(j_{1},j_{2})$ iff $i_1<j_1$ or $i_1=j_1$ and $i_2\leq j_2$.

In order to find $c_{k}$, let us first show a general formula for a multi-index derivative of a compositum of functions:
\begin{align}
\text{d}^{\nu}\left(f\circ g\right)&=\sum_{p=1}^{|\nu|}f^{(p)}\circ g\sideset{}{^{'}}\sum_{\begin{array}{c}
P=\left\{ \nu_{1},\nu_{2},\ldots,\nu_{p}\right\} \subset\mathcal{P}(\nu)\\
\nu_{1}\cup\nu_{2}\cup\ldots\cup\nu_{p}=\nu\\
\nu_i\neq\oslash,\,
\nu_{i}\cap\nu_{j}=\begin{cases}
\oslash & i\neq j\\
\nu_{i} & i=j
\end{cases}
\end{array}}\text{d}^{\nu_{1}}g\cdots\text{d}^{\nu_{p}}g \label{eqApp:derivative_general}
\end{align}
Here, the inner sum $\sideset{}{^{'}}\sum$ runs over all partitions $P$ of the set $\nu$ of cardinality $|P|=p$. 
A partition is collection of non-empty sets, a subset of the power set $\mathcal{P}(\nu)$, such that its union equals to the whole $\nu$ and that the each two sets in the collection are mutually disjoint, which is the last notation line below the sum. 
The outer sum runs over all the possible cardinalities of partitions. 
The two extremes, the partition with $p=1$ is the whole $\nu$ and $p=|\nu|$ is a collection sets consisting of single singletons. 
The notation $f^{(p)}$ is the standard notation for the $p$-th derivative of $f$.

We can prove the formula \eqref{eqApp:derivative_general} by induction. 
For $|\nu|=1$, this is just the chain rule for derivation. 
For $|\nu|+1$, we take another derivative by a singleton and use the product rule for derivation:
\begin{align}
\text{d}^{\nu\cup e_{i}}\left(f\circ g\right)&=\sum_{p=1}^{|\nu|}\left(f^{(p+1)}\circ g\right)\text{d}^{e_{i}}g\sideset{}{^{'}}\sum_{\text{part. of }\nu }\text{d}^{\nu_{1}}g\cdots\text{d}^{\nu_{p}}g\nonumber\\
&\hspace{1em}+\sum_{p=1}^{|\nu|}\left(f^{(p)}\circ g\right)\sideset{}{^{'}}\sum_{\text{part. of }\nu}\text{d}^{\nu_{1}\cup e_{i}}g\cdots\text{d}^{\nu_{p}}g+\text{d}^{\nu_{1}}g\text{d}^{\nu_{2}\cup e_{i}}g\cdots\text{d}^{\nu_{p}}g+\ldots+\text{d}^{\nu_{1}}g\cdots\text{d}^{\nu_{p}\cup e_{i}}g\nonumber\\
&=\sum_{p=1}^{|\nu|+1}\left(f^{(p)}\circ g\right)\sideset{}{^{'}}\sum_{\text{part. of }\nu\cup e_{i}}\text{d}^{\mu_{1}}g\cdots\text{d}^{\mu_{p}}g
\end{align}
We recognise the first line of the first equality as precisely the addition of an (extra) singleton set to each of the partitions of $\nu$. 
This increases the cardinality of the corresponding partition by one. 
The second line of the first equality consists of the singleton $e_{i}$ added to each possible choice of an already existing subset of a partition of $\nu$. 
In particular, the later operation does not increase the number of subsets in the collection, thus the derivative $f^{(p)}$ still matches the cardinality of the corresponding collections. 
The collection of sets from the first and the second line thus precisely constitute all the possible partitions of the set $\nu\cup e_{i}$ and the expression can be put in the exactly same form as \eqref{eqApp:derivative_general}. 
This is written out in the last equality of the expression above. 

Going back to evaluating the derivatives of the generating functional, eq.~\eqref{eqApp:derivative_functional}, we can now use the above derived formula~\eqref{eqApp:derivative_general} for $f=\exp(\bullet)$ and $g=T(J_1,J_2,\ldots,J_N)$. 
Using the quadratic form of $T$, resulting in the derivative rules in eq.~\eqref{eqApp:Tderivatives_rules}, the nonzero contributions are only going to come from those partitions of $\nu$ where every subset contains exactly two elements:
\begin{align}
    \left.\text{d}^{\nu}e^{T}\right|_{J_{i}=0\,\forall i}&=\sum_{\begin{array}{c}
P\text{ partition of }\nu\\
|P|=\frac{|\nu|}{2},\,|\nu_{i}|=2,\,\forall\nu_{i}\in P
\end{array}}T[\nu_{1}]T[\nu_{2}]\cdots T[\nu_{|\nu|/2}]\\&=\sum_{\begin{array}{c}
\text{lexicograph. ord. index sets}\\
I_{k}=(i_{1},i_{2};\ldots;i_{|\nu|-1},i_{|\nu|})
\end{array}}c_{k}\,T[i_{1},i_{2}]T[i_{3},i_{4}]\cdots T[i_{|\nu|-1},i_{|\nu|}]\nonumber
\end{align}
where we have copied the second line of eq.~\eqref{eqApp:derivative_functional} for comparison. 

It now remains to count how many partitions in the sum in the top line correspond to the same lexicographically ordered index set in the bottom line. 
Since on the partition side we distinguish singletons $e_i^{(j)}$ from $e_i^{(k)}$ for $j\neq k$, we can get different partitions by permuting between themselves $e_i^{(1)}$, $e_i^{(2)}$, ..., $e_i^{(n_i)}$ (equivalent to mutually equivalent indices in the index set $i_{j_1}=i_{j_2}=\cdots=i_{j_{n_i}}$). 
This gives a multiplicity
\begin{equation}
    \prod_{i}\left(n_{i}!\right).
\end{equation}
However, not all the partitions that we obtain with this procedure are distinct. 
Because the partitions are insensitive to the order of elements in subsets $\nu_j$, we have over counted by a factor of 2 every diagonal $\nu_j$. 
These are subsets containing two equal elements $\nu_j=\left\{e_i^{(j_1)},e_i^{(j_2)}\right\}$. 
This is indeed a manifestation of the commutativity of derivatives. 
We thus overall over counted the number of partitions by a factor of
\begin{equation}
    2^{N_{\text{diag}}}
\end{equation}
where $N_{\text{diag},k}$ is the number of diagonal $\nu_j$ (diagonal bi-indices $(i,i)$) in the index set $I_k$. 
Furthermore, partitions are insensitive to the order of the subsets $\nu_i$. 
We have therefore over counted by the factor $p_{kl}!$ for all those bi-indices $(i_{l_1},i_{l_2})$ that repeat $p_{kl}$ number of times in the index set $I_k$. 
This gives an overall factor of 
\begin{equation}
    \prod_{l}\left(p_{kl}!\right)
\end{equation}
of overcounted partitions. 

Combining all these together, the number of distinct partitions that each lexicographically ordered index set $I_k$ corresponds to is
\begin{align}
    c_{k}=\frac{\prod_{i}\left(n_{i}!\right)}{2^{N_{\text{diag},k}}\prod_{l}\left(p_{kl}!\right)}.
\end{align}
This is the sought for combinatorial multiplicity.

\section{Hamiltonian truncation matrix elements\label{sec:app_ht_matrix_elements}}

We use the following matrix elements from the methods developed in~\cite{bajnok_finite_2002,kukuljan_correlation_2018} to perform the HT calculations of ground, thermal and non-equilibrium states.

For an operator $O$, we list matrix elements 
\begin{align}
    O^{\vec{n}',\vec{n}}=\bra{\vec{n}'}O\ket{\vec{n}}.
\end{align}
They are computed between states spanning the computational basis, the Hilbert space of the massless free boson:
\begin{align}
    \ket{\vec{n}}\equiv \ket{n_1,n_2,\ldots}\equiv \frac{1}{N_{\vec{n}}}\prod_{k>0}\left( A_k^\dagger\right)^{n_k}\ket{0}.
\end{align}
Here, $A_k$, $k=1,2,\ldots$ are the bosonic modes fulfilling the canonical commutation relations $\comm{A_k}{A_l}=\comm{A_k^\dagger}{A_l^\dagger}=0$ and $\comm{A_k}{A_l^\dagger}=\delta_{k,l}$, the normalisation is $N_{\vec{n}}=\prod_{k>0}\sqrt{n_k!}$ and $A_k\ket{0}=0$ $\forall k$ is the vacuum of the massless free boson theory. 

The massless free boson Hamiltonian for Dirichlet boundary conditions
\begin{align}
    H_{\text{0FB}}=\frac{1}{2}\int_{0}^{L}\dd x \left[\left(\partial_t \phi\right)^2+\left(\partial_x \phi\right)^2\right].
\end{align}
is diagonal with matrix elements
\begin{align}
    H_{\text{0FB}}^{\vec{n}',\vec{n}}=\frac{\pi}{L}\left(
    \sum_{k=1}^{\infty}kn_k-\frac{1}{24}\right)\delta_{\vec{n}',\vec{n}}.\label{eq:FBmatrixElements}
\end{align}

The Hamiltonian of the massive free boson
\begin{align}
    H_{\text{mFB}}=\frac{1}{2}\int_{0}^{L}\dd x \left[\left(\partial_t \phi\right)^2+\left(\partial_x \phi\right)^2+m^2\phi^2\right].
\end{align}
has the following matrix elements
\begin{eqnarray}
    H_{\text{mFB}}^{\vec{n}',\vec{n}}&=&\frac{\pi}{L}\left\{\delta_{\vec{n}',\vec{n}}\left(\sum_{k=1}^{\infty}\left(1+\frac{m^2L^2}{2\pi^2k^2}\right)kn_k-\frac{1}{24}\right)+\right.\\
    &&\left.+\frac{m^2L^2}{4\pi^2}\sum_{k=1}^{\infty}\left(\prod_{j=1\atop j\neq k}^{\infty}\delta_{n'_j,n_j}\right)\frac{1}{k^2}\left(\sqrt{n_k k}\sqrt{(n_k-1) k}\,\delta_{n'_k+2,n_k}
    +\sqrt{(n_k+2) k}\sqrt{(n_k+1) k}\,\delta_{n'_k-2,n_k}
    \right)\right\}.\nonumber
\end{eqnarray}

The Hamiltonian of the sine-Gordon model can be expressed as
\begin{eqnarray}
    H_{\text{sG}}&=&\int_{0}^{L}\dd x \left[\frac{1}{2}\left\{\left(\partial_t \phi(x)\right)^2+\left(\partial_x \phi(x)\right)^2\right\}-2\kappa(\Delta)M_{S}^{2-2\Delta}\cos\left(\beta\phi(x)\right)\right]\nonumber\\
    &=&H_{\text{0FB}}-\kappa(\Delta)M_{S}^{2-2\Delta}\int_{0}^{L}\dd x \left(V_{1}(x)+V_{-1}(x)\right).
\end{eqnarray}
Here
\begin{align}
    V_{p}(x)\equiv e^{i q\phi(x)},\quad p\in\mathbb{Z}
\end{align}
for $q\equiv p \beta$ is the \textit{vertex operator},  $M$ is the semi-classical soliton mass, the interaction related coefficient $\Delta$ is defined as
$
    \Delta\equiv\frac{\beta^2}{8\pi}
$
and the coupling-mass ratio $\kappa(\Delta)$~\cite{zamolodchikov_mass_1995} is
\begin{align}
    \kappa(\Delta)=\frac{1}{\pi}\frac{\gamma_b\left(\Delta\right)}{\gamma_b\left(1-\Delta\right)}\left[\frac{\sqrt{\pi}\gamma_b\left(\frac{1}{2-2\Delta}\right)}{2\gamma_b\left(\frac{\Delta}{2-2\Delta}\right)}\right]^{2-2\Delta}.
\end{align}

The vertex operator can be written in normal ordered form as
\begin{align}
    V_p(z,\bar{z})=e^{iq\phi(z,\bar{z})}=\left|z-\bar{z}\right|^{-q^2/(4\pi)}:e^{iq\phi(z,\bar{z})}:
\end{align}
where for convenience we have introduced $z\equiv e^{i\frac{\pi}{L}x}$. 
The matrix elements are
\begin{align}
    V_p^{\psi',\psi}\left(z,\bar{z}\right)=N_{\vec{n}'}^{-1}N_{\vec{n}}^{-1}\left[2\sin\left(\frac{\pi x}{L}\right)\right]^{-q^2/(4\pi)}\prod_{k=1}^{\infty}\left<0\right|A_k^{n'_k}e^{- q \frac{A_k^\dagger}{\sqrt{4\pi k}}(z^k-\bar{z}^k)}e^{ q \frac{A_k}{\sqrt{4\pi k}}(z^{-k}-\bar{z}^{-k})}\left( A_k^\dagger\right)^{n_k}\left|0\right>,
\end{align}
with
\begin{eqnarray}
    &\left<0\right|A_k^{n'_k}e^{- q \frac{A_k^\dagger}{\sqrt{4\pi k}}(z^k-\bar{z}^k)}e^{ q \frac{A_k}{\sqrt{4\pi k}}(z^{-k}-\bar{z}^{-k})}\left( A_k^\dagger\right)^{n_k}\left|0\right>=&\nonumber\\
    &=\sum_{j'=0}^{\infty}\sum_{j=0}^{\infty}\frac{(-1)^{j'}}{j'!j!}\left(\frac{2q}{\sqrt{4\pi k}}\right)^{j'+j}\left[\frac{\bar{z}^k-z^k}{2}\right]^{j'+j}\left<0\right|A_k^{n'_k}\left( A_k^\dagger\right)^{j'}A_k^{j}\left( A_k^\dagger\right)^{n_k}\left|0\right>&
\end{eqnarray}
and
\begin{align}
   \left<0\right|A_k^{n'_k}\left( A_k^\dagger\right)^{j'}A_k^{j}\left( A_k^\dagger\right)^{n_k}\left|0\right>=\left(\begin{array}{c}n'_k\\j'\end{array}\right)\left(\begin{array}{c}n_k\\j\end{array}\right)j'!j!(n_k-j)!\delta_{n'_k-j',n_k-j}\Theta(n_k\geq j)
   \label{eq:CombinatoricTerm2}
\end{align}
with the Heaviside step function $\Theta(\bullet)$.

To get the matrix elements of the spatially integrated vertex operator that appears in the sine-Gordon Hamiltonian, the following relation is useful
\begin{align}
    \int_{0}^{\pi}\dd u \left[2\sin\left(u\right)\right]^{-q^2/(4\pi)}e^{-i k u}=\frac{e^{-i\frac{\pi}{2}k}\pi}{(1-\frac{q^2}{4\pi})B\left(\frac{1}{2}(2-\frac{q^2}{4\pi}-k),\frac{1}{2}(2-\frac{q^2}{4\pi}+k)\right)}.
\end{align}
Here, $B(x,y)=\frac{\Gamma(x)\Gamma(y)}{\Gamma(x+y)}$ is the beta function.

\section{Covariance matrix formalism for entanglement entropies of free theories\label{sec:app_covmat}}

Eigen, thermal and time-evolved states of free theories are Gaussian and the computation of their entanglement properties can be simply achieved by means of the covariance matrix formalism~\cite{serafini_quantum_2017}.

A Gaussian state is completely determined by its covariance matrix
\begin{align}
    \Gamma=\mqty[ Q & R\\ R^T & P]
\end{align}
with
\begin{align}
    Q_{mn}&=\left<\phi_m\phi_n\right>\nonumber\\
    P_{mn}&=\left<\pi_m\pi_n\right>\nonumber\\
    R_{mn}&=\left<\frac{1}{2}\left\{\phi_m,\pi_n\right\}\right>
\end{align}
where $[\phi_m,\phi_n]=[\pi_m,\pi_n]=0$, $[\phi_m,\pi_n]=i\delta_{m,n}$ are harmonic oscillator conjugate pairs. 
All higher order correlations are given by Wick's theorem. 

In the case of a bosonic field theory, we have to introduce an IR (finite volume $L$) and UV (maximal momentum mode kept, $s_F$) cutoff in order to keep the covariance matrix finite. 
Then, the harmonic oscillators are finitely many and we can treat the covariance matrix either in momentum space or position space. 
For covariance matrix calculations it is convenient to go back and forth between those representations using a discrete sine transform. 
The field expansion can be written as
\begin{align}
    \phi(x_n,t)&=\sqrt{\frac{2}{L}}\sum_{k=1}^{s_F}\phi_k(t)\sin\left(p_k x_n\right)\nonumber\\
    \pi(x_n,t)&=\sqrt{\frac{2}{L}}\sum_{k=1}^{s_F}\pi_k(t)\sin\left(p_k x_n\right)
\end{align}
with $p_k=\frac{k\pi}{L}$ and $x_n=n a$ for $k,n=1,\ldots,s_F$ with the lattice spacing $a=\frac{L}{s_F+1}$. The inverse discrete sine transform is achieved by
\begin{align}
    \phi_k(t)&=a\sqrt{\frac{2}{L}}\sum_{n=1}^{s_F}\phi(x_n,t)\sin\left(p_k x_n\right)\nonumber\\
    \pi_k(t)&=a\sqrt{\frac{2}{L}}\sum_{n=1}^{s_F}\pi(x_n,t)\sin\left(p_k x_n\right)\, .
\end{align}
Such definitions of fields correspond to approximating the field theory with a lattice system. 
We shall be keeping the relativistic dispersion, however. 

For the calculation of entanglement entropy it is most convenient to take the position space covariance matrix of the bosonic degrees multiplied by the lattice spacing $a$, that is $\phi_n=a\phi(x_n)$ and $\pi_n=a\pi(x_n)$ to keep the correct dimensions. 
In position space, the covariance matrix of a reduced density matrix corresponding to a subsystem is achieved by taking only those matrix elements corresponding to the lattice points that lie within the subsystem. 
For example if we are interested in the entanglement between the interval $[0,\ell]$ and the rest of the system, we take the covariance matrix of the lattice sites $x_n\in[0,\ell]$. 

Then, the von Neumann entanglement entropy is computed by calculating the symplectic spectrum of the covariance matrix $\Gamma$. 
This is achieved by diagonalising
\begin{align}
    iJ\Gamma
\end{align}
with the symplectic unit
\begin{align}
    \Gamma=\mqty[ 0 & \id\\ -\id & 0] \, .
\end{align}
The eigenvalues appear in pairs $\pm\gamma_k$, $k=1,\ldots,s_F$. 
This maps the problem to computing the entropy of $s_F$ harmonic oscillators at inverse temperatures
\begin{align}
    \beta=\log\frac{\gamma_k+\frac{1}{2}}{\gamma_k-\frac{1}{2}}\, .
\end{align}
The von Neumann entropy is then
\begin{align}
    S(\Gamma)=\sum_{k=1}^{s_F}\left[\left(\gamma_{k}+\frac{1}{2}\right)\log\left(\gamma_{k}+\frac{1}{2}\right)-\left(\gamma_{k}-\frac{1}{2}\right)\log\left(\gamma_{k}-\frac{1}{2}\right)\right]
\end{align}
and the Rényi entropies are
\begin{align}
    S^{\alpha}(\Gamma)=\frac{1}{\alpha-1}\sum_{k=1}^{s_F}\log\left[\left(\gamma_{k}+\frac{1}{2}\right)^{\alpha}-\left(\gamma_{k}-\frac{1}{2}\right)^{\alpha}\right]\, .
\end{align}
The covariance matrix approach to computing the entanglement entropies is convenient because it lets us model also the truncation effects by taking finite $s_F$. 
The results are not exactly comparable to the HT cutoff at the same maximal momentum, because in the HT case, we have an energy cutoff which implies also a maximal occupation number for a mode. 
But it is the closest approximation of the cutoff effect that we can get. 
Taking $s_F$ large, we can recover the exact analytical results in the continuum limit. 

The concrete results presented in the main text for the thermal states of the Klein-Gordon (KG) model 
\begin{align}
H_{\text{mFB}}=\frac{1}{2}\int_0^{L}\dd x \left[ \Pi^2(x) + (\partial_x\phi(x))^2 +m^2 \phi^2(x)\right]
\end{align}
can be recovered using the thermal covariance matrix of the model
\begin{align}
    \left\langle \phi(x_m)\phi(x_n)\right\rangle &=\frac{1}{L}\sum_{k=1}^{s_F}\frac{1}{\epsilon_{k}}\coth\left(\frac{\epsilon_{k}}{2T}\right)\sin(p_k x_m)\sin(p_k x_n)\nonumber\\
    \left\langle \pi(x_m)\pi(x_n)\right\rangle &=\frac{1}{L}\sum_{k=1}^{s_F}\epsilon_{k}\coth\left(\frac{\epsilon_{k}}{2T}\right)\sin(p_k x_m)\sin(p_k x_n)\nonumber\\
    \frac{1}{2}\left\langle \left\{ \phi(x_m),\pi(x_n)\right\} \right\rangle &=0\label{eqSup:KGthermal}
\end{align}
with the dispersion relation $\epsilon_{k}=\sqrt{p_k^{2}+m^{2}}$.

The quench dynamics can be computed using the equations of motion $\dot{O}=i\comm{H}{O}$, yielding for the KG model
\begin{align}
    \phi_{k}(t)&=\cos\left(\epsilon_{k}t\right)\phi_{k}(0)+\frac{\sin\left(\epsilon_{k}t\right)}{\epsilon_{k}}\pi_{k}(0)\\
    \pi_{k}(t)&=-\epsilon_{k}\sin\left(\epsilon_{k}t\right)\phi_{k}(0)+\cos\left(\epsilon_{k}t\right)\pi_{k}(0) \, .
    \label{eqSup:KGeqsMotion}
\end{align}
The procedure for a KG mass quench from the prequench mass $m_0$ to the postquench mass $m$ is the following:  take the momentum space representation of the thermal correlations~\eqref{eqSup:KGthermal} for the prequench mass $m_0$ and propagate them using the equations of motion~\eqref{eqSup:KGeqsMotion} for the postquench mass $m$. 
The covariance matrix is transformed back to position space, the reduced covariance matrix corresponding to the subsystem taken as described above and the entanglement entropies computed. 

\section{Finite size sine-Gordon breather masses\label{sec:app_sg_breather}}

The correction to infinite size sine-Gordon breather masses to obtain their finite volume counterparts can be computed using the boundary bootstrap \cite{ghoshal_boundary_1994,ghoshal_bound_1994,mattsson_boundary_2000,bajnok_finite_2002}. This gives the finite size energy $E_n$ of the (excited) $n$-the breather in the parametric form:
\begin{eqnarray}
\left(ML(\theta),\epsilon(\theta)-\epsilon_0\right)=\left(\frac{2\pi I_n+2i\log \left(R^{(n)}(\theta)\right)}{2\frac{m_{n}}{M}\sinh(\theta)},\frac{m_{n}}{M}\cosh(\theta)\right),\label{eqApp:BreatherEnergies}
\end{eqnarray}
where  $\theta\in[0,\infty)$ is the parameter of the parametrisation and $I_n\in\mathbb{Z}$ is the quantum number labeling the breather lines. Each value of $I_n$ corresponds to one of the excited states of the $n$-the breather. These are moving breathers whose momentum has a discrete set of possible values due to the finite volume. In the $L\rightarrow\infty$ limit, all those lines converge to the infinite volume breather mass $m_{n}$.  Further, $\lambda$ is the sine-Gordon interaction parameter (eq. \eqref{eq:sg_lambda} in the main text), $M$ is the soliton mass and the infinite volume breather mass of the $n$-th breather $m_{n}$ is given by
\begin{align}
    m_n=2M\sin\left(\frac{n\pi}{2\lambda}\right).
\end{align} 
The functions $R^{(n)}$ are the boundary breather reflection factors.

In case of Dirichlet boundary conditions the reflection factors are given by \cite{ghoshal_boundary_1994,ghoshal_bound_1994,mattsson_boundary_2000,bajnok_finite_2002}
\begin{align}
R^{(n)}(\theta)=R_0^{(n)}(\theta)S^{(n)}(0,\theta),
\end{align}
with
\begin{eqnarray}
R_0^{(n)}(\theta)&=&\frac{\left(\frac{1}{2}\right)_{\theta}
	\left(\frac{n}{2\lambda}+1\right)_{\theta}
	}{
	\left(\frac{n}{2\lambda}+\frac{3}{2}\right)_{\theta}
	}
	\prod_{k=1}^{n-1}
	\frac{\left(\frac{k}{2\lambda}\right)_{\theta}
		\left(\frac{k}{2\lambda}+1\right)_{\theta}
	}{
	\left(\frac{k}{2\lambda}+\frac{3}{2}\right)_{\theta}^2
},\nonumber\\
S^{(n)}(x,\theta)&=&\prod_{k=0}^{n-1}
\frac{
	\left(
	\frac{x}{\pi\lambda}-\frac{1}{2}+\frac{n-2k-1}{2\lambda}
	\right)_{\theta}
	}{
	\left(
	\frac{x}{\pi\lambda}+\frac{1}{2}+\frac{n-2k-1}{2\lambda}
	\right)_{\theta}
	}.
\end{eqnarray}
For compactness we have introduced the following notation
\begin{align}
\left(x\right)_{\theta}\equiv\frac{\sin\left[\frac{i\theta}{2}-\frac{\pi x}{2}\right]}{\sin\left[\frac{i\theta}{2}+\frac{\pi x}{2}\right]}.
\end{align}

\end{document}

%% file: figures/derivation_tree_222.tikz
\begin{tikzpicture}
[
  grow                    = right,
  edge from parent/.style = {draw, -latex},
  every node/.style       = {font=\footnotesize},
  sloped
]
\tikzset{
  level 1/.style={level distance=7.5em, sibling distance=4em},
  level 2/.style={level distance=7.5em, sibling distance=2em},
  level 3/.style={level distance=6em, sibling distance=2em},
}
\node[]{$(2,2,2)$}
child { node {$[1,1]:(0,2,2)$}
  child { node {$[2,2]:(0,0,2)$}
    child { node {$[3,3]$}}
  }
  child { node {$[2,3]:(0,1,1)$}	
    child { node{$[2,3]$}} 
  }
}
child{node {$[1,2]:(1,1,2)$}
  child { node {$[1,2]:(0,0,2)$}
    child { node {$[3,3]$}
    }
  }
  child { node {$[1,3]:(0,1,1)$}
    child { node {$[2,3]$}
    }
  }
}
child{node {$[1,3]:(1,2,1)$}
  child { node {$[1,2]\phantom{:(0,0,0)}$} }
  child { node {$[1,3]:(0,2,0)$}
    child { node {$[2,2]$}
    }
  }
};
\end{tikzpicture}

%% file: main.bbl
\begin{thebibliography}{79}%
\makeatletter
\providecommand \@ifxundefined [1]{%
 \@ifx{#1\undefined}
}%
\providecommand \@ifnum [1]{%
 \ifnum #1\expandafter \@firstoftwo
 \else \expandafter \@secondoftwo
 \fi
}%
\providecommand \@ifx [1]{%
 \ifx #1\expandafter \@firstoftwo
 \else \expandafter \@secondoftwo
 \fi
}%
\providecommand \natexlab [1]{#1}%
\providecommand \enquote  [1]{``#1''}%
\providecommand \bibnamefont  [1]{#1}%
\providecommand \bibfnamefont [1]{#1}%
\providecommand \citenamefont [1]{#1}%
\providecommand \href@noop [0]{\@secondoftwo}%
\providecommand \href [0]{\begingroup \@sanitize@url \@href}%
\providecommand \@href[1]{\@@startlink{#1}\@@href}%
\providecommand \@@href[1]{\endgroup#1\@@endlink}%
\providecommand \@sanitize@url [0]{\catcode `\\12\catcode `\$12\catcode
  `\&12\catcode `\#12\catcode `\^12\catcode `\_12\catcode `\%12\relax}%
\providecommand \@@startlink[1]{}%
\providecommand \@@endlink[0]{}%
\providecommand \url  [0]{\begingroup\@sanitize@url \@url }%
\providecommand \@url [1]{\endgroup\@href {#1}{\urlprefix }}%
\providecommand \urlprefix  [0]{URL }%
\providecommand \Eprint [0]{\href }%
\providecommand \doibase [0]{http://dx.doi.org/}%
\providecommand \selectlanguage [0]{\@gobble}%
\providecommand \bibinfo  [0]{\@secondoftwo}%
\providecommand \bibfield  [0]{\@secondoftwo}%
\providecommand \translation [1]{[#1]}%
\providecommand \BibitemOpen [0]{}%
\providecommand \bibitemStop [0]{}%
\providecommand \bibitemNoStop [0]{.\EOS\space}%
\providecommand \EOS [0]{\spacefactor3000\relax}%
\providecommand \BibitemShut  [1]{\csname bibitem#1\endcsname}%
\let\auto@bib@innerbib\@empty
\bibitem [{\citenamefont {Chen}(2021)}]{chen_review_2021}%
  \BibitemOpen
  \bibfield  {author} {\bibinfo {author} {\bibfnamefont {J.}~\bibnamefont
  {Chen}},\ }\href {\doibase 10.1088/1742-6596/1865/2/022008} {\bibfield
  {journal} {\bibinfo  {journal} {Journal of Physics: Conference Series}\
  }\textbf {\bibinfo {volume} {1865}},\ \bibinfo {pages} {022008} (\bibinfo
  {year} {2021})}\BibitemShut {NoStop}%
\bibitem [{\citenamefont {Verstraete}\ and\ \citenamefont
  {Cirac}(2004)}]{verstraete_renormalization_2004}%
  \BibitemOpen
  \bibfield  {author} {\bibinfo {author} {\bibfnamefont {F.}~\bibnamefont
  {Verstraete}}\ and\ \bibinfo {author} {\bibfnamefont {J.}~\bibnamefont
  {Cirac}},\ }\href@noop {} {\bibfield  {journal} {\bibinfo  {journal}
  {arXiv:cond-mat/0407066 [cond-mat.str-el]}\ } (\bibinfo {year} {2004})},\
  \Eprint {http://arxiv.org/abs/cond-mat/0407066} {arXiv:cond-mat/0407066}
  \BibitemShut {NoStop}%
\bibitem [{\citenamefont {Or{\'u}s}(2014)}]{orus_practical_2014}%
  \BibitemOpen
  \bibfield  {author} {\bibinfo {author} {\bibfnamefont {R.}~\bibnamefont
  {Or{\'u}s}},\ }\href {\doibase 10.1016/j.aop.2014.06.013} {\bibfield
  {journal} {\bibinfo  {journal} {Annals of Physics}\ }\textbf {\bibinfo
  {volume} {349}},\ \bibinfo {pages} {117} (\bibinfo {year}
  {2014})}\BibitemShut {NoStop}%
\bibitem [{\citenamefont {Bridgeman}\ and\ \citenamefont
  {Chubb}(2017)}]{bridgeman_hand-waving_2017}%
  \BibitemOpen
  \bibfield  {author} {\bibinfo {author} {\bibfnamefont {J.~C.}\ \bibnamefont
  {Bridgeman}}\ and\ \bibinfo {author} {\bibfnamefont {C.~T.}\ \bibnamefont
  {Chubb}},\ }\href {\doibase 10.1088/1751-8121/aa6dc3} {\bibfield  {journal}
  {\bibinfo  {journal} {Journal of Physics A: Mathematical and Theoretical}\
  }\textbf {\bibinfo {volume} {50}},\ \bibinfo {pages} {223001} (\bibinfo
  {year} {2017})}\BibitemShut {NoStop}%
\bibitem [{\citenamefont {Polchinski}(2017)}]{polchinski_black_2017}%
  \BibitemOpen
  \bibfield  {author} {\bibinfo {author} {\bibfnamefont {J.}~\bibnamefont
  {Polchinski}},\ }in\ \href {\doibase 10.1142/9789813149441_0006} {\emph
  {\bibinfo {booktitle} {New {{Frontiers}} in {{Fields}} and {{Strings}}}}}\
  (\bibinfo  {publisher} {{WORLD SCIENTIFIC}},\ \bibinfo {address} {{Boulder,
  Colorado}},\ \bibinfo {year} {2017})\ pp.\ \bibinfo {pages}
  {353--397}\BibitemShut {NoStop}%
\bibitem [{\citenamefont {Ryu}\ and\ \citenamefont
  {Takayanagi}(2006)}]{ryu_holographic_2006}%
  \BibitemOpen
  \bibfield  {author} {\bibinfo {author} {\bibfnamefont {S.}~\bibnamefont
  {Ryu}}\ and\ \bibinfo {author} {\bibfnamefont {T.}~\bibnamefont
  {Takayanagi}},\ }\href {\doibase 10.1103/PhysRevLett.96.181602} {\bibfield
  {journal} {\bibinfo  {journal} {Physical Review Letters}\ }\textbf {\bibinfo
  {volume} {96}},\ \bibinfo {pages} {181602} (\bibinfo {year}
  {2006})}\BibitemShut {NoStop}%
\bibitem [{\citenamefont {Horodecki}\ \emph {et~al.}(2009)\citenamefont
  {Horodecki}, \citenamefont {Horodecki}, \citenamefont {Horodecki},\ and\
  \citenamefont {Horodecki}}]{horodecki_quantum_2009}%
  \BibitemOpen
  \bibfield  {author} {\bibinfo {author} {\bibfnamefont {R.}~\bibnamefont
  {Horodecki}}, \bibinfo {author} {\bibfnamefont {P.}~\bibnamefont
  {Horodecki}}, \bibinfo {author} {\bibfnamefont {M.}~\bibnamefont
  {Horodecki}}, \ and\ \bibinfo {author} {\bibfnamefont {K.}~\bibnamefont
  {Horodecki}},\ }\href {\doibase 10.1103/RevModPhys.81.865} {\bibfield
  {journal} {\bibinfo  {journal} {Reviews of Modern Physics}\ }\textbf
  {\bibinfo {volume} {81}},\ \bibinfo {pages} {865} (\bibinfo {year}
  {2009})}\BibitemShut {NoStop}%
\bibitem [{\citenamefont {Laflorencie}(2016)}]{laflorencie_quantum_2016}%
  \BibitemOpen
  \bibfield  {author} {\bibinfo {author} {\bibfnamefont {N.}~\bibnamefont
  {Laflorencie}},\ }\href {\doibase 10.1016/j.physrep.2016.06.008} {\bibfield
  {journal} {\bibinfo  {journal} {Physics Reports}\ }\textbf {\bibinfo {volume}
  {646}},\ \bibinfo {pages} {1} (\bibinfo {year} {2016})}\BibitemShut {NoStop}%
\bibitem [{\citenamefont {Calabrese}\ and\ \citenamefont
  {Cardy}(2009)}]{calabrese_entanglement_2009}%
  \BibitemOpen
  \bibfield  {author} {\bibinfo {author} {\bibfnamefont {P.}~\bibnamefont
  {Calabrese}}\ and\ \bibinfo {author} {\bibfnamefont {J.}~\bibnamefont
  {Cardy}},\ }\href {\doibase 10.1088/1751-8113/42/50/504005} {\bibfield
  {journal} {\bibinfo  {journal} {Journal of Physics A: Mathematical and
  Theoretical}\ }\textbf {\bibinfo {volume} {42}},\ \bibinfo {pages} {504005}
  (\bibinfo {year} {2009})}\BibitemShut {NoStop}%
\bibitem [{\citenamefont {Casini}\ and\ \citenamefont
  {Huerta}(2009)}]{casini_entanglement_2009}%
  \BibitemOpen
  \bibfield  {author} {\bibinfo {author} {\bibfnamefont {H.}~\bibnamefont
  {Casini}}\ and\ \bibinfo {author} {\bibfnamefont {M.}~\bibnamefont
  {Huerta}},\ }\href {\doibase 10.1088/1751-8113/42/50/504007} {\bibfield
  {journal} {\bibinfo  {journal} {Journal of Physics A: Mathematical and
  Theoretical}\ }\textbf {\bibinfo {volume} {42}},\ \bibinfo {pages} {504007}
  (\bibinfo {year} {2009})}\BibitemShut {NoStop}%
\bibitem [{\citenamefont {Calabrese}\ and\ \citenamefont
  {Cardy}(2004)}]{calabrese_entanglement_2004}%
  \BibitemOpen
  \bibfield  {author} {\bibinfo {author} {\bibfnamefont {P.}~\bibnamefont
  {Calabrese}}\ and\ \bibinfo {author} {\bibfnamefont {J.}~\bibnamefont
  {Cardy}},\ }\href {\doibase 10.1088/1742-5468/2004/06/P06002} {\bibfield
  {journal} {\bibinfo  {journal} {Journal of Statistical Mechanics: Theory and
  Experiment}\ }\textbf {\bibinfo {volume} {2004}},\ \bibinfo {pages} {P06002}
  (\bibinfo {year} {2004})}\BibitemShut {NoStop}%
\bibitem [{\citenamefont {Cardy}\ \emph {et~al.}(2007)\citenamefont {Cardy},
  \citenamefont {{Castro-Alvaredo}},\ and\ \citenamefont
  {Doyon}}]{cardy_form_2007}%
  \BibitemOpen
  \bibfield  {author} {\bibinfo {author} {\bibfnamefont {J.~L.}\ \bibnamefont
  {Cardy}}, \bibinfo {author} {\bibfnamefont {O.~A.}\ \bibnamefont
  {{Castro-Alvaredo}}}, \ and\ \bibinfo {author} {\bibfnamefont
  {B.}~\bibnamefont {Doyon}},\ }\href {\doibase 10.1007/s10955-007-9422-x}
  {\bibfield  {journal} {\bibinfo  {journal} {Journal of Statistical Physics}\
  }\textbf {\bibinfo {volume} {130}},\ \bibinfo {pages} {129} (\bibinfo {year}
  {2007})}\BibitemShut {NoStop}%
\bibitem [{\citenamefont {Doyon}(2009)}]{doyon_bipartite_2009}%
  \BibitemOpen
  \bibfield  {author} {\bibinfo {author} {\bibfnamefont {B.}~\bibnamefont
  {Doyon}},\ }\href {\doibase 10.1103/PhysRevLett.102.031602} {\bibfield
  {journal} {\bibinfo  {journal} {Physical Review Letters}\ }\textbf {\bibinfo
  {volume} {102}},\ \bibinfo {pages} {031602} (\bibinfo {year}
  {2009})}\BibitemShut {NoStop}%
\bibitem [{\citenamefont {Serafini}(2017)}]{serafini_quantum_2017}%
  \BibitemOpen
  \bibfield  {author} {\bibinfo {author} {\bibfnamefont {A.}~\bibnamefont
  {Serafini}},\ }\href@noop {} {\emph {\bibinfo {title} {Quantum Continuous
  Variables: A Primer of Theoretical Methods}}}\ (\bibinfo  {publisher} {{CRC
  Press, Taylor \& Francis Group, CRC Press is an imprint of the Taylor \&
  Francis Group, an informa business}},\ \bibinfo {address} {{Boca Raton}},\
  \bibinfo {year} {2017})\BibitemShut {NoStop}%
\bibitem [{\citenamefont
  {Schollw{\"o}ck}(2011)}]{schollwock_density-matrix_2011}%
  \BibitemOpen
  \bibfield  {author} {\bibinfo {author} {\bibfnamefont {U.}~\bibnamefont
  {Schollw{\"o}ck}},\ }\href {\doibase 10.1016/j.aop.2010.09.012} {\bibfield
  {journal} {\bibinfo  {journal} {Annals of Physics}\ }\textbf {\bibinfo
  {volume} {326}},\ \bibinfo {pages} {96} (\bibinfo {year} {2011})}\BibitemShut
  {NoStop}%
\bibitem [{\citenamefont {Carmen~Ba{\~n}uls}\ and\ \citenamefont
  {Cichy}(2020)}]{carmen_banuls_review_2020}%
  \BibitemOpen
  \bibfield  {author} {\bibinfo {author} {\bibfnamefont {M.}~\bibnamefont
  {Carmen~Ba{\~n}uls}}\ and\ \bibinfo {author} {\bibfnamefont {K.}~\bibnamefont
  {Cichy}},\ }\href {\doibase 10.1088/1361-6633/ab6311} {\bibfield  {journal}
  {\bibinfo  {journal} {Reports on Progress in Physics}\ }\textbf {\bibinfo
  {volume} {83}},\ \bibinfo {pages} {024401} (\bibinfo {year}
  {2020})}\BibitemShut {NoStop}%
\bibitem [{\citenamefont {James}\ \emph {et~al.}(2018)\citenamefont {James},
  \citenamefont {Konik}, \citenamefont {Lecheminant}, \citenamefont
  {Robinson},\ and\ \citenamefont {Tsvelik}}]{james_non-perturbative_2018}%
  \BibitemOpen
  \bibfield  {author} {\bibinfo {author} {\bibfnamefont {A.~J.~A.}\
  \bibnamefont {James}}, \bibinfo {author} {\bibfnamefont {R.~M.}\ \bibnamefont
  {Konik}}, \bibinfo {author} {\bibfnamefont {P.}~\bibnamefont {Lecheminant}},
  \bibinfo {author} {\bibfnamefont {N.~J.}\ \bibnamefont {Robinson}}, \ and\
  \bibinfo {author} {\bibfnamefont {A.~M.}\ \bibnamefont {Tsvelik}},\ }\href
  {\doibase 10.1088/1361-6633/aa91ea} {\bibfield  {journal} {\bibinfo
  {journal} {Reports on Progress in Physics}\ }\textbf {\bibinfo {volume}
  {81}},\ \bibinfo {pages} {046002} (\bibinfo {year} {2018})}\BibitemShut
  {NoStop}%
\bibitem [{\citenamefont {L{\"a}ssig}\ \emph {et~al.}(1991)\citenamefont
  {L{\"a}ssig}, \citenamefont {Mussardo},\ and\ \citenamefont
  {Cardy}}]{lassig_scaling_1991}%
  \BibitemOpen
  \bibfield  {author} {\bibinfo {author} {\bibfnamefont {M.}~\bibnamefont
  {L{\"a}ssig}}, \bibinfo {author} {\bibfnamefont {G.}~\bibnamefont
  {Mussardo}}, \ and\ \bibinfo {author} {\bibfnamefont {J.~L.}\ \bibnamefont
  {Cardy}},\ }\href {\doibase 10.1016/0550-3213(91)90206-D} {\bibfield
  {journal} {\bibinfo  {journal} {Nuclear Physics B}\ }\textbf {\bibinfo
  {volume} {348}},\ \bibinfo {pages} {591} (\bibinfo {year}
  {1991})}\BibitemShut {NoStop}%
\bibitem [{\citenamefont {Feverati}\ \emph {et~al.}(1998)\citenamefont
  {Feverati}, \citenamefont {Ravanini},\ and\ \citenamefont
  {Tak{\'a}cs}}]{feverati_scaling_1998}%
  \BibitemOpen
  \bibfield  {author} {\bibinfo {author} {\bibfnamefont {G.}~\bibnamefont
  {Feverati}}, \bibinfo {author} {\bibfnamefont {F.}~\bibnamefont {Ravanini}},
  \ and\ \bibinfo {author} {\bibfnamefont {G.}~\bibnamefont {Tak{\'a}cs}},\
  }\href {\doibase 10.1016/S0370-2693(98)01406-3} {\bibfield  {journal}
  {\bibinfo  {journal} {Physics Letters B}\ }\textbf {\bibinfo {volume}
  {444}},\ \bibinfo {pages} {442} (\bibinfo {year} {1998})}\BibitemShut
  {NoStop}%
\bibitem [{\citenamefont {Bajnok}\ \emph {et~al.}(2001)\citenamefont {Bajnok},
  \citenamefont {Palla},\ and\ \citenamefont
  {Tak{\'a}cs}}]{bajnok_boundary_2001}%
  \BibitemOpen
  \bibfield  {author} {\bibinfo {author} {\bibfnamefont {Z.}~\bibnamefont
  {Bajnok}}, \bibinfo {author} {\bibfnamefont {L.}~\bibnamefont {Palla}}, \
  and\ \bibinfo {author} {\bibfnamefont {G.}~\bibnamefont {Tak{\'a}cs}},\
  }\href {\doibase 10.1016/S0550-3213(01)00391-1} {\bibfield  {journal}
  {\bibinfo  {journal} {Nuclear Physics B}\ }\textbf {\bibinfo {volume}
  {614}},\ \bibinfo {pages} {405} (\bibinfo {year} {2001})}\BibitemShut
  {NoStop}%
\bibitem [{\citenamefont {Bajnok}\ \emph {et~al.}(2002)\citenamefont {Bajnok},
  \citenamefont {Palla},\ and\ \citenamefont
  {Tak{\'a}cs}}]{bajnok_finite_2002}%
  \BibitemOpen
  \bibfield  {author} {\bibinfo {author} {\bibfnamefont {Z.}~\bibnamefont
  {Bajnok}}, \bibinfo {author} {\bibfnamefont {L.}~\bibnamefont {Palla}}, \
  and\ \bibinfo {author} {\bibfnamefont {G.}~\bibnamefont {Tak{\'a}cs}},\
  }\href {\doibase 10.1016/S0550-3213(01)00616-2} {\bibfield  {journal}
  {\bibinfo  {journal} {Nuclear Physics B}\ }\textbf {\bibinfo {volume}
  {622}},\ \bibinfo {pages} {565} (\bibinfo {year} {2002})}\BibitemShut
  {NoStop}%
\bibitem [{\citenamefont {Rychkov}\ and\ \citenamefont
  {Vitale}(2015)}]{rychkov_hamiltonian_2015}%
  \BibitemOpen
  \bibfield  {author} {\bibinfo {author} {\bibfnamefont {S.}~\bibnamefont
  {Rychkov}}\ and\ \bibinfo {author} {\bibfnamefont {L.~G.}\ \bibnamefont
  {Vitale}},\ }\href {\doibase 10.1103/PhysRevD.91.085011} {\bibfield
  {journal} {\bibinfo  {journal} {Physical Review D}\ }\textbf {\bibinfo
  {volume} {91}},\ \bibinfo {pages} {085011} (\bibinfo {year} {2015})},\
  \Eprint {http://arxiv.org/abs/1412.3460} {arXiv:1412.3460} \BibitemShut
  {NoStop}%
\bibitem [{\citenamefont {Rychkov}\ and\ \citenamefont
  {Vitale}(2016)}]{rychkov_hamiltonian_2016}%
  \BibitemOpen
  \bibfield  {author} {\bibinfo {author} {\bibfnamefont {S.}~\bibnamefont
  {Rychkov}}\ and\ \bibinfo {author} {\bibfnamefont {L.~G.}\ \bibnamefont
  {Vitale}},\ }\href {\doibase 10.1103/PhysRevD.93.065014} {\bibfield
  {journal} {\bibinfo  {journal} {Physical Review D}\ }\textbf {\bibinfo
  {volume} {93}},\ \bibinfo {pages} {065014} (\bibinfo {year}
  {2016})}\BibitemShut {NoStop}%
\bibitem [{\citenamefont {{Elias-Mir{\'o}}}\ \emph {et~al.}(2017)\citenamefont
  {{Elias-Mir{\'o}}}, \citenamefont {Rychkov},\ and\ \citenamefont
  {Vitale}}]{elias-miro_nlo_2017}%
  \BibitemOpen
  \bibfield  {author} {\bibinfo {author} {\bibfnamefont {J.}~\bibnamefont
  {{Elias-Mir{\'o}}}}, \bibinfo {author} {\bibfnamefont {S.}~\bibnamefont
  {Rychkov}}, \ and\ \bibinfo {author} {\bibfnamefont {L.~G.}\ \bibnamefont
  {Vitale}},\ }\href {\doibase 10.1103/PhysRevD.96.065024} {\bibfield
  {journal} {\bibinfo  {journal} {Physical Review D}\ }\textbf {\bibinfo
  {volume} {96}},\ \bibinfo {pages} {065024} (\bibinfo {year}
  {2017})}\BibitemShut {NoStop}%
\bibitem [{\citenamefont {Konik}\ \emph {et~al.}(2021)\citenamefont {Konik},
  \citenamefont {L{\'a}jer},\ and\ \citenamefont
  {Mussardo}}]{konik_approaching_2021}%
  \BibitemOpen
  \bibfield  {author} {\bibinfo {author} {\bibfnamefont {R.}~\bibnamefont
  {Konik}}, \bibinfo {author} {\bibfnamefont {M.}~\bibnamefont {L{\'a}jer}}, \
  and\ \bibinfo {author} {\bibfnamefont {G.}~\bibnamefont {Mussardo}},\ }\href
  {\doibase 10.1007/JHEP01(2021)014} {\bibfield  {journal} {\bibinfo  {journal}
  {Journal of High Energy Physics}\ }\textbf {\bibinfo {volume} {2021}},\
  \bibinfo {pages} {14} (\bibinfo {year} {2021})}\BibitemShut {NoStop}%
\bibitem [{\citenamefont {Horvath}\ \emph {et~al.}(2022)\citenamefont
  {Horvath}, \citenamefont {Hodsagi},\ and\ \citenamefont
  {Takacs}}]{horvath_chirally_2022}%
  \BibitemOpen
  \bibfield  {author} {\bibinfo {author} {\bibfnamefont {D.~X.}\ \bibnamefont
  {Horvath}}, \bibinfo {author} {\bibfnamefont {K.}~\bibnamefont {Hodsagi}}, \
  and\ \bibinfo {author} {\bibfnamefont {G.}~\bibnamefont {Takacs}},\
  }\href@noop {} {\bibfield  {journal} {\bibinfo  {journal} {arXiv:2201.06509
  [cond-mat, physics:hep-th, physics:physics]}\ } (\bibinfo {year} {2022})},\
  \Eprint {http://arxiv.org/abs/2201.06509} {arXiv:2201.06509 [cond-mat,
  physics:hep-th, physics:physics]} \BibitemShut {NoStop}%
\bibitem [{\citenamefont {Konik}\ \emph {et~al.}(2015)\citenamefont {Konik},
  \citenamefont {P{\'a}lmai}, \citenamefont {Tak{\'a}cs},\ and\ \citenamefont
  {Tsvelik}}]{konik_studying_2015}%
  \BibitemOpen
  \bibfield  {author} {\bibinfo {author} {\bibfnamefont {R.}~\bibnamefont
  {Konik}}, \bibinfo {author} {\bibfnamefont {T.}~\bibnamefont {P{\'a}lmai}},
  \bibinfo {author} {\bibfnamefont {G.}~\bibnamefont {Tak{\'a}cs}}, \ and\
  \bibinfo {author} {\bibfnamefont {A.}~\bibnamefont {Tsvelik}},\ }\href
  {\doibase 10.1016/j.nuclphysb.2015.08.016} {\bibfield  {journal} {\bibinfo
  {journal} {Nuclear Physics B}\ }\textbf {\bibinfo {volume} {899}},\ \bibinfo
  {pages} {547} (\bibinfo {year} {2015})}\BibitemShut {NoStop}%
\bibitem [{\citenamefont {Azaria}\ \emph {et~al.}(2016)\citenamefont {Azaria},
  \citenamefont {Konik}, \citenamefont {Lecheminant}, \citenamefont
  {P{\'a}lmai}, \citenamefont {Tak{\'a}cs},\ and\ \citenamefont
  {Tsvelik}}]{azaria_particle_2016}%
  \BibitemOpen
  \bibfield  {author} {\bibinfo {author} {\bibfnamefont {P.}~\bibnamefont
  {Azaria}}, \bibinfo {author} {\bibfnamefont {R.~M.}\ \bibnamefont {Konik}},
  \bibinfo {author} {\bibfnamefont {P.}~\bibnamefont {Lecheminant}}, \bibinfo
  {author} {\bibfnamefont {T.}~\bibnamefont {P{\'a}lmai}}, \bibinfo {author}
  {\bibfnamefont {G.}~\bibnamefont {Tak{\'a}cs}}, \ and\ \bibinfo {author}
  {\bibfnamefont {A.~M.}\ \bibnamefont {Tsvelik}},\ }\href {\doibase
  10.1103/PhysRevD.94.045003} {\bibfield  {journal} {\bibinfo  {journal}
  {Physical Review D}\ }\textbf {\bibinfo {volume} {94}},\ \bibinfo {pages}
  {045003} (\bibinfo {year} {2016})}\BibitemShut {NoStop}%
\bibitem [{\citenamefont {Kukuljan}(2021)}]{kukuljan_continuum_2021}%
  \BibitemOpen
  \bibfield  {author} {\bibinfo {author} {\bibfnamefont {I.}~\bibnamefont
  {Kukuljan}},\ }\href {\doibase 10.1103/PhysRevD.104.L021702} {\bibfield
  {journal} {\bibinfo  {journal} {Physical Review D}\ }\textbf {\bibinfo
  {volume} {104}},\ \bibinfo {pages} {L021702} (\bibinfo {year}
  {2021})}\BibitemShut {NoStop}%
\bibitem [{\citenamefont {Kukuljan}\ \emph {et~al.}(2018)\citenamefont
  {Kukuljan}, \citenamefont {Sotiriadis},\ and\ \citenamefont
  {Takacs}}]{kukuljan_correlation_2018}%
  \BibitemOpen
  \bibfield  {author} {\bibinfo {author} {\bibfnamefont {I.}~\bibnamefont
  {Kukuljan}}, \bibinfo {author} {\bibfnamefont {S.}~\bibnamefont
  {Sotiriadis}}, \ and\ \bibinfo {author} {\bibfnamefont {G.}~\bibnamefont
  {Takacs}},\ }\href {\doibase 10.1103/PhysRevLett.121.110402} {\bibfield
  {journal} {\bibinfo  {journal} {Physical Review Letters}\ }\textbf {\bibinfo
  {volume} {121}},\ \bibinfo {pages} {110402} (\bibinfo {year}
  {2018})}\BibitemShut {NoStop}%
\bibitem [{\citenamefont {Rakovszky}\ \emph {et~al.}(2016)\citenamefont
  {Rakovszky}, \citenamefont {Mesty{\'a}n}, \citenamefont {Collura},
  \citenamefont {Kormos},\ and\ \citenamefont
  {Tak{\'a}cs}}]{rakovszky_hamiltonian_2016}%
  \BibitemOpen
  \bibfield  {author} {\bibinfo {author} {\bibfnamefont {T.}~\bibnamefont
  {Rakovszky}}, \bibinfo {author} {\bibfnamefont {M.}~\bibnamefont
  {Mesty{\'a}n}}, \bibinfo {author} {\bibfnamefont {M.}~\bibnamefont
  {Collura}}, \bibinfo {author} {\bibfnamefont {M.}~\bibnamefont {Kormos}}, \
  and\ \bibinfo {author} {\bibfnamefont {G.}~\bibnamefont {Tak{\'a}cs}},\
  }\href {\doibase 10.1016/j.nuclphysb.2016.08.024} {\bibfield  {journal}
  {\bibinfo  {journal} {Nuclear Physics B}\ }\textbf {\bibinfo {volume}
  {911}},\ \bibinfo {pages} {805} (\bibinfo {year} {2016})}\BibitemShut
  {NoStop}%
\bibitem [{\citenamefont {H{\'o}ds{\'a}gi}\ \emph {et~al.}(2018)\citenamefont
  {H{\'o}ds{\'a}gi}, \citenamefont {Kormos},\ and\ \citenamefont
  {Tak{\'a}cs}}]{hodsagi_quench_2018}%
  \BibitemOpen
  \bibfield  {author} {\bibinfo {author} {\bibfnamefont {K.}~\bibnamefont
  {H{\'o}ds{\'a}gi}}, \bibinfo {author} {\bibfnamefont {M.}~\bibnamefont
  {Kormos}}, \ and\ \bibinfo {author} {\bibfnamefont {G.}~\bibnamefont
  {Tak{\'a}cs}},\ }\href {\doibase 10.21468/SciPostPhys.5.3.027} {\bibfield
  {journal} {\bibinfo  {journal} {SciPost Physics}\ }\textbf {\bibinfo {volume}
  {5}},\ \bibinfo {pages} {027} (\bibinfo {year} {2018})}\BibitemShut {NoStop}%
\bibitem [{\citenamefont {Horv{\'a}th}\ \emph {et~al.}(2019)\citenamefont
  {Horv{\'a}th}, \citenamefont {Lovas}, \citenamefont {Kormos}, \citenamefont
  {Tak{\'a}cs},\ and\ \citenamefont
  {Zar{\'a}nd}}]{horvath_nonequilibrium_2019}%
  \BibitemOpen
  \bibfield  {author} {\bibinfo {author} {\bibfnamefont {D.~X.}\ \bibnamefont
  {Horv{\'a}th}}, \bibinfo {author} {\bibfnamefont {I.}~\bibnamefont {Lovas}},
  \bibinfo {author} {\bibfnamefont {M.}~\bibnamefont {Kormos}}, \bibinfo
  {author} {\bibfnamefont {G.}~\bibnamefont {Tak{\'a}cs}}, \ and\ \bibinfo
  {author} {\bibfnamefont {G.}~\bibnamefont {Zar{\'a}nd}},\ }\href {\doibase
  10.1103/PhysRevA.100.013613} {\bibfield  {journal} {\bibinfo  {journal}
  {Physical Review A}\ }\textbf {\bibinfo {volume} {100}},\ \bibinfo {pages}
  {013613} (\bibinfo {year} {2019})}\BibitemShut {NoStop}%
\bibitem [{\citenamefont {Horv{\'a}th}\ \emph {et~al.}(2021)\citenamefont
  {Horv{\'a}th}, \citenamefont {Kormos}, \citenamefont {Sotiriadis},\ and\
  \citenamefont {Tak{\'a}cs}}]{horvath_inhomogeneous_2021}%
  \BibitemOpen
  \bibfield  {author} {\bibinfo {author} {\bibfnamefont {D.~X.}\ \bibnamefont
  {Horv{\'a}th}}, \bibinfo {author} {\bibfnamefont {M.}~\bibnamefont {Kormos}},
  \bibinfo {author} {\bibfnamefont {S.}~\bibnamefont {Sotiriadis}}, \ and\
  \bibinfo {author} {\bibfnamefont {G.}~\bibnamefont {Tak{\'a}cs}},\
  }\href@noop {} {\bibfield  {journal} {\bibinfo  {journal} {arXiv:2109.06869
  [cond-mat, physics:hep-th]}\ } (\bibinfo {year} {2021})},\ \Eprint
  {http://arxiv.org/abs/2109.06869} {arXiv:2109.06869 [cond-mat,
  physics:hep-th]} \BibitemShut {NoStop}%
\bibitem [{\citenamefont {H{\'o}ds{\'a}gi}\ and\ \citenamefont
  {Kormos}(2020)}]{hodsagi_kibblezurek_2020}%
  \BibitemOpen
  \bibfield  {author} {\bibinfo {author} {\bibfnamefont {K.}~\bibnamefont
  {H{\'o}ds{\'a}gi}}\ and\ \bibinfo {author} {\bibfnamefont {M.}~\bibnamefont
  {Kormos}},\ }\href {\doibase 10.21468/SciPostPhys.9.4.055} {\bibfield
  {journal} {\bibinfo  {journal} {SciPost Physics}\ }\textbf {\bibinfo {volume}
  {9}},\ \bibinfo {pages} {055} (\bibinfo {year} {2020})}\BibitemShut {NoStop}%
\bibitem [{\citenamefont {Brandino}\ \emph {et~al.}(2010)\citenamefont
  {Brandino}, \citenamefont {Konik},\ and\ \citenamefont
  {Mussardo}}]{brandino_energy_2010}%
  \BibitemOpen
  \bibfield  {author} {\bibinfo {author} {\bibfnamefont {G.~P.}\ \bibnamefont
  {Brandino}}, \bibinfo {author} {\bibfnamefont {R.~M.}\ \bibnamefont {Konik}},
  \ and\ \bibinfo {author} {\bibfnamefont {G.}~\bibnamefont {Mussardo}},\
  }\href {\doibase 10.1088/1742-5468/2010/07/P07013} {\bibfield  {journal}
  {\bibinfo  {journal} {Journal of Statistical Mechanics: Theory and
  Experiment}\ }\textbf {\bibinfo {volume} {2010}},\ \bibinfo {pages} {P07013}
  (\bibinfo {year} {2010})}\BibitemShut {NoStop}%
\bibitem [{\citenamefont {Srdin{\v s}ek}\ \emph {et~al.}(2021)\citenamefont
  {Srdin{\v s}ek}, \citenamefont {Prosen},\ and\ \citenamefont
  {Sotiriadis}}]{srdinsek_signatures_2021}%
  \BibitemOpen
  \bibfield  {author} {\bibinfo {author} {\bibfnamefont {M.}~\bibnamefont
  {Srdin{\v s}ek}}, \bibinfo {author} {\bibfnamefont {T.}~\bibnamefont
  {Prosen}}, \ and\ \bibinfo {author} {\bibfnamefont {S.}~\bibnamefont
  {Sotiriadis}},\ }\href {\doibase 10.1103/PhysRevLett.126.121602} {\bibfield
  {journal} {\bibinfo  {journal} {Physical Review Letters}\ }\textbf {\bibinfo
  {volume} {126}},\ \bibinfo {pages} {121602} (\bibinfo {year}
  {2021})}\BibitemShut {NoStop}%
\bibitem [{\citenamefont {Lencs{\'e}s}\ \emph {et~al.}(2021)\citenamefont
  {Lencs{\'e}s}, \citenamefont {Mussardo},\ and\ \citenamefont
  {Tak{\'a}cs}}]{lencses_confinement_2021}%
  \BibitemOpen
  \bibfield  {author} {\bibinfo {author} {\bibfnamefont {M.}~\bibnamefont
  {Lencs{\'e}s}}, \bibinfo {author} {\bibfnamefont {G.}~\bibnamefont
  {Mussardo}}, \ and\ \bibinfo {author} {\bibfnamefont {G.}~\bibnamefont
  {Tak{\'a}cs}},\ }\href@noop {} {\bibfield  {journal} {\bibinfo  {journal}
  {arXiv:2111.05360 [cond-mat, physics:hep-th]}\ } (\bibinfo {year} {2021})},\
  \Eprint {http://arxiv.org/abs/2111.05360} {arXiv:2111.05360 [cond-mat,
  physics:hep-th]} \BibitemShut {NoStop}%
\bibitem [{\citenamefont {Cubero}\ \emph {et~al.}(2022)\citenamefont {Cubero},
  \citenamefont {Konik}, \citenamefont {Lencs{\'e}s}, \citenamefont
  {Mussardo},\ and\ \citenamefont {Tak{\'a}cs}}]{cubero_duality_2022}%
  \BibitemOpen
  \bibfield  {author} {\bibinfo {author} {\bibfnamefont {A.~C.}\ \bibnamefont
  {Cubero}}, \bibinfo {author} {\bibfnamefont {R.~M.}\ \bibnamefont {Konik}},
  \bibinfo {author} {\bibfnamefont {M.}~\bibnamefont {Lencs{\'e}s}}, \bibinfo
  {author} {\bibfnamefont {G.}~\bibnamefont {Mussardo}}, \ and\ \bibinfo
  {author} {\bibfnamefont {G.}~\bibnamefont {Tak{\'a}cs}},\ }\href@noop {}
  {\bibfield  {journal} {\bibinfo  {journal} {arXiv:2109.09767 [cond-mat,
  physics:hep-th]}\ } (\bibinfo {year} {2022})},\ \Eprint
  {http://arxiv.org/abs/2109.09767} {arXiv:2109.09767 [cond-mat,
  physics:hep-th]} \BibitemShut {NoStop}%
\bibitem [{\citenamefont {Kukuljan}\ \emph {et~al.}(2020)\citenamefont
  {Kukuljan}, \citenamefont {Sotiriadis},\ and\ \citenamefont
  {Tak{\'a}cs}}]{kukuljan_out--horizon_2020}%
  \BibitemOpen
  \bibfield  {author} {\bibinfo {author} {\bibfnamefont {I.}~\bibnamefont
  {Kukuljan}}, \bibinfo {author} {\bibfnamefont {S.}~\bibnamefont
  {Sotiriadis}}, \ and\ \bibinfo {author} {\bibfnamefont {G.}~\bibnamefont
  {Tak{\'a}cs}},\ }\href {\doibase 10.1007/JHEP07(2020)224} {\bibfield
  {journal} {\bibinfo  {journal} {Journal of High Energy Physics}\ }\textbf
  {\bibinfo {volume} {2020}},\ \bibinfo {pages} {224} (\bibinfo {year}
  {2020})}\BibitemShut {NoStop}%
\bibitem [{\citenamefont {Hogervorst}\ \emph {et~al.}(2015)\citenamefont
  {Hogervorst}, \citenamefont {Rychkov},\ and\ \citenamefont {{van
  Rees}}}]{hogervorst_truncated_2015}%
  \BibitemOpen
  \bibfield  {author} {\bibinfo {author} {\bibfnamefont {M.}~\bibnamefont
  {Hogervorst}}, \bibinfo {author} {\bibfnamefont {S.}~\bibnamefont {Rychkov}},
  \ and\ \bibinfo {author} {\bibfnamefont {B.~C.}\ \bibnamefont {{van Rees}}},\
  }\href {\doibase 10.1103/PhysRevD.91.025005} {\bibfield  {journal} {\bibinfo
  {journal} {Physical Review D}\ }\textbf {\bibinfo {volume} {91}},\ \bibinfo
  {pages} {025005} (\bibinfo {year} {2015})}\BibitemShut {NoStop}%
\bibitem [{\citenamefont {{Elias-Mir{\'o}}}\ and\ \citenamefont
  {Hardy}(2020)}]{elias-miro_exploring_2020}%
  \BibitemOpen
  \bibfield  {author} {\bibinfo {author} {\bibfnamefont {J.}~\bibnamefont
  {{Elias-Mir{\'o}}}}\ and\ \bibinfo {author} {\bibfnamefont {E.}~\bibnamefont
  {Hardy}},\ }\href {\doibase 10.1103/PhysRevD.102.065001} {\bibfield
  {journal} {\bibinfo  {journal} {Physical Review D}\ }\textbf {\bibinfo
  {volume} {102}},\ \bibinfo {pages} {065001} (\bibinfo {year}
  {2020})}\BibitemShut {NoStop}%
\bibitem [{\citenamefont {Palmai}(2016)}]{palmai_entanglement_2016}%
  \BibitemOpen
  \bibfield  {author} {\bibinfo {author} {\bibfnamefont {T.}~\bibnamefont
  {Palmai}},\ }\href {\doibase 10.1016/j.physletb.2016.06.012} {\bibfield
  {journal} {\bibinfo  {journal} {Physics Letters B}\ }\textbf {\bibinfo
  {volume} {759}},\ \bibinfo {pages} {439} (\bibinfo {year}
  {2016})}\BibitemShut {NoStop}%
\bibitem [{\citenamefont {Murciano}\ \emph {et~al.}(2021)\citenamefont
  {Murciano}, \citenamefont {Calabrese},\ and\ \citenamefont
  {Konik}}]{murciano_post-quantum_2021}%
  \BibitemOpen
  \bibfield  {author} {\bibinfo {author} {\bibfnamefont {S.}~\bibnamefont
  {Murciano}}, \bibinfo {author} {\bibfnamefont {P.}~\bibnamefont {Calabrese}},
  \ and\ \bibinfo {author} {\bibfnamefont {R.~M.}\ \bibnamefont {Konik}},\
  }\href@noop {} {\bibfield  {journal} {\bibinfo  {journal} {arXiv:2112.04412
  [cond-mat, physics:quant-ph]}\ } (\bibinfo {year} {2021})},\ \Eprint
  {http://arxiv.org/abs/2112.04412} {arXiv:2112.04412 [cond-mat,
  physics:quant-ph]} \BibitemShut {NoStop}%
\bibitem [{\citenamefont {Yurov}\ and\ \citenamefont
  {Zamolodchikov}(1990)}]{yurov_truncated_1990}%
  \BibitemOpen
  \bibfield  {author} {\bibinfo {author} {\bibfnamefont {V.~P.}\ \bibnamefont
  {Yurov}}\ and\ \bibinfo {author} {\bibfnamefont {A.~B.}\ \bibnamefont
  {Zamolodchikov}},\ }\href {\doibase 10.1142/S0217751X9000218X} {\bibfield
  {journal} {\bibinfo  {journal} {International Journal of Modern Physics A}\
  }\textbf {\bibinfo {volume} {05}},\ \bibinfo {pages} {3221} (\bibinfo {year}
  {1990})}\BibitemShut {NoStop}%
\bibitem [{\citenamefont {Yurov}\ and\ \citenamefont
  {Zamolodchikov}(1991)}]{yurov_truncated-fermionic-space_1991}%
  \BibitemOpen
  \bibfield  {author} {\bibinfo {author} {\bibfnamefont {V.}~\bibnamefont
  {Yurov}}\ and\ \bibinfo {author} {\bibfnamefont {A.}~\bibnamefont
  {Zamolodchikov}},\ }\href {\doibase 10.1142/S0217751X91002161} {\bibfield
  {journal} {\bibinfo  {journal} {International Journal of Modern Physics A}\
  }\textbf {\bibinfo {volume} {06}},\ \bibinfo {pages} {4557} (\bibinfo {year}
  {1991})}\BibitemShut {NoStop}%
\bibitem [{\citenamefont {Konik}\ and\ \citenamefont
  {Adamov}(2007)}]{konik_numerical_2007}%
  \BibitemOpen
  \bibfield  {author} {\bibinfo {author} {\bibfnamefont {R.~M.}\ \bibnamefont
  {Konik}}\ and\ \bibinfo {author} {\bibfnamefont {Y.}~\bibnamefont {Adamov}},\
  }\href {\doibase 10.1103/PhysRevLett.98.147205} {\bibfield  {journal}
  {\bibinfo  {journal} {Physical Review Letters}\ }\textbf {\bibinfo {volume}
  {98}},\ \bibinfo {pages} {147205} (\bibinfo {year} {2007})}\BibitemShut
  {NoStop}%
\bibitem [{\citenamefont {Qin}\ \emph {et~al.}(2001)\citenamefont {Qin},
  \citenamefont {Wang},\ and\ \citenamefont {Li}}]{qin_general_2001}%
  \BibitemOpen
  \bibfield  {author} {\bibinfo {author} {\bibfnamefont {G.}~\bibnamefont
  {Qin}}, \bibinfo {author} {\bibfnamefont {K.-l.}\ \bibnamefont {Wang}}, \
  and\ \bibinfo {author} {\bibfnamefont {T.-z.}\ \bibnamefont {Li}},\
  }\href@noop {} {\bibfield  {journal} {\bibinfo  {journal}
  {arXiv:quant-ph/0109020}\ } (\bibinfo {year} {2001})},\ \Eprint
  {http://arxiv.org/abs/quant-ph/0109020} {arXiv:quant-ph/0109020} \BibitemShut
  {NoStop}%
\bibitem [{\citenamefont {Trembinska}(1985)}]{trembinska_variations_1985}%
  \BibitemOpen
  \bibfield  {author} {\bibinfo {author} {\bibfnamefont {A.}~\bibnamefont
  {Trembinska}},\ }\href@noop {} {\emph {\bibinfo {title} {Variations on
  Carlson's Theorem}}}\ (\bibinfo  {publisher} {{Northwestern University}},\
  \bibinfo {year} {1985})\BibitemShut {NoStop}%
\bibitem [{\citenamefont {Titchmarsh}(2002)}]{titchmarsh_theory_2002}%
  \BibitemOpen
  \bibfield  {author} {\bibinfo {author} {\bibfnamefont {E.~C.}\ \bibnamefont
  {Titchmarsh}},\ }\href@noop {} {\emph {\bibinfo {title} {The Theory of
  Functions}}},\ \bibinfo {edition} {2nd}\ ed.,\ Oxford Science Publications\
  (\bibinfo  {publisher} {{Oxford Univ. Press}},\ \bibinfo {address}
  {{Oxford}},\ \bibinfo {year} {2002})\BibitemShut {NoStop}%
\bibitem [{\citenamefont {Hausdorff}(1906)}]{hausdorff_symbolische_1906}%
  \BibitemOpen
  \bibfield  {author} {\bibinfo {author} {\bibfnamefont {F.}~\bibnamefont
  {Hausdorff}},\ }\href@noop {} {\bibfield  {journal} {\bibinfo  {journal}
  {Ber. Verh. Kgl. S\"achs. Ges. Wiss. Leipzig., Math.-phys. Kl.}\ }\textbf
  {\bibinfo {volume} {58}},\ \bibinfo {pages} {19} (\bibinfo {year}
  {1906})}\BibitemShut {NoStop}%
\bibitem [{\citenamefont {Wengert}(1964)}]{wengert_simple_1964}%
  \BibitemOpen
  \bibfield  {author} {\bibinfo {author} {\bibfnamefont {R.~E.}\ \bibnamefont
  {Wengert}},\ }\href {\doibase 10.1145/355586.364791} {\bibfield  {journal}
  {\bibinfo  {journal} {Communications of the ACM}\ }\textbf {\bibinfo {volume}
  {7}},\ \bibinfo {pages} {463} (\bibinfo {year} {1964})}\BibitemShut {NoStop}%
\bibitem [{\citenamefont {{Bartholomew-Biggs}}\ \emph
  {et~al.}(2000)\citenamefont {{Bartholomew-Biggs}}, \citenamefont {Brown},
  \citenamefont {Christianson},\ and\ \citenamefont
  {Dixon}}]{bartholomew-biggs_automatic_2000}%
  \BibitemOpen
  \bibfield  {author} {\bibinfo {author} {\bibfnamefont {M.}~\bibnamefont
  {{Bartholomew-Biggs}}}, \bibinfo {author} {\bibfnamefont {S.}~\bibnamefont
  {Brown}}, \bibinfo {author} {\bibfnamefont {B.}~\bibnamefont {Christianson}},
  \ and\ \bibinfo {author} {\bibfnamefont {L.}~\bibnamefont {Dixon}},\ }\href
  {\doibase 10.1016/S0377-0427(00)00422-2} {\bibfield  {journal} {\bibinfo
  {journal} {Journal of Computational and Applied Mathematics}\ }\textbf
  {\bibinfo {volume} {124}},\ \bibinfo {pages} {171} (\bibinfo {year}
  {2000})}\BibitemShut {NoStop}%
\bibitem [{\citenamefont {Affleck}\ and\ \citenamefont
  {Ludwig}(1991)}]{affleck_universal_1991}%
  \BibitemOpen
  \bibfield  {author} {\bibinfo {author} {\bibfnamefont {I.}~\bibnamefont
  {Affleck}}\ and\ \bibinfo {author} {\bibfnamefont {A.~W.~W.}\ \bibnamefont
  {Ludwig}},\ }\href {\doibase 10.1103/PhysRevLett.67.161} {\bibfield
  {journal} {\bibinfo  {journal} {Physical Review Letters}\ }\textbf {\bibinfo
  {volume} {67}},\ \bibinfo {pages} {161} (\bibinfo {year} {1991})}\BibitemShut
  {NoStop}%
\bibitem [{\citenamefont {Alba}\ and\ \citenamefont
  {Calabrese}(2018)}]{alba_entanglement_2018}%
  \BibitemOpen
  \bibfield  {author} {\bibinfo {author} {\bibfnamefont {V.}~\bibnamefont
  {Alba}}\ and\ \bibinfo {author} {\bibfnamefont {P.}~\bibnamefont
  {Calabrese}},\ }\href {\doibase 10.21468/SciPostPhys.4.3.017} {\bibfield
  {journal} {\bibinfo  {journal} {SciPost Physics}\ }\textbf {\bibinfo {volume}
  {4}},\ \bibinfo {pages} {017} (\bibinfo {year} {2018})}\BibitemShut {NoStop}%
\bibitem [{\citenamefont {Mussardo}(2020)}]{mussardo_statistical_2020}%
  \BibitemOpen
  \bibfield  {author} {\bibinfo {author} {\bibfnamefont {G.}~\bibnamefont
  {Mussardo}},\ }\href {\doibase 10.1093/oso/9780198788102.001.0001} {\emph
  {\bibinfo {title} {Statistical {{Field Theory}}: {{An Introduction}} to
  {{Exactly Solved Models}} in {{Statistical Physics}}}}},\ \bibinfo {edition}
  {2nd}\ ed.\ (\bibinfo  {publisher} {{Oxford University Press}},\ \bibinfo
  {year} {2020})\BibitemShut {NoStop}%
\bibitem [{\citenamefont {Ghoshal}\ and\ \citenamefont
  {Zamolodchikov}(1994)}]{ghoshal_boundary_1994}%
  \BibitemOpen
  \bibfield  {author} {\bibinfo {author} {\bibfnamefont {S.}~\bibnamefont
  {Ghoshal}}\ and\ \bibinfo {author} {\bibfnamefont {A.}~\bibnamefont
  {Zamolodchikov}},\ }\href {\doibase 10.1142/S0217751X94001552} {\bibfield
  {journal} {\bibinfo  {journal} {International Journal of Modern Physics A}\
  }\textbf {\bibinfo {volume} {09}},\ \bibinfo {pages} {3841} (\bibinfo {year}
  {1994})}\BibitemShut {NoStop}%
\bibitem [{\citenamefont {Ghoshal}(1994)}]{ghoshal_bound_1994}%
  \BibitemOpen
  \bibfield  {author} {\bibinfo {author} {\bibfnamefont {S.}~\bibnamefont
  {Ghoshal}},\ }\href {\doibase 10.1142/S0217751X94001941} {\bibfield
  {journal} {\bibinfo  {journal} {International Journal of Modern Physics A}\
  }\textbf {\bibinfo {volume} {09}},\ \bibinfo {pages} {4801} (\bibinfo {year}
  {1994})}\BibitemShut {NoStop}%
\bibitem [{\citenamefont {Mattsson}\ and\ \citenamefont
  {Dorey}(2000)}]{mattsson_boundary_2000}%
  \BibitemOpen
  \bibfield  {author} {\bibinfo {author} {\bibfnamefont {P.}~\bibnamefont
  {Mattsson}}\ and\ \bibinfo {author} {\bibfnamefont {P.}~\bibnamefont
  {Dorey}},\ }\href {\doibase 10.1088/0305-4470/33/49/304} {\bibfield
  {journal} {\bibinfo  {journal} {Journal of Physics A: Mathematical and
  General}\ }\textbf {\bibinfo {volume} {33}},\ \bibinfo {pages} {9065}
  (\bibinfo {year} {2000})}\BibitemShut {NoStop}%
\bibitem [{\citenamefont {{Castro-Alvaredo}}\ and\ \citenamefont
  {Doyon}(2008)}]{castro-alvaredo_bi-partite_2008}%
  \BibitemOpen
  \bibfield  {author} {\bibinfo {author} {\bibfnamefont {O.~A.}\ \bibnamefont
  {{Castro-Alvaredo}}}\ and\ \bibinfo {author} {\bibfnamefont {B.}~\bibnamefont
  {Doyon}},\ }\href {\doibase 10.1088/1751-8113/41/27/275203} {\bibfield
  {journal} {\bibinfo  {journal} {Journal of Physics A: Mathematical and
  Theoretical}\ }\textbf {\bibinfo {volume} {41}},\ \bibinfo {pages} {275203}
  (\bibinfo {year} {2008})}\BibitemShut {NoStop}%
\bibitem [{\citenamefont {Ercolessi}\ \emph {et~al.}(2010)\citenamefont
  {Ercolessi}, \citenamefont {Evangelisti},\ and\ \citenamefont
  {Ravanini}}]{ercolessi_exact_2010}%
  \BibitemOpen
  \bibfield  {author} {\bibinfo {author} {\bibfnamefont {E.}~\bibnamefont
  {Ercolessi}}, \bibinfo {author} {\bibfnamefont {S.}~\bibnamefont
  {Evangelisti}}, \ and\ \bibinfo {author} {\bibfnamefont {F.}~\bibnamefont
  {Ravanini}},\ }\href {\doibase 10.1016/j.physleta.2010.03.014} {\bibfield
  {journal} {\bibinfo  {journal} {Physics Letters A}\ }\textbf {\bibinfo
  {volume} {374}},\ \bibinfo {pages} {2101} (\bibinfo {year}
  {2010})}\BibitemShut {NoStop}%
\bibitem [{\citenamefont {{Castro-Alvaredo}}\ and\ \citenamefont
  {Horvath}(2021)}]{castro-alvaredo_branch_2021}%
  \BibitemOpen
  \bibfield  {author} {\bibinfo {author} {\bibfnamefont {O.}~\bibnamefont
  {{Castro-Alvaredo}}}\ and\ \bibinfo {author} {\bibfnamefont {D.}~\bibnamefont
  {Horvath}},\ }\href {\doibase 10.21468/SciPostPhys.10.6.132} {\bibfield
  {journal} {\bibinfo  {journal} {SciPost Physics}\ }\textbf {\bibinfo {volume}
  {10}},\ \bibinfo {pages} {132} (\bibinfo {year} {2021})}\BibitemShut
  {NoStop}%
\bibitem [{\citenamefont {Bisognano}\ and\ \citenamefont
  {Wichmann}(1975)}]{bisognano_duality_1975}%
  \BibitemOpen
  \bibfield  {author} {\bibinfo {author} {\bibfnamefont {J.~J.}\ \bibnamefont
  {Bisognano}}\ and\ \bibinfo {author} {\bibfnamefont {E.~H.}\ \bibnamefont
  {Wichmann}},\ }\href {\doibase 10.1063/1.522605} {\bibfield  {journal}
  {\bibinfo  {journal} {Journal of Mathematical Physics}\ }\textbf {\bibinfo
  {volume} {16}},\ \bibinfo {pages} {985} (\bibinfo {year} {1975})}\BibitemShut
  {NoStop}%
\bibitem [{\citenamefont {Bisognano}(1976)}]{bisognano_duality_1976}%
  \BibitemOpen
  \bibfield  {author} {\bibinfo {author} {\bibfnamefont {J.~J.}\ \bibnamefont
  {Bisognano}},\ }\href {\doibase 10.1063/1.522898} {\bibfield  {journal}
  {\bibinfo  {journal} {Journal of Mathematical Physics}\ }\textbf {\bibinfo
  {volume} {17}},\ \bibinfo {pages} {303} (\bibinfo {year} {1976})}\BibitemShut
  {NoStop}%
\bibitem [{\citenamefont {Cardy}\ and\ \citenamefont
  {Tonni}(2016)}]{cardy_entanglement_2016}%
  \BibitemOpen
  \bibfield  {author} {\bibinfo {author} {\bibfnamefont {J.}~\bibnamefont
  {Cardy}}\ and\ \bibinfo {author} {\bibfnamefont {E.}~\bibnamefont {Tonni}},\
  }\href {\doibase 10.1088/1742-5468/2016/12/123103} {\bibfield  {journal}
  {\bibinfo  {journal} {Journal of Statistical Mechanics: Theory and
  Experiment}\ }\textbf {\bibinfo {volume} {2016}},\ \bibinfo {pages} {123103}
  (\bibinfo {year} {2016})}\BibitemShut {NoStop}%
\bibitem [{\citenamefont {Wen}\ \emph {et~al.}(2018)\citenamefont {Wen},
  \citenamefont {Ryu},\ and\ \citenamefont {Ludwig}}]{wen_entanglement_2018}%
  \BibitemOpen
  \bibfield  {author} {\bibinfo {author} {\bibfnamefont {X.}~\bibnamefont
  {Wen}}, \bibinfo {author} {\bibfnamefont {S.}~\bibnamefont {Ryu}}, \ and\
  \bibinfo {author} {\bibfnamefont {A.~W.~W.}\ \bibnamefont {Ludwig}},\ }\href
  {\doibase 10.1088/1742-5468/aae84e} {\bibfield  {journal} {\bibinfo
  {journal} {Journal of Statistical Mechanics: Theory and Experiment}\ }\textbf
  {\bibinfo {volume} {2018}},\ \bibinfo {pages} {113103} (\bibinfo {year}
  {2018})}\BibitemShut {NoStop}%
\bibitem [{\citenamefont {Giudici}\ \emph {et~al.}(2018)\citenamefont
  {Giudici}, \citenamefont {{Mendes-Santos}}, \citenamefont {Calabrese},\ and\
  \citenamefont {Dalmonte}}]{giudici_entanglement_2018}%
  \BibitemOpen
  \bibfield  {author} {\bibinfo {author} {\bibfnamefont {G.}~\bibnamefont
  {Giudici}}, \bibinfo {author} {\bibfnamefont {T.}~\bibnamefont
  {{Mendes-Santos}}}, \bibinfo {author} {\bibfnamefont {P.}~\bibnamefont
  {Calabrese}}, \ and\ \bibinfo {author} {\bibfnamefont {M.}~\bibnamefont
  {Dalmonte}},\ }\href {\doibase 10.1103/PhysRevB.98.134403} {\bibfield
  {journal} {\bibinfo  {journal} {Physical Review B}\ }\textbf {\bibinfo
  {volume} {98}},\ \bibinfo {pages} {134403} (\bibinfo {year}
  {2018})}\BibitemShut {NoStop}%
\bibitem [{\citenamefont {Roy}\ \emph {et~al.}(2020)\citenamefont {Roy},
  \citenamefont {Pollmann},\ and\ \citenamefont
  {Saleur}}]{roy_entanglement_2020}%
  \BibitemOpen
  \bibfield  {author} {\bibinfo {author} {\bibfnamefont {A.}~\bibnamefont
  {Roy}}, \bibinfo {author} {\bibfnamefont {F.}~\bibnamefont {Pollmann}}, \
  and\ \bibinfo {author} {\bibfnamefont {H.}~\bibnamefont {Saleur}},\ }\href
  {\doibase 10.1088/1742-5468/aba498} {\bibfield  {journal} {\bibinfo
  {journal} {Journal of Statistical Mechanics: Theory and Experiment}\ }\textbf
  {\bibinfo {volume} {2020}},\ \bibinfo {pages} {083104} (\bibinfo {year}
  {2020})}\BibitemShut {NoStop}%
\bibitem [{\citenamefont {Dalmonte}\ \emph {et~al.}(2018)\citenamefont
  {Dalmonte}, \citenamefont {Vermersch},\ and\ \citenamefont
  {Zoller}}]{dalmonte_quantum_2018}%
  \BibitemOpen
  \bibfield  {author} {\bibinfo {author} {\bibfnamefont {M.}~\bibnamefont
  {Dalmonte}}, \bibinfo {author} {\bibfnamefont {B.}~\bibnamefont {Vermersch}},
  \ and\ \bibinfo {author} {\bibfnamefont {P.}~\bibnamefont {Zoller}},\ }\href
  {\doibase 10.1038/s41567-018-0151-7} {\bibfield  {journal} {\bibinfo
  {journal} {Nature Physics}\ }\textbf {\bibinfo {volume} {14}},\ \bibinfo
  {pages} {827} (\bibinfo {year} {2018})}\BibitemShut {NoStop}%
\bibitem [{\citenamefont {Kokail}\ \emph {et~al.}(2021)\citenamefont {Kokail},
  \citenamefont {{van Bijnen}}, \citenamefont {Elben}, \citenamefont
  {Vermersch},\ and\ \citenamefont {Zoller}}]{kokail_entanglement_2021}%
  \BibitemOpen
  \bibfield  {author} {\bibinfo {author} {\bibfnamefont {C.}~\bibnamefont
  {Kokail}}, \bibinfo {author} {\bibfnamefont {R.}~\bibnamefont {{van
  Bijnen}}}, \bibinfo {author} {\bibfnamefont {A.}~\bibnamefont {Elben}},
  \bibinfo {author} {\bibfnamefont {B.}~\bibnamefont {Vermersch}}, \ and\
  \bibinfo {author} {\bibfnamefont {P.}~\bibnamefont {Zoller}},\ }\href
  {\doibase 10.1038/s41567-021-01260-w} {\bibfield  {journal} {\bibinfo
  {journal} {Nature Physics}\ }\textbf {\bibinfo {volume} {17}},\ \bibinfo
  {pages} {936} (\bibinfo {year} {2021})}\BibitemShut {NoStop}%
\bibitem [{\citenamefont {Goldstein}\ and\ \citenamefont
  {Sela}(2018)}]{goldstein_symmetry-resolved_2018}%
  \BibitemOpen
  \bibfield  {author} {\bibinfo {author} {\bibfnamefont {M.}~\bibnamefont
  {Goldstein}}\ and\ \bibinfo {author} {\bibfnamefont {E.}~\bibnamefont
  {Sela}},\ }\href {\doibase 10.1103/PhysRevLett.120.200602} {\bibfield
  {journal} {\bibinfo  {journal} {Physical Review Letters}\ }\textbf {\bibinfo
  {volume} {120}},\ \bibinfo {pages} {200602} (\bibinfo {year}
  {2018})}\BibitemShut {NoStop}%
\bibitem [{\citenamefont {Lukin}\ \emph {et~al.}(2019)\citenamefont {Lukin},
  \citenamefont {Rispoli}, \citenamefont {Schittko}, \citenamefont {Tai},
  \citenamefont {Kaufman}, \citenamefont {Choi}, \citenamefont {Khemani},
  \citenamefont {L{\'e}onard},\ and\ \citenamefont
  {Greiner}}]{lukin_probing_2019}%
  \BibitemOpen
  \bibfield  {author} {\bibinfo {author} {\bibfnamefont {A.}~\bibnamefont
  {Lukin}}, \bibinfo {author} {\bibfnamefont {M.}~\bibnamefont {Rispoli}},
  \bibinfo {author} {\bibfnamefont {R.}~\bibnamefont {Schittko}}, \bibinfo
  {author} {\bibfnamefont {M.~E.}\ \bibnamefont {Tai}}, \bibinfo {author}
  {\bibfnamefont {A.~M.}\ \bibnamefont {Kaufman}}, \bibinfo {author}
  {\bibfnamefont {S.}~\bibnamefont {Choi}}, \bibinfo {author} {\bibfnamefont
  {V.}~\bibnamefont {Khemani}}, \bibinfo {author} {\bibfnamefont
  {J.}~\bibnamefont {L{\'e}onard}}, \ and\ \bibinfo {author} {\bibfnamefont
  {M.}~\bibnamefont {Greiner}},\ }\href {\doibase 10.1126/science.aau0818}
  {\bibfield  {journal} {\bibinfo  {journal} {Science}\ }\textbf {\bibinfo
  {volume} {364}},\ \bibinfo {pages} {256} (\bibinfo {year}
  {2019})}\BibitemShut {NoStop}%
\bibitem [{\citenamefont {Bonsignori}\ \emph {et~al.}(2019)\citenamefont
  {Bonsignori}, \citenamefont {Ruggiero},\ and\ \citenamefont
  {Calabrese}}]{bonsignori_symmetry_2019}%
  \BibitemOpen
  \bibfield  {author} {\bibinfo {author} {\bibfnamefont {R.}~\bibnamefont
  {Bonsignori}}, \bibinfo {author} {\bibfnamefont {P.}~\bibnamefont
  {Ruggiero}}, \ and\ \bibinfo {author} {\bibfnamefont {P.}~\bibnamefont
  {Calabrese}},\ }\href {\doibase 10.1088/1751-8121/ab4b77} {\bibfield
  {journal} {\bibinfo  {journal} {Journal of Physics A: Mathematical and
  Theoretical}\ }\textbf {\bibinfo {volume} {52}},\ \bibinfo {pages} {475302}
  (\bibinfo {year} {2019})}\BibitemShut {NoStop}%
\bibitem [{\citenamefont {Fraenkel}\ and\ \citenamefont
  {Goldstein}(2020)}]{fraenkel_symmetry_2020}%
  \BibitemOpen
  \bibfield  {author} {\bibinfo {author} {\bibfnamefont {S.}~\bibnamefont
  {Fraenkel}}\ and\ \bibinfo {author} {\bibfnamefont {M.}~\bibnamefont
  {Goldstein}},\ }\href {\doibase 10.1088/1742-5468/ab7753} {\bibfield
  {journal} {\bibinfo  {journal} {Journal of Statistical Mechanics: Theory and
  Experiment}\ }\textbf {\bibinfo {volume} {2020}},\ \bibinfo {pages} {033106}
  (\bibinfo {year} {2020})}\BibitemShut {NoStop}%
\bibitem [{\citenamefont {Azses}\ and\ \citenamefont
  {Sela}(2020)}]{azses_symmetry-resolved_2020}%
  \BibitemOpen
  \bibfield  {author} {\bibinfo {author} {\bibfnamefont {D.}~\bibnamefont
  {Azses}}\ and\ \bibinfo {author} {\bibfnamefont {E.}~\bibnamefont {Sela}},\
  }\href {\doibase 10.1103/PhysRevB.102.235157} {\bibfield  {journal} {\bibinfo
   {journal} {Physical Review B}\ }\textbf {\bibinfo {volume} {102}},\ \bibinfo
  {pages} {235157} (\bibinfo {year} {2020})}\BibitemShut {NoStop}%
\bibitem [{\citenamefont {Parez}\ \emph {et~al.}(2021)\citenamefont {Parez},
  \citenamefont {Bonsignori},\ and\ \citenamefont
  {Calabrese}}]{parez_exact_2021}%
  \BibitemOpen
  \bibfield  {author} {\bibinfo {author} {\bibfnamefont {G.}~\bibnamefont
  {Parez}}, \bibinfo {author} {\bibfnamefont {R.}~\bibnamefont {Bonsignori}}, \
  and\ \bibinfo {author} {\bibfnamefont {P.}~\bibnamefont {Calabrese}},\ }\href
  {\doibase 10.1088/1742-5468/ac21d7} {\bibfield  {journal} {\bibinfo
  {journal} {Journal of Statistical Mechanics: Theory and Experiment}\ }\textbf
  {\bibinfo {volume} {2021}},\ \bibinfo {pages} {093102} (\bibinfo {year}
  {2021})}\BibitemShut {NoStop}%
\bibitem [{\citenamefont {Weisenberger}\ \emph {et~al.}(2021)\citenamefont
  {Weisenberger}, \citenamefont {Zhao}, \citenamefont {Northe},\ and\
  \citenamefont {Meyer}}]{weisenberger_symmetry-resolved_2021}%
  \BibitemOpen
  \bibfield  {author} {\bibinfo {author} {\bibfnamefont {K.}~\bibnamefont
  {Weisenberger}}, \bibinfo {author} {\bibfnamefont {S.}~\bibnamefont {Zhao}},
  \bibinfo {author} {\bibfnamefont {C.}~\bibnamefont {Northe}}, \ and\ \bibinfo
  {author} {\bibfnamefont {R.}~\bibnamefont {Meyer}},\ }\href {\doibase
  10.1007/JHEP12(2021)104} {\bibfield  {journal} {\bibinfo  {journal} {Journal
  of High Energy Physics}\ }\textbf {\bibinfo {volume} {2021}},\ \bibinfo
  {pages} {104} (\bibinfo {year} {2021})}\BibitemShut {NoStop}%
\bibitem [{\citenamefont {Calabrese}\ \emph {et~al.}(2021)\citenamefont
  {Calabrese}, \citenamefont {Dubail},\ and\ \citenamefont
  {Murciano}}]{calabrese_symmetry-resolved_2021}%
  \BibitemOpen
  \bibfield  {author} {\bibinfo {author} {\bibfnamefont {P.}~\bibnamefont
  {Calabrese}}, \bibinfo {author} {\bibfnamefont {J.}~\bibnamefont {Dubail}}, \
  and\ \bibinfo {author} {\bibfnamefont {S.}~\bibnamefont {Murciano}},\ }\href
  {\doibase 10.1007/JHEP10(2021)067} {\bibfield  {journal} {\bibinfo  {journal}
  {Journal of High Energy Physics}\ }\textbf {\bibinfo {volume} {2021}},\
  \bibinfo {pages} {67} (\bibinfo {year} {2021})}\BibitemShut {NoStop}%
\bibitem [{\citenamefont {Zamolodchikov}(1995)}]{zamolodchikov_mass_1995}%
  \BibitemOpen
  \bibfield  {author} {\bibinfo {author} {\bibfnamefont {A.~B.}\ \bibnamefont
  {Zamolodchikov}},\ }\href {\doibase 10.1142/S0217751X9500053X} {\bibfield
  {journal} {\bibinfo  {journal} {International Journal of Modern Physics A}\
  }\textbf {\bibinfo {volume} {10}},\ \bibinfo {pages} {1125} (\bibinfo {year}
  {1995})}\BibitemShut {NoStop}%
\end{thebibliography}%
